\documentclass[aps,10pt,eqsecnum,preprint,nofootinbib,superscriptaddress]{revtex4}
\usepackage{amssymb,amsmath,amsthm,graphicx,amscd,mathtools}
\usepackage[mathscr]{eucal}
\usepackage{upgreek,enumerate,color,verbatim,comment,ulem,bigints,stackrel,tabstackengine}
\usepackage{bm,pbox,cancel}
\usepackage[hidelinks]{hyperref}
\usepackage{orcidlink}

\stackMath

\begin{document}
\title{Fluctuations-Induced Quantum Radiation and Reaction  \\ from an Atom in a Squeezed Quantum Field}   
\author{Matthew Bravo}
\email{mbravo@terpmail.umd.edu}
\affiliation{Department of Physics, University of Maryland, College Park, Maryland 20742, USA}
\author{Jen-Tsung Hsiang\orcidlink{0000-0002-9801-208X}}
\email{cosmology@gmail.com}
\affiliation{Center for High Energy and High Field Physics, National Central University, Taoyuan 320317, Taiwan, R.O.C.}
\author{Bei-Lok Hu\orcidlink{0000-0003-2489-9914}}
\email{blhu@umd.edu}
\affiliation{Maryland Center for Fundamental Physics and Joint Quantum Institute,  University of Maryland, College Park, Maryland 20742, USA}
\date{\small \today}
%\\ Invited refereed paper for Physics -- Special Issue on Vacuum Fluctuations eds G. J. Maclay and R. Passante}}
\begin{abstract}
In this third of a series on quantum radiation, we further explore the feasibility of using the memories (nonMarkovianity) kept in a quantum field to decipher certain information about the early universe.  As a model study, we let a massless quantum field be subjected to a parametric process for  a finite time interval such that the mode frequency of the field transits from one constant value to another. This configuration thus mimics a statically-bounded universe,  where there is an in and out state with the scale factor approaching constant, not a continuously evolving one.  The field subjected to squeezing by this process should contain some information of the process itself. If an atom is coupled to the field after the parametric process, its response will depend on the squeezing, and any quantum radiation emitted by the atom will carry this information away so that an observer at a much later time may still identify it. Our analyses show that 1) a remote observer cannot measure the generated squeezing via the radiation energy flux from the atom because the net radiation energy flux is canceled due to the correlation between the radiation field from the atom and the free field at the observer's location. However, 2) there is a chance to identify squeezing by measuring the constant radiation energy density at late times. The only restriction is that this energy density is of the near-field nature and only an observer close to the atom can use it to unravel the information of squeezing. The second part of this paper focuses on 3) the dependence of squeezing on the functional form of the parametric process. By explicitly working out several examples we demonstrate that the behavior of
squeezing does reflect essential properties of the parametric process. In fact, striking features may show up in more complicated processes involving various scales. These analyses allow us to establish the connection between properties of a squeezed quantum field and details of the parametric process which does the squeezing. Therefore,  4) one can construct templates  to reconstitute the unknown parametric processes from the data of measurable quantities subjected to squeezing. In a sequel paper these results will be applied to a study of quantum radiations in cosmology.
\end{abstract}
\maketitle

\hypersetup{linktoc=all}
\baselineskip=18pt
\allowdisplaybreaks

\section{Introduction}
This paper is the third of a series on {\it quantum radiation}, in the form of emitted radiation with an energy flux (distinct from thermal radiance felt by an accelerated atom, as in the Unruh effect \cite{Unr76}) from the internal degrees of freedom (idf) of an atom, tracing its origin to the {\it vacuum fluctuations} of a quantum field, including its back-reaction on the idf dynamics of the atom, in the form of {\it quantum dissipation}. In \cite{QRadVac} two of us considered how vacuum fluctuations in the field act on the idf of an atom (we may call this the `emitter'), modeled by a harmonic oscillator. We first showed how a stochastic component of the internal dynamics of the atom arises from the vacuum fluctuations of the field, resulting in the emittance of quantum radiation.  We then show  how the back-reaction of this quantum radiation induces   quantum dissipation  in the atom's idf dynamics.  We  explicitly identified  the different terms representing these processes  in the Langevin equations of motion.  Then, using the example of a stationary atom, we show how in this simple case the absence of radiation  at a far away observation point where a probe (or detector -- note the so-called  Unruh-DeWitt `detector' \cite{Unr76,DeW79} is an emitter in the present context) is located is actually a result of complex cancellations of the interference between emitted radiation from the atom's idf  and the local fluctuations in the free field. In so doing we point out that the entity which enters into the duality relation with vacuum fluctuations is not {\it radiation reaction}\footnote{In the quantum optics literature, e.g., \cite{RadReact,RadReact1,RadReact2,RadReact3}, the relation between quantum fluctuations and radiation reaction is often mentioned without emphasizing the difference between classical radiation reaction~\cite{Jackson, Rohrich} and quantum dissipation, which exist at two separate theoretical levels. Only quantum dissipation enters in the fluctuations-dissipation relation with quantum fluctuations, not classical radiation reaction.}, which can exist as a classical entity, but quantum dissipation \cite{JH1,JHUnr}.  In the second paper \cite{QRadCoh} we considered the idf of the atom interacting with a quantum scalar field initially in a coherent state. We   showed  how the deterministic mean field drives the internal classical mean component to emit classical radiation and receive classical radiation reaction. Both components are statistically distinct and fully decoupled. It is clearly seen that the effects of the vacuum fluctuations of the field are matched with those of quantum radiation reaction, not with classical radiation reaction, as the folklore goes, even promulgated in some textbooks. Furthermore, we identified the reason why quantum radiation from a stationary emitter is not observed, and a probe  located far away only sees classical radiation.

\subsection{Quantum Radiation from an Atom in a Squeezed Quantum Field}

In this paper we treat quantum radiation from an atom's idf interacting with a quantum field in a squeezed state. The squeezed state is probably the most important quantum state,  next to the vacuum state, with both rich theoretical meanings and broad practical applications, as well-known in quantum optics. (See, e.g.,  \cite{SqBooks}). We are particularly interested in its role in cosmology of the early universe.   

 \paragraph{Quantum field squeezed by an expanding universe}
 
Squeezed states of a quantum field  are naturally produced in an expanding universe in fundamental processes  which involve the parametric amplification  of quantum fluctuations, such as particle creation \cite{GriSid,HKM94}, either spontaneous production from the vacuum  or stimulated production from $n$-particle states, and structure formation \cite{StrFor}  from quantum fluctuations of the inflaton field. From relics such as primordial radiation and matter contents observed today, with the help of theoretical models governing their evolution, one attempts to deduce the state of the early universe at different stages of development. In addition to particle creation and structure formation we add here another fundamental quantum process, namely, quantum radiation. One can ask questions like, if such radiation of a quantum nature is detected, how can one use it to reveal certain quantum aspects of the early universe?  Revealing secrets of the early universe by digging out information  stored in the quantum field is 
%in parallel to questions related to quantum processes such as particle creation or structure formation, and 
in a similar spirit as the quest we initiated about the nonMarkovianity of the universe through the  memory kept in the quantum field \cite{nMCos}.  

\subsection{Three components: Radiation, Squeeze, Drive}

Towards such a goal we carry out this investigation which involves three components: 1) Quantum radiation 2) Squeezed state,  3) Driven dynamics, either under some external force  or as provided by an expanding universe, which describes how the squeeze parameter changes in time. Part 1)  was initiated in \cite{QRadVac} and continued in \cite{QRadCoh},  where the required technical tools for the present investigation can be found.

Part 2) is performed here:  we  consider a quantum field in a squeezed state with a fixed squeeze parameter, regarded as the end state of the quantum field after being squeezed under some drive protocol or cosmological evolution. For the description of squeezed states we shall follow the descriptions in our earlier paper \cite{FDRSq}.  Under the classification of the three types of squeezing described there, we shall adhere to the first type, namely, with fixed squeezing.  The second type of dynamical parametric  squeezing will be treated in our sequel cosmological papers. We shall set aside the third type of squeezing due to finite coupling between the system (here the atom's idf) and the environment (here the quantum field) completely.  
  
 To see more explicitly what Part 3) entails, consider a situation where the atom at the initial time $t_i$ is in a quantum field in a squeezed state with a fixed squeeze parameter $\zeta_i$  and the same atom  at the final time $t_f$  in a squeezed field with a different squeeze parameter $\zeta_f$. If an atom emits quantum radiation at either or both the initial and the final times, comparing the signals from both should tell us something about how the drive had affected the atom through the quantum radiation emitted from the atom, or in cosmology, how the universe had evolved as measured by the parametric squeezing of the field. The more challenging situation is if it turns out that there is no emitted quantum radiation from an atom in a squeezed field.  This is what our next work intends to find out.

\subsection{Our objectives, in two stages}

{We wish to ask questions in the same spirit as in \cite{nMCos}: can we extract information about the history of the quantum field which had undergone a parametric process from the responses of an atom coupled to it  in the epoch after  the parametric process?  In particular,  for observers (receivers) situated far away from the atom (emitter) whether they can detect radiations emitted  from the atom of a quantum nature. Quantum radiation is of special interest as it originates from the vacuum fluctuations of a quantum field, and is expected to keep some memories of the parametric process it went through. In a cosmological context it acts as a carrier of  quantum information about the early universe.} 

 We divide our program of quantum radiation in cosmology into two stages, phrased as two questions. A) Is there emitted quantum radiation from a stationary atom in a  quantum field with a fixed squeeze parameter? If there is no emitted quantum radiation, then question B) Is there emitted quantum radiation from a stationary atom in a quantum field subjected to a changing squeeze parameter, such as in an expanding universe?  If so, what kind of evolutionary dynamics would produce what types of emitted radiation and with what magnitudes?  The first stage addresses components 1) and 2) listed above, the second stage, components 2) and 3), {which will be continued in a follow-up paper.}
 
 \paragraph{Radiation pattern as template for squeezing}
 
 This paper operates in the first stage, with a set up  {meant for} a statically-bounded universe, not the continuously evolving type such as in the Friedmann–Lema{\^i}tre–Robertson–Walker (FLRW) or the inflationary universe. The first half answers the first question A), 
 %{in the negative -- there is no radiation flux even though the free component of the field is squeezed and nonstationary. Nevertheless, the radiation energy density contains useful information.} 
 and the second half of this paper examines the response of the atom coupled to a field that had undergone a parametric process earlier.   This evokes the template idea, i.e.,  one can use the response of the atom, coupled to the field in the out-region, to identify the dependence of the squeeze parameter on the earlier parametric process before the out-region.  In the cosmology context, this offers a way for a late time observer of such quantum phenomena to uncover how the universe had evolved in much earlier times. 
   
\paragraph{Stress-energy tensor of squeezed field}

To answer the central question A),  {we use a simplified model in which the quantum field undergoes a parametric process of finite duration, such as under an external drive for a definite period of time, or in an asymptotically stationary (statically bounded) universe. The field could be a quantized matter or graviton field, or an inflaton field whose quantum fluctuations engender cosmological structures.} 
{Then we calculate the expectation value of the stress-energy tensor of the radiation field emitted by an atom coupled to a free field in a fixed two-mode squeeze state. In order to identify the unambiguous signals and conform to the typical settings, we focus on the late-time results. We learned from our earlier work \cite{QRadVac} for a quantum field in the vacuum state  that the procedure for checking this is quite involved,  as it entails both the radiation flux emitted from the atom as detected at the spacetime point of the probe, and an incoming flux from infinity.} 

\subsection{Key steps and major findings}

{The answer we found after calculating all relevant contributions for an atom in a quantum field in a squeezed state shows that there is no net radiation flux, same as in~\cite{QRadVac}. This is a  consequence of relaxation dynamics of the atom's internal dynamics when it is coupled to the squeezed field~\cite{AHH22}. However, if the probe can measure the radiation energy density, it should obtain a residual constant radiation energy density at late time. Its value will fall off like the inverse cubic power of the distance between the probe and the atom. Thus it is more like a near-field effect. Nonetheless this energy density has an interesting characteristic: it depends on the squeeze parameter, which is what we are after.}

{Thus, our investigation turns to whether and how the squeeze parameter of the field after the parametric process would depend on the details of the process. We first show that the squeeze parameter can formally be expressed by the fundamental solutions of the parametrically driven field, the latter contains useful information about the process. {Then by working out several examples numerically we can make the following observations}:

\begin{enumerate}[1)]
    \item For a monotonically varying process, the squeeze parameter has a monotonic dependence on the duration of this process; it does not depend on when the process starts, if we fix the duration.
    
    \item The magnitude of squeezing is related to the rate of change in the process. That is, large squeezing can be induced from a non-adiabatic transition. This is consistent with our understanding of spontaneous particle creation from parametric amplification of  vacuum fluctuations~\cite{Parker} and that copious particles can be produced at the Planck time under rapid expansion of the universe~\cite{Zeldovich}. Thus we expect that  non-adiabatic processes may contribute to larger residual radiation energy density around the atom. 
    
    \item For a non-monotonic parametric process, various scales in the process induces richer structure to the behavior of the squeeze parameter. In particular, if the parametric process changes with time sinusoidally {at some frequency range, it may induce parametric resonance and yield exceptionally large squeezing in the out-region}.
\end{enumerate}
Similar considerations can likewise be  applied to the frequency spectrum of the squeeze parameter of the field in the out-region. One can then examine its dependence on the parametric process which the field has gone through. This illustrates the way to obtain templates in how the squeeze parameter is related to the parametric process, and how certain information of the unknown parametric process can be inferred from these templates.

\subsection{Organization} 
The paper is organized as follows. In Sec.~\ref{S:hgddfd}, as a prerequisite, we briefly summarize our earlier results on the relaxation process of the atom-field interaction, and pose the questions we would like to answer in this paper. In Sec.~\ref{S:eoired}, we lay out the formalism and the essential tools for detailing the nonequilibrium evolution of the atom's internal dynamics and the squeezed field when they are coupled together. In Sec.~\ref{S:eiutie} we study the general spatial-temporal behavior of the energy flux and the energy density of the radiation field and examine their late-time behaviors. In Sec.~\ref{S:rthgsdfg} we turn to the functional dependence of the squeeze parameter on the functional form of the parametric process. Several  examples are provided to illustrate the formal analysis. In Sec.~\ref{S:etudguei} we give some concluding remarks. Appendix~\ref{S:eoueisf} offers a succinct summary of the two-mode squeezed state. In Appendices~\ref{S:vbkdf} and \ref{S:evsfdjg}, we offer more details on the late-time, large distance behaviors of the energy flux and the energy density of the radiation field. In Appendices~\ref{S:eerudj}, we discuss the time-translational invariance of the squeeze parameter in the out-region.

%\section{Discussions and Prospects} 

%The grand challenge we pose here,  at least theoretically, is to find ways to enable one at a later time to  decipher the quantum state of the field at an earlier time?  We study quantum radiation emitted from atoms in a squeezed quantum field as a means to do so, similar to particle creation and structure formation being treated as a messenger or encoder of the primordial universe.  
 
%Our focus here has been quantum radiation from a stationary atom, like our earlier two papers,  but  in a squeezed quantum field. This work can be considered as a necessary preparation for the treatment of quantum radiation from atoms which have undergone dynamical evolution as in cosmology. 

% We know that the emitted quantum radiation from the atom is zero at both the initial and the final times. However, one can calculate  the changes in the atom in the intermediate times between the initial and the final field configurations.  The first part of our studies in this paper deals with $\xi$ at  $t_f$, while the second part we studied three different scenarios of how the squeeze parameter  $\xi$ varies from  $t_i$ and  $t_f$.   ...

\begin{figure}
	\centering
	\includegraphics[width=0.5\textwidth]{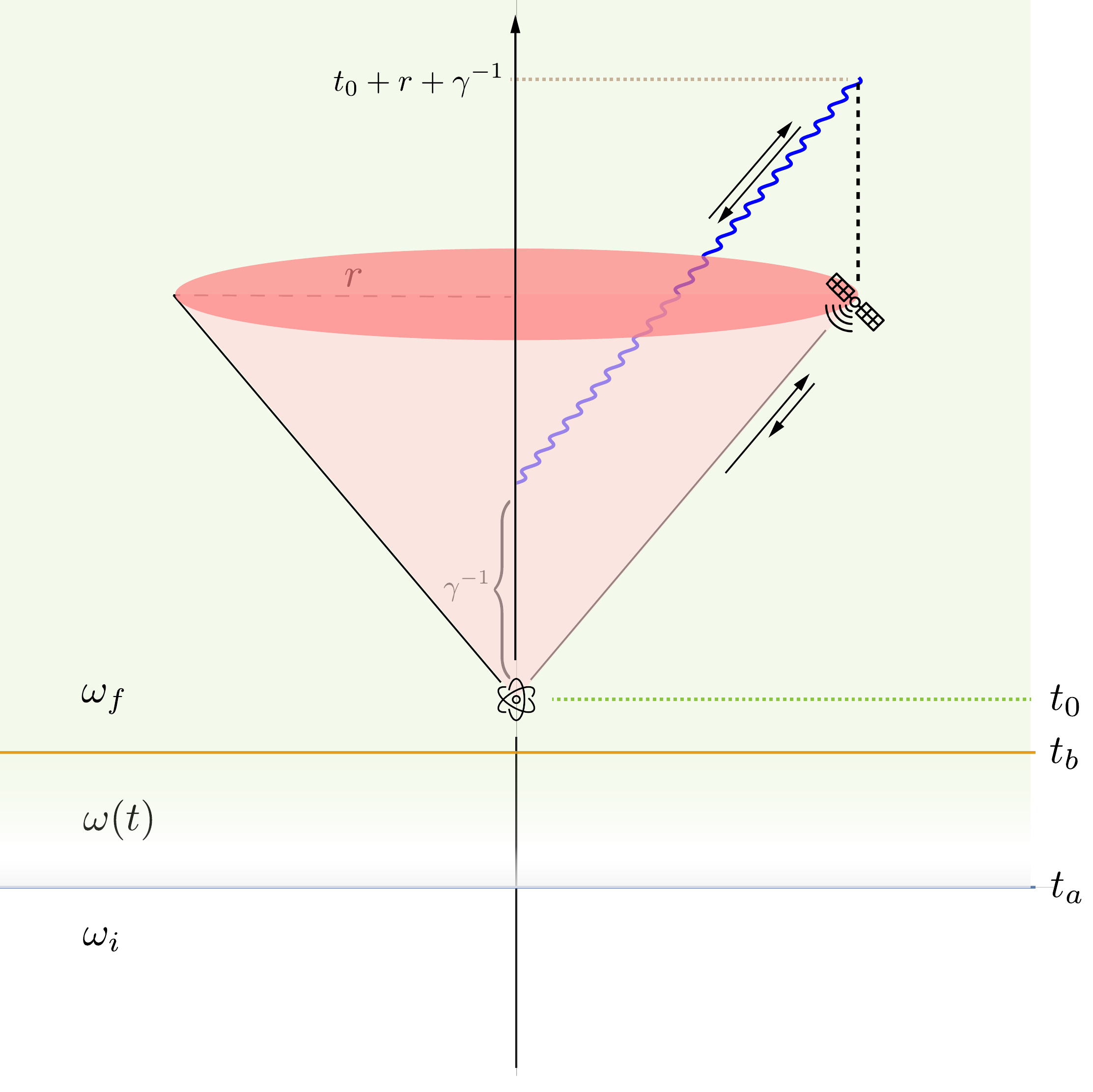}
	\caption{The configuration of the parametric process of the field. The process changes one of the mode frequencies from $\omega_{i}$ at time $t\leq t_{a}$ to $\omega_{f}$ at $t\geq t_{b}$. The transition is assumed to happen continuously. The interaction between an atom and the field is turned on at $t_{0}>t_{a}$ and radiation is generated by the internal motion of the atom. For a receiver at distance $r$ away from the atom, it will take another time $r$ to meet the radiation. The relaxation time of the atom's internal motion is of the order of $\gamma^{-1}$.}\label{Fi:config}
\end{figure}

\section{Scenario: Quantum Radiation in Atom-Field Systems}\label{S:hgddfd}

{Classical radiation is a familiar subject. But what is its quantum field origin? Can one trace back its link all the way to quantum fluctuations? We obtained partial  answers  in our last two papers to this question for a harmonic  atom interacting with a quantum field in a vacuum and in a coherent states. This paper deals with a squeezed field, necessary for treating cosmological quantum processes.  Since fluctuation-induced quantum radiation is not a household topic yet, it might be useful to first provide a physical picture of the global landscape of quantum radiation based on our understanding from  previous works, setting the stage for the current work.}

It has been shown that~\cite{AHH22} when the internal degree of freedom, modeled as a harmonic oscillator, of an atom (called a harmonic atom) in any initial state, is linearly coupled to a  massless linear scalar field in the stationary state, its motion will settle down to an equilibrium state. The presence of this equilibrium state, from the perspective of the atom, implies a balance of energy exchange between the atom and the environmental field. Expectantly,  the radiation generated by the nonequilibrium motion of the atom's internal degree of freedom propagates outward to spatial infinity. It is  lesser known whether and how, from the perspective of the field, the radiation energy from the atom is balanced. In our previous works~\cite{QRadVac}, we have demonstrated that the correlation between the outward radiation field and the local free field at spatial infinity constitutes an inward energy flow to balance the outward radiation energy flow. Furthermore this inward flux serves to supplement the needed energy around the atom for the field to drive the atom's internal motion. Thus we clearly see how energy flows from the atom to spatial infinity and then back-flows to the atom. This is a consequence of the nonequilibrium fluctuation-dissipation relations~\cite{QRadVac,HCSH18} for the atom's internal degree of freedom and the free field. This is a stronger condition than the conservation of energy. This point is better appreciated once we take the global view of the total system, composed of the atom and field. The entire system is closed and the total energy is conserved, but this does not guarantee the energy exchange between two subsystems is balanced unless the reduced dynamics is fully relaxed to an equilibrium state.

If, instead, the internal degree of freedom of the atom is initially coupled to the quantum field in a nonstationary state, like the squeezed state, then will the nonstationary nature of the field prevent the internal degree of freedom from approaching an equilibrium state? We have shown that \cite{AHH22} when the (mode-dependent) squeeze parameters is time-independent, the atom's internal degree of freedom still settles down to an equilibrium state in the long run. The stationary components of the covariance matrix elements or the energy flows decay exponentially fast to time-independent constants as in the previous case, but the additional nonstationary components  fall off to zero. The mechanisms that account for the late-time behavior of the stationary and the nonstationary components seem to be quite distinct. For the stationary components, it is a consequence of dissipation dynamically adapting to the driving fluctuations from the environment, but for  the nonstationary components, cancellation due to fast phase variations in the nonstationary terms plays the decisive role. In contrast, from the viewpoint of the squeezed field, what is the nature of radiation emitted from the atom, and can this outward energy flux at distances far away from the atom find a corresponding inward flux at late times such that there is no net energy output to spatial infinity, as proven for the case of a stationary field? If so, what makes it possible?

These questions are of particular interest  in a cosmological setting for the consideration of fundamental issues described in the Introduction.  There,  evolution of the universe parametrically drives the ambient quantum field into a squeezed state. The extent of squeezing may depend on the characteristics of the parametric process actuated by the evolution of the background universe. When an atom is coupled to such a field, emitted radiation from the atom should carry   the information of squeezing of the field, which in turn may reveal the  evolution history of the universe.

As a prerequisite to addressing these issues, in this paper we will examine a simpler configuration that may cover the essential features. Consider in Minkowski spacetime, a massless scalar field undergoes a parametric process such that each mode frequency changes smoothly from one constant value, say $\omega_i$, to another $\omega_f$, for example, as shown in Fig.~\ref{Fi:config}. The parametric process occurs during the time interval $t_a\leq t\leq t_b$. In the out-region $(t\geq t_b)$ of the process, an atom is coupled to the squeezed field at time $t_0$. The nonequilibrium evolution of the internal degree of freedom of the atom then generates outward radiation, which after time $r$ will reach a detector (like a satellite around the earth) far distance $r$ away from the atom. The detector may measure the time variation of the energy flux. The corresponding signal will be extracted and amplified to sieve out the information regarding the parametric process of the field occurred earlier.

\section{Massless Scalar Field interacting with a Harmonic Atom}\label{S:eoired}

For the case of a massless scalar field $\phi$ coupled to a static atom, with its internal degree of freedom $\chi$ represented by a harmonic oscillator, we have the following Heisenberg equations of motion 
\begin{align}
    \ddot{\hat{\chi}}(t) + \omega_{\textsc{b}}^2 \hat{\chi}(t) &= \frac{e}{m}\hat{\phi}(\bm{0}, t)\,,\label{E:dkddhn1}
    \\
    \bigl(\partial_t^2 - \nabla^2  \bigr)\hat{\phi}(\bm{x}, t) &=e\, \delta^{(3)}(\bm{x}) \hat{\chi}(t)\,,\label{E:dkddhn2}
\end{align}
where the atom is located at the origin of the spatial coordinates $\bm{x}$ in a $3+1$ Minkowski spacetime. The parameter $e$ denotes the coupling strength between the internal degree of freedom of the atom and the field, and $\omega_{\textsc{b}}$ is the bare frequency of the oscillator.  Solving the equation of  motion of the field \eqref{E:dkddhn1} yields
\begin{equation}\label{E:dkbsieus}
    \hat{\phi}(\bm{x}, t) = \hat{\phi}_{\mathrm{h}}(\bm{x}, t) + e \int_0^t\!ds \; G^{(\phi)}_{\text{R}, 0}(\bm{x}, t ; \bm{0}, s) \hat{\chi}(s)\,,
\end{equation}
where the homogeneous part $\hat{\phi}_{\mathrm{h}}(\bm{x}, t)$ gives the free field while the integral expression represents the radiation field emitted by the atom. The retarded Green's function $G^{(\phi)}_{\text{R}, 0}(x, x')$ of the free field is defined by 
\begin{equation}
    G^{(\phi)}_{\text{R}, 0}(x, x') =i\,\theta(t-t')\,\bigl[\hat{\phi}_{\mathrm{h}}(x),\,\hat{\phi}_{\mathrm{h}}(x')\bigr]=\frac{1}{2\pi} \theta(\tau) \delta (\tau^2 - R^2)\,,
\end{equation}
where $x=(\bm{x},t)$, $\tau=t-t'$, $R=\lvert\bm{x}-\bm{x}'\rvert$, and $\theta(\tau)$ is the Heaviside unit-step function. In the special case $R \to 0$ which is met for a single Brownian oscillator, we instead have 
\begin{equation}
    G^{(\phi)}_{\text{R}, 0}(t, t')= - \frac{1}{2\pi} \theta(\tau) \delta'(\tau)\,.
\end{equation}
Here we have suppressed the trivial spatial coordinates $\bm{0}$. The superscript of the Green's function indicates the operator by which the Green's function is constructed, while the $0$ in the subscript reminds that the Green's function of interest is associated with a free operator; otherwise it is an interacting operator. For example, $G^{(\phi)}_{\text{R}, 0}(\tau)$ refers to the retarded Green's function of the free field while $G^{(\chi)}_{\text{R}}(\tau)$ is the retarded Green's function of the interacting $\chi$.

Plugging Eq.~\eqref{E:dkbsieus} into the equation of motion \eqref{E:dkddhn2} for the atom's internal degree of freedom, we get 
\begin{equation}
    \ddot{\hat{\chi}}(t) + \omega_{\textsc{b}}^2 \hat{\chi}(t) - \frac{e^2}{m} \int_0^t\!ds \; G^{(\phi)}_{\text{R}, 0}(t, s) \hat{\chi}(s) = \frac{e}{m}\hat{\phi}_{\mathrm{h}}(\bm{0}, t)\,.\label{E:kfbsut}
\end{equation}
This is the generalized quantum Langevin equation for $\hat{\chi}(t)$  -- it has included backreactions of the environmental field in terms of the noise force on the righthand side and the integral expression on the left hand side. As has been identified in \eqref{E:dkbsieus}, the latter is the reaction force due to emitted radiation which retards the motion of the atom's internal degree of freedom.

Suppose the free field has a Markovian spectrum, we can reduce \eqref{E:kfbsut} into a local form
\begin{equation}\label{E:bvkesfds}
      \ddot{\hat{\chi}}(t) + 2\gamma \dot{\hat{\chi}}(t) + \omega^2_{\textsc{r}}\, \hat{\chi}(t) = \frac{e}{m}\hat{\phi}_{\mathrm{h}}(\bm{0}, t)\,,
\end{equation}
and the oscillation frequency is renormalized to a physical value $\omega_{\textsc{r}}$. Since we want to examine the nature of radiation and its late time behavior in relation to the atom's internal dynamics in a self-consistent way, we need the field expression \eqref{E:dkbsieus} written in terms of the solution to the equation of the reduced dynamics \eqref{E:bvkesfds}
\begin{equation}\label{E:kdgfjvgcs}
        \hat{\chi}(t) = \hat{\chi}_{\mathrm{h}}(t) + e\int_0^t\!ds \; G^{(\chi)}_{\textsc{R}}(t-s) \hat{\phi}_{\mathrm{h}}(\bm{0}, s)\,,
\end{equation}
where the retarded Green's function associated with $\hat{\chi}$ has the form
\begin{equation}
    G^{(\chi)}_{\textsc{R}}(\tau)=\frac{1}{m\Omega}\,e^{-\gamma\tau}\sin\Omega\tau\,,
\end{equation}
with $\Omega=\sqrt{\omega_{\textsc{r}}^2-\gamma^2}$, and $\hat{\chi}_{\mathrm{h}}(t)$ is the homogeneous part. Eq.~\eqref{E:bvkesfds} always implies
\begin{equation}\label{E:fdgeiuier}
    \frac{d}{dt}\Bigl[\frac{m}{2}\langle\dot{\hat{\chi}}(t)\rangle+\frac{m\omega_{\textsc{r}}}{2}\langle\hat{\chi}^2(t)\rangle\Bigr]=P_{\xi}(t)+P_{\gamma}(t)\,,
\end{equation}
where $P_{\xi}(t)$ is the power delivered by the free field fluctuations
\begin{equation}
    P_{\xi}(t)=\frac{e}{2}\langle\bigl\{\hat{\phi}_{\mathrm{h}}(\bm{0},t),\,\dot{\chi}(t)\bigr\}\rangle\,,
\end{equation}
and $P_{\gamma}(t)$ represents the rate of energy lost due to dissipation
\begin{equation}
    P_{\gamma}(t)=-2m\gamma\,\langle\dot{\chi}^2(t)\rangle\,.
\end{equation}
We have shown~\cite{AHH22} that even if the scalar field is initially in a nonstationary squeezed thermal state, the internal dynamics of the atom will still settle down to an asymptotic equilibrium state, where, in particular,  
\begin{equation}
    P_{\xi}(t)+P_{\gamma}(t)=0\,.
\end{equation}
This is a stronger condition than \eqref{E:fdgeiuier}, which reflects   energy conservation of the reduced dynamics. 

From \eqref{E:dkbsieus} and \eqref{E:kdgfjvgcs}, the full interacting field is then given by
\begin{align}
        \hat{\phi}(x) & = \hat{\phi}_{\mathrm{h}}(x) + e \int_0^t\! dt' \; G^{(\phi)}_{\textsc{R}, 0}(\bm{x}, t ; \bm{0}, t')\biggl[\hat{\chi}_{\mathrm{h}}(t') + e\int_0^{t'}\! ds \; G^{(\chi)}_{\textsc{R}}(t'-s) \hat{\phi}_{\mathrm{h}}(\bm{0}, s)  \biggr]\,.
\end{align}
It will turn out convenient to split the full interacting field into three physically distinct components, $\hat{\phi}(x)=\hat{\phi}_{\mathrm{h}}(x)+\hat{\phi}_{\textsc{tr}}(x)+\hat{\phi}_{\textsc{br}}(x)$ with
\begin{align}
    \hat{\phi}_{\textsc{tr}}(x)&=e \int_0^t\! dt' \; G^{(\phi)}_{\textsc{R}, 0}(\bm{x}, t ; \bm{0}, t')\,\hat{\chi}_{\mathrm{h}}(t')\,,\label{E:dkkdg1}\\
    \hat{\phi}_{\textsc{br}}(x)&=e^2\int_0^t\! dt' \; G^{(\phi)}_{\textsc{R}, 0}(\bm{x}, t ; \bm{0}, t')\!\int_0^{t'}\! ds \; G^{(\chi)}_{\textsc{R}}(t'-s) \hat{\phi}_{\mathrm{h}}(\bm{0}, s)\,,\label{E:dkkdg2}
\end{align}
where $\hat{\phi}_{\textsc{tr}}(x)$ is the transient term associated with the atom's transient internal dynamics and $\hat{\phi}_{\textsc{br}}(x)$ is the backreaction field correlated with $\hat{\phi}_{\mathrm{h}}(x)$ everywhere.

\subsection{Hadamard function}
At this point, we have not specified the state of the field. For the reason that will be explained in Sec.~\ref{S:rthgsdfg}, we assume the initial state of the field the atom interacts with is a time-independent two-mode squeezed thermal state, for which the field's density matrix has the form 
\begin{equation}
    \hat{\rho}_{\textsc{tmst}} = \prod_{\bm{k}} \hat{S}_2^{\vphantom{\dagger}}(\zeta_{\bm{k}}) \hat{\rho}_{\beta}\hat{S}^{\dagger}_2(\zeta_{\bm{k}}) 
\end{equation}
where $\hat{\rho}_{\beta}$ is the thermal density matrix of the free field at temperature $T_{\textsc{b}}=\beta^{-1}$, and $\hat{S}_{2}(\zeta_{\bm{k}})$ is the two-mode squeeze operator, with squeeze parameter $\zeta_{\bm{k}} = \eta_{\bm{k}} e^{i\theta_{\bm{k}}}$, defined by
\begin{equation}
    \hat{S}_{2}(\zeta_{\bm{k}})=\exp\Bigl[\zeta^{*}_{\bm{k}}\,\hat{a}_{+\bm{k}}^{\vphantom{\dagger}}\hat{a}_{-\bm{k}}^{\vphantom{\dagger}}-\zeta_{\bm{k}}\,\hat{a}_{+\bm{k}}^{\dagger}a_{-\bm{k}}^{\dagger}\Bigr]\,.
\end{equation}
For the reader's convenience we have collected the essential properties of the two-mode squeezed state in Appendix~\ref{S:eoueisf}. Here we will assume that the squeeze parameter is mode-independent to simplify the arguments.

If the free field has the plane-wave expansion 
\begin{equation}
    \hat{\phi}_{\mathrm{h}}(x) = \int\! \frac{d^3 \bm{k}}{(2\pi)^{\frac{3}{2}}}\; \frac{1}{\sqrt{2\omega}} \Bigl[\hat{a}_{\bm{k}} e^{ik\cdot x} + \hat{a}^{\dagger}_{\bm{k}} e^{-ik\cdot x}  \Bigr]\,,
\end{equation}
with $k\cdot x = -\omega t + \bm{k} \cdot \bm{x}$ and $\omega=\lvert\bm{k}\rvert$, then the free field's Hadamard function for this two-mode squeezed thermal state has the form 
{\begin{equation}
    G^{(\phi)}_{\mathrm{H}, 0}(x, x') = \int\! \frac{d^3 \bm{k}}{(2\pi)^{3}}\;\frac{1}{4\omega}\,\coth\frac{\beta \omega}{2}\,e^{i\bm{k}\cdot(\bm{x}-\bm{x}')}\, \Bigl[\cosh2\eta\,e^{-i\omega(t-t')} -\sinh2\eta\,e^{i\theta}e^{-i\omega(t+t')} \Bigr]+ \textrm{C.C.}\,.\label{E:ugdfj1}
\end{equation}}
After carrying out the angular integration, we arrive at the following decomposition
\begin{equation}\label{E:ugdfj2}
    G^{(\phi)}_{\mathrm{H}, 0}(x, x') = G^{(\phi),\textsc{st}}_{\mathrm{H}, 0}(x, x') + G^{(\phi),\textsc{ns}}_{\mathrm{H}, 0}(x, x')\,,
\end{equation}
with
\begin{align}
    G^{(\phi),\textsc{st}}_{\mathrm{H}, 0}(x, x')&= \int_{-\infty}^{\infty}\!\frac{d\omega}{2\pi}\; \cosh2\eta\,\coth\frac{\beta \omega }{2}\,\frac{\sin(\omega\lvert\bm{x}-\bm{x}'\rvert)}{4\pi\lvert\bm{x} - \bm{x}'\rvert}\,e^{-i\omega(t-t')}\,,\\
    G^{(\phi),\textsc{ns}}_{\mathrm{H}, 0}(x, x')&=-\int_{0}^{\infty}\!\frac{d\omega}{2\pi} \;\sinh2\eta \,\coth\frac{\beta \omega}{2}\,\frac{\sin(\omega\lvert\bm{x}-\bm{x}'\rvert)}{4 \pi\lvert\bm{x}-\bm{x}'\rvert}\,e^{-i\omega(t+t')+i \theta} + \text{C.C.}\,.
\end{align}
Observe that the Hadamard function has an extra component that is not invariant under time translation, so we call it the nonstationary component of the Hadamard function. This is distinct from the case we consider in~\cite{QRadVac}. The emergence of this component is a consequence of the field being squeezed. We shall discuss its dynamical origin  in Sec.~\ref{S:rthgsdfg}. This nonstationary nature of the free field's Hadamard function will impact on the dynamical evolution of the atom's internal dynamics because the Hadamard function governs the statistics of the noise force  sourcing \eqref{E:kfbsut}, which naturally raises the concern whether the internal dynamics can ever equilibrate. As was mentioned earlier, it turns out the atom's internal motion can still relax to an  equilibrium, meaning time-translation invariant, state. Details of derivations and discussions on this point can be found in Ref.~\cite{AHH22}.

Eq.~\eqref{E:ugdfj1} or \eqref{E:ugdfj2} is the essential ingredient used to express the Hadamard function $ G^{(\phi)}_{\mathrm{H}}(x, x')$ of the interacting field, which we will need to evaluate the stress-energy tensor. Following the decomposition laid out before \eqref{E:dkkdg1}, the interacting field's Hadamard function is given by
\begin{align}
    G^{(\phi)}_{\mathrm{H}} (x,x') & = \frac{1}{2}\langle \bigl\{ \hat{\phi}_{\mathrm{h}}(x), \hat{\phi}_{\mathrm{h}}(x') \bigr\}\rangle + \frac{1}{2} \biggl[\langle \bigl\{ \hat{\phi}_{\mathrm{h}}(x), \hat{\phi}_{\textsc{br}}(x')\bigr\}\rangle + \langle \bigl\{ \hat{\phi}_{\textsc{br}}(x), \hat{\phi}_{\mathrm{h}}(x')\big\}\rangle \biggr]\notag\\
    & \qquad\qquad\qquad + \frac{1}{2}\langle \bigl\{ \hat{\phi}_{\textsc{br}}(x), \hat{\phi}_{\textsc{br}}(x') \bigr\}\rangle + \frac{1}{2}\langle \bigl\{ \hat{\phi}_{\textsc{tr}}(x), \hat{\phi}_{\textsc{tr}}(x') \bigr\} \rangle\,,\label{E:kfdgbdf}
\end{align}
where
\begin{align}
     \frac{1}{2}\langle \bigl\{ \hat{\phi}_{\mathrm{h}}(x), \hat{\phi}_{\mathrm{h}}(x') \bigr\}\rangle& = G^{(\phi)}_{\mathrm{H}, 0}(x,x')\,, \\
     \frac{1}{2}\langle \bigl\{ \hat{\phi}_{\mathrm{h}}(x), \hat{\phi}_{\textsc{br}}(x')\bigr\}\rangle&=e^2\int_0^{t'}\! ds_1'\; G_{\mathrm{R}, 0}^{(\phi)}(\bm{x}', t' ;\bm{0}, s_1')\! \int_0^{s_1'}\! ds_2' G_{\mathrm{R}}^{(\chi)}(s_1' - s_2') G_{\mathrm{H}, 0}^{(\phi)} (\bm{x}, t; \bm{0}, s_2')\,,\\
    \frac{1}{2}\langle \bigl\{ \hat{\phi}_{\textsc{br}}(x), \hat{\phi}_{\textsc{br}}(x') \bigr\}\rangle&=e^4\int_0^t\! ds_1\! \int_0^{t'}\!ds_1'\; G_{\mathrm{R}, 0}^{(\phi)}(\bm{x},t;\bm{0}, s_1) G^{(\phi)}_{\mathrm{R}, 0} (\bm{x}', t'; \bm{0}, s_1')\notag\\
    & \qquad \qquad  \times \int_0^{s_1}\! ds_2\! \int_0^{s_1'}\! ds_2'\; G_{\mathrm{R}}^{(\chi)}(s_1 - s_2)G_{\mathrm{R}}^{(\chi)}(s_1' - s_2') G^{(\phi)}_{\mathrm{H}, 0}(\bm{0}, s_2 ; \bm{0}, s_2')\,,    \\
    \frac{1}{2}\langle \bigl\{ \hat{\phi}_{\textsc{tr}}(x), \hat{\phi}_{\textsc{tr}}(x') \bigr\} \rangle&=\frac{e^2}{2}\int_0^t \!ds\!\int_0^{t'}\! ds'\; G_{\mathrm{R}, 0}^{(\phi)}(\bm{x},t; \bm{0} , s) G_{\mathrm{R}, 0}^{(\phi)}(\bm{x}', t'; \bm{0} , s') \,\langle \bigl\{ \hat{\chi}_{\mathrm{h}}(s) , \hat{\chi}_{\mathrm{h}} (s') \bigr\}\rangle\,.\label{E:dkgbsk4}
\end{align}
The second group inside the square brackets is of special interest because it describes the correlation between the radiation field and the free field at any location outside the atom. We do not see such counterpart in classical electrodynamics because in classical field theory, {there is no vacuum state to establish any correlation with the radiation field}. Here the correlation must be present because 1) the internal dynamics of the atom, which emits the radiation, is driven by the free field at the atom's location and 2) the free field has nonvanishing correlation among any spacetime interval. Note that we are in the Heisenberg picture, so the expectation values are evaluated with respect to the initial state of both subsystems. The contribution in \eqref{E:dkgbsk4} can be ignored at late times due to its transient nature. In addition, we also note that these four components are at most linearly proportional to $G^{(\phi)}_{\mathrm{H}, 0}(x,x')$. Thus, at least before $t-r\gg\gamma^{-1}$, the interaction field is expected to behave like a squeezed field to some extent.

For later convenience, we introduce a shorthand notation for an expression which comes up frequently,
\begin{equation}
    L_{\omega}(\bm{x}, t) =  \int_0^t\! ds_1 \; G^{(\phi)}_{\mathrm{R}, 0}(\bm{x}, t-s_1)\! \int_0^{s_1}\! ds_2 \; G^{(\chi)}_{\mathrm{R}}(s_1 - s_2) e^{-i\omega s_2}\,.
\end{equation}
After we carry out the integrations, we find its explicit form is given by
\begin{align}
    L_{\omega}(\bm{x}, t)&=\theta(t - r)\,\widetilde{G}_{\mathrm{R}}^{(\chi)}(\omega) \Bigl\{ \widetilde{G}_{\mathrm{R}, 0}^{(\phi)}(\bm{x}; \omega)\,e^{-i\omega t} - \frac{e^{-\gamma(t - r)}}{4\pi\Omega r} \Bigl[(\omega+i\gamma)\, \cos\Omega(r-t) + i\,\Omega\,\sin\Omega(r-t)\Bigr] \Big\}\,,\label{E:dfhgsiuer}
\end{align}
where we have used the identity for the Fourier transform of $G_{\mathrm{R}, 0}^{(\phi)}(\bm{x},\tau;\bm{0}, 0)$
\begin{equation}
    \widetilde{G}_{R, 0}^{(\phi)}(\bm{x}; \omega)=\frac{e^{i\omega r}}{4\pi r}
\end{equation}    
with $r=\lvert\bm{x}\rvert$. The convention of the Fourier transformation we adopt is
\begin{equation}
    f(t)=\int_{-\infty}^{\infty}\!\frac{d\omega}{2\pi}\;\tilde{f}(\omega)\,e^{-i\omega t}\,.
\end{equation}
At times $t\gg r$, we find that the dominant term in \eqref{E:dfhgsiuer} is given by
\begin{equation}
    L_{\omega}(\bm{x}, t) =\theta(t - r)\,\widetilde{G}_{\mathrm{R}}^{(\chi)}(\omega)\,\widetilde{G}_{\mathrm{R}, 0}^{(\phi)}(\bm{x}; \omega)\, e^{-i\omega t}\,.
\end{equation} 
These are convenient snippets that will greatly simplify the analysis of the late-time behavior of the expectation values of the energy-momentum stress tensor of the interacting field. Henceforth, we will proceed with the analysis separately for the contributions due to the stationary and the nonstationary components of $G^{(\phi)}_{\mathrm{H}, 0}(x,x')$.

At late times, the explicit expressions for the stationary and nonstationary components of the Hadamard functions of the free field and the interacting field $\phi$ become relatively simple. The interacting field's Hadamard function can likewise be decomposed into a stationary and a nonstationary component, $G^{(\phi)}_{\mathrm{H}}(x, x') = G^{(\phi),\textsc{st}}_{\mathrm{H}}(x, x') + G^{(\phi),\textsc{ns}}_{\mathrm{H}}(x, x')$, and
\begin{align}
    G^{(\phi),\textsc{st}}_{\mathrm{H}}(x, x')&= G^{(\phi),\textsc{st}}_{\mathrm{H}, 0}(x, x')+e^2\int_{-\infty}^{\infty}\!\frac{d\omega}{2\pi}\;\biggl\{\widetilde{G}^{(\phi),\textsc{st}}_{\mathrm{H}, 0}(\bm{x}, \bm{0};\omega)\, \widetilde{G}_{\mathrm{R}}^{(\chi)*}(\omega)\,\widetilde{G}_{\mathrm{R}, 0}^{(\phi)*}(\bm{x}'; \omega)\biggr. \\
    &\qquad\qquad\qquad\qquad\qquad\qquad\qquad\quad+\biggl. \widetilde{G}^{(\phi),\textsc{st}}_{\mathrm{H}, 0}(\bm{x}', \bm{0};\omega)\,\widetilde{G}_{\mathrm{R}}^{(\chi)}(\omega)\,\widetilde{G}_{\mathrm{R}, 0}^{(\phi)}(\bm{x}; \omega)\biggr\}\, e^{-i\omega (t-t')}+\text{C.C.} \notag\\
    &\qquad\quad+e^2\int_{-\infty}^{\infty}\frac{d\omega}{2\pi}\; \cosh2\eta\, \coth\frac{\beta \omega}{2} \,\widetilde{G}^{(\phi)}_{\mathrm{R}, 0}(\bm{x}, \omega) \,\widetilde{G}^{(\phi) *}_{\mathrm{R}, 0}(\bm{x}', \omega)\, \operatorname{Im}\widetilde{G}^{(\chi)}_R(\omega)\,e^{-i\omega(t-t')}+\text{C.C.}\,,\notag
    \intertext{and}
    G^{(\phi),\textsc{ns}}_{\mathrm{H}}(x, x')&=G^{(\phi), \textsc{ns}}_{\mathrm{H}, 0}(x, x')-e^2\int_{0}^{\infty}\!\frac{d\omega}{2\pi}\;\sinh2\eta \, \coth\frac{\beta \omega}{2}\, \widetilde{G}^{(\chi)}_{\mathrm{R}}(\omega)\\
    &\qquad\qquad\qquad\qquad\times\biggl\{\frac{\sin(\omega\lvert\bm{x}\rvert)}{4 \pi \lvert\bm{x}\rvert}\, \widetilde{G}^{(\phi)}_{\mathrm{R}, 0}(\bm{x}'; \omega)+\frac{\sin(\omega\lvert\bm{x}'\rvert)}{4 \pi \lvert\bm{x}'\rvert}\, \widetilde{G}^{(\phi)}_{\mathrm{R}, 0}(\bm{x}; \omega)\biggr\}\,e^{-i\omega (t+t')+i\theta}+\text{C.C.}\notag\\
    &\qquad-e^4\int_{0}^{\infty}\!\frac{d\omega}{2\pi}\;\frac{\omega}{4\pi}\,\sinh2\eta\, \coth\frac{\beta \omega}{2}\,  \widetilde{G}^{(\chi) 2}_{\mathrm{R}} (\omega)\,\widetilde{G}^{(\phi)}_{\mathrm{R}, 0}(\bm{x}; \omega) \,\widetilde{G}^{(\phi)}_{\mathrm{R}, 0}(\bm{x}'; \omega)\,e^{ -i(\omega(t+t')+i\theta}+\text{C.C.}\,,\notag
\end{align}
where
\begin{align}
    \widetilde{G}^{(\phi),\textsc{st}}_{\mathrm{H}, 0}(\bm{x}, \bm{0};\omega)\equiv\widetilde{G}^{(\phi),\textsc{st}}_{\mathrm{H}, 0}(\bm{x};\omega)=\cosh2\eta\,\coth\frac{\beta \omega }{2}\,\frac{\sin(\omega\lvert\bm{x}\rvert)}{4\pi\lvert\bm{x}\rvert}=\cosh2\eta\,\coth\frac{\beta \omega }{2}\,\operatorname{Im}\widetilde{G}_{\mathrm{R}, 0}^{(\phi)}(\bm{x}; \omega)\,.
\end{align}
The energy momentum stress tensor of the interacting field is then constructed via the Hadamard function $G^{(\phi)}_{\mathrm{H}}(x, x')$.  Analyses of the late-time energy flow of the field at distances far away from the atom resulting from the atom-field interaction are contained in Appendices~\ref{S:vbkdf} and \ref{S:evsfdjg}.  A general discussion of the spatial-temporal behavior of the stress-energy tensor due to the radiation field emitted from the atom follows in the next section.

\section{stress-energy tensor due to the radiation field}\label{S:eiutie}
The expectation value of the energy-momentum stress tensor due to the radiation field is defined by
\begin{equation}
    \langle\Delta \hat{T}_{\mu \nu}(x)\rangle = \lim_{x'\to x} \biggl\{ \frac{\partial^2}{\partial x^{\mu} \partial x'^{\nu}} -  \frac{1}{2}g_{\mu \nu} g^{\alpha \beta} \frac{\partial^2}{\partial x^{\alpha} \partial x'^{\beta}} \biggr\} \Bigl[  G^{(\phi)}_{\mathrm{H}} (x,x') - G^{(\phi)}_{\mathrm{H}, 0}(x,x') \Bigr]
\end{equation}
with the signature of $g_{\mu\nu}$ being $(-,+,+,+)$. Here we have subtracted off the contribution purely from the free field, which is irrelevant to the atom-field interaction. Due to symmetry, it has no dependence on the azimuthal angle $\vartheta$ and the polar angle $\varphi$ of the spherical coordinate system, so the stress-energy tensor already reduces to
\begin{equation}
    \setstackgap{L}{1.0\baselineskip}
    \fixTABwidth{T}
    \langle\Delta\hat{T}_{\mu \nu} \rangle = 
    \parenMatrixstack{
        \langle\Delta\hat{T}_{tt} \rangle &  \langle\Delta\hat{T}_{tr} \rangle & 0 & 0 \\
        \langle\Delta\hat{T}_{rt} \rangle &  \langle\Delta\hat{T}_{rr} \rangle & 0 & 0 \\
        0 & 0 &  0 & 0 \\
        0 & 0 & 0 &  0 
    }\,.
\end{equation}
Then we will focus on the components $\langle\Delta\hat{T}_{tr} \rangle$ and $\langle\Delta\hat{T}_{tt} \rangle$ that are relevant to our following discussions.

Here the analysis is not limited to the large-distance and the late-time limits, so that we can have a more complete, global view of the energy flow and energy density of the field caused by the atom-field interaction. We will put the analysis in those specific regimes in Appendices~\ref{S:vbkdf} and \ref{S:evsfdjg}.

Since in our configuration the coupling between the harmonic atom and the massless scalar field takes a different form from that between the point charge and the electromagnetic fields, it would be more illustrative if we first quickly go over the elementary derivations of the continuity equation and the total atom-field Hamiltonian in the classical regime to highlight their differences.

From the equation of motion of the scalar field \eqref{E:dkddhn2}, if we multiply both sides by $\partial_t\phi$, we obtain a local form of the continuity equation
\begin{equation}\label{E:kdkkde}
    \partial_{t}u_{\phi}+\bm{\nabla}\cdot\bm{S}_{\phi}=\mathsf{p}\,\partial_{t}\phi\,,
\end{equation}
where $\mathsf{p}(\bm{x},t)=e\,\delta^{(3)}(\bm{x}) \chi(t)$ serves as the point dipole associated with the atom, and we identify the field energy density $u_{\phi}$ and the field momentum density $\bm{S}_{\phi}$ respectively by
\begin{align}
    u_{\phi}(\bm{x},t)&=\frac{1}{2}\Bigl[\partial_{t}\phi(\bm{x},t)\Bigr]^{2}+\frac{1}{2}\Bigl[\bm{\nabla}\phi(\bm{x},t)\Bigr]^{2}\,,&\bm{S}_{\phi}(\bm{x},t)&=-\partial_{t}\phi(\bm{x},t)\,\bm{\nabla}\phi(\bm{x},t)\,.
\end{align}
The unusual issue with the right hand side of the continuity equation \eqref{E:kdkkde} is best appreciated when we compare with the corresponding equation for the electromagnetic fields
\begin{align}\label{E:kieturd}
	\partial_t u+\bm{\nabla}\cdot\bm{S}&=-\bm{J}\cdot\bm{E}\,,&&\text{with} &u&=\frac{1}{2}\bm{E}^{2}+\frac{1}{2}\bm{B}^{2}\,,  &\bm{S}&=\bm{E}\times\bm{B}\,,
\end{align}
where $\bm{J}$ is the current density and the corresponding current is equal to $e\dot{\bm{x}}$ for a point charge at the location $\bm{x}$.  The electric field $\bm{E}$ field is defined by $\bm{E}=-\bm{\nabla}A^0-\partial_t\bm{A}$ and the magnetic field $\bm{B}$ by $\bm{B}=\bm{\nabla}\times\bm{A}$. Here $A^0$ is the scalar potential and $\bm{A}$ is the vector potential. Thus the right hand side of \eqref{E:kdkkde} should bear a similar interpretation as $-\bm{J}\cdot\bm{E}$, which is usually understood as the dissipation of the field energy due to the work done by the field on the charge. That is, it is the opposite of the power delivered by the Lorentz force. However, there are a few subtle catches: 1) In classical field theory, we seldom consider the free field component of the field equation, and we never have vacuum field fluctuations, so the electromagnetic fields in \eqref{E:kieturd} are usually assumed to be the field generated by the charge. In contrast, the quantum scalar field $\phi$ under our consideration is a full interacting field, comprising of the free field and the radiation field emitted by the atom. Thus the right hand side of \eqref{E:kdkkde} will contain an extra contribution associated with the free field fluctuations. 2) The right hand side of \eqref{E:kdkkde} does not correspond to the net power delivered by the field to the internal degree of freedom of the atom, which is $e\,\phi(\bm{0},t)\,\partial_t\chi(t)$, apart from a renormalization contribution, according to the right hand side of \eqref{E:dkddhn1}. Then the interpretation of the right hand of \eqref{E:kdkkde} is not so straightforward, even though it has a form of the dipole interaction, similar to the right hand side of \eqref{E:kieturd}.

We may trace such a feature to the Hamiltonian. The Hamiltonian $H_{\chi\phi}$ associated with the Lagrangian of the atom-field interacting system
\begin{align}
	L_{\chi\phi}=\frac{m}{2}\,\dot{\chi}^{2}-\frac{m\omega_{\textsc{b}}^{2}}{2}\,\dot{\chi}^{2}+\int_{V}\!d\mathcal{V}\;e\,\delta(\bm{x})\chi(t)\phi(\bm{x},t)+\int_{V}\!d\mathcal{V}\;\Bigl\{\frac{1}{2}\bigl[\partial_{t}\phi(\bm{x},t)\bigr]^{2}-\frac{1}{2}\bigl[\bm{\nabla}\phi(\bm{x},t)\bigr]^{2}\Bigr\}\,,
\end{align}
under our consideration is given by
\begin{equation}\label{E:djnfawiwe}
	H_{\chi\phi}=\frac{m}{2}\,\dot{\chi}^{2}+\frac{m\omega_{\textsc{b}}^{2}}{2}\,\dot{\chi}^{2}-e\,\chi(t)\phi(\bm{0},t)+\int_{V}\!d\mathcal{V}\;\Bigl\{\frac{1}{2}\bigl[\partial_{t}\phi(\bm{x},t)\bigr]^{2}+\frac{1}{2}\bigl[\bm{\nabla}\phi(\bm{x},t)\bigr]^{2}\Bigr\}\,, 
\end{equation}
where we implicitly assume that $\dot{\chi}(t)$ and $\dot{\phi}(\bm{x},t)$ are functions of the respective canonical momenta, $p(t)$ and $\pi(\bm{x},t)$. The quantity $V$ denotes the whole spatial volume on a fixed time slice. Since the whole atom-field system forms a closed system, the total energy is conserved, so we have $dH/dt=0$,
\begin{align}\label{E:egdfhys}
	\frac{d}{dt}H_{\chi\phi}(t)=0&=\dot{\chi}(t)\biggl\{m\ddot{\chi}(t)+m\omega_{\textsc{b}}^{2}\chi(t)-e\,\phi(\bm{x},0)\biggr\}\\
	&\qquad\qquad\qquad\qquad+\int_{V}\!d\mathcal{V}\;\biggl[-e\,\delta(\bm{x})\chi(t)\,\partial_{t}\phi(\bm{x},t)+\frac{d}{dt}\biggl\{\frac{1}{2}\bigl[\partial_{t}\phi(\bm{x},t)\bigr]^{2}+\frac{1}{2}\bigl[\bm{\nabla}\phi(\bm{x},t)\bigr]^{2}\biggr\}\biggr]\,.\notag
\end{align}
Note how the derivative of the interaction term is distributed, and further observe that the contribution inside of the first pair of curly brackets gives zero, following the equation of motion \eqref{E:dkddhn1}. It actually has a nice interpretation. Following the derivations of \eqref{E:kfbsut} and \eqref{E:bvkesfds}, it accounts for the reduced dynamics of the atom's internal degree of freedom, and, from the atom's perspective, describes the energy exchange between the atom and the surrounding quantum field.

The integral on the right hand side of  \eqref{E:egdfhys}, on the other hand, describes the energy exchange from the field's perspective, and thus gives zero too. In the integrand, we see the presence of the same dipole interaction term $e\,\delta(\bm{x})\chi(t)\,\partial_{t}\phi(\bm{x},t)$ that appears on the righthand side of \eqref{E:kdkkde}. At first sight it may seem odd that the dipole interaction term is solely responsible for the change of the field energy. However, we can further show
\begin{equation}
	\frac{d}{dt}H_{\chi\phi}(t)=0=\int_{V}\!d\mathcal{V}\;\bm{\nabla}\cdot[\bm{\nabla}\phi(\bm{x},t)\,\partial_{t}\phi(\bm{x},t)\bigr]=-\oint_{\partial V}\!d\bm{\mathcal{A}}\cdot\bm{S}(\bm{x},t)\,.\label{E:kfeuusiwe}
\end{equation}
It says that there is no flux entering from or leaving to spatial infinity because the boundary $\partial V$ of the space volume $V$ is at spatial infinity. It is a rephrase of energy conservation in a closed system, so there is no inconsistency. On the other hand, Eq.~\eqref{E:kfeuusiwe} is too restrictive because $V$ is the total volume. The differential form of the continuity equation \eqref{E:kdkkde} is more suitable to see the energy distribution in the field.

In contrast to \eqref{E:djnfawiwe}, the Hamiltonian for the interacting system of a point charge and the electromagnetic field is
\begin{align}
    H_{xA}=\frac{m_{\textsc{b}}}{2}\,\dot{\bm{x}}^{2}+\frac{m_{\textsc{b}}\omega^{2}}{2}\,\bm{x}^{2}+\frac{1}{2}\int_{V}\!d\mathcal{V}\;\bigl(\bm{E}^{2}+\bm{B}^{2}\bigr)\,.
\end{align}
It does not have the interaction term manifestly if the Hamiltonian is not expressed in terms of the canonical momentum of the charge. However, it is consistent with the observation that $-\bm{J}\cdot\bm{E}$ is opposite to the power delivered by the Lorentz force.

Now we discuss the general result of $\langle\Delta\hat{T}_{\mu\nu} \rangle$ that will be valid for $t>r$ at any location distance $r$ away from the atom, so we come back to the manipulation of quantum operators. We first compute $\langle\Delta\hat{T}_{tr}\rangle$.

\subsection{General behavior of field energy flux \texorpdfstring{$\langle\Delta\hat{T}_{tr} \rangle$}{}}
If we write $\hat{\phi}_{\textsc{r}}(\bm{x},t)$ as
\begin{equation}\label{E:ieusgloet}
	\hat{\phi}_{\textsc{r}}(\bm{x},t)=e\int_{0}^{t}\!ds\;G_{\mathrm{R},0}^{(\phi)}(\bm{x},t;\bm{0},s)\,\hat{\chi}(s)=\frac{e}{4\pi r}\,\theta(t-r)\,\hat{\chi}(t-r)\,,
\end{equation}
which explicitly shows that the radiation field is a retarded Coulomb-like field emitted by a source at an earlier time $t-r$. Note that the simple expression of the radiation field like \eqref{E:ieusgloet} is not obtainable if the field has a non-Markovian spectrum.

With the help of \eqref{E:ieusgloet}, we obtain
\begin{align}
	\frac{1}{2}\langle\bigl\{\partial_{t}\hat{\phi}_{\textsc{r}}(\bm{x},t),\,\partial_{r}\hat{\phi}_{\textsc{r}}(\bm{x},t)\bigr\}\rangle&=-\biggl(\frac{e}{4\pi r}\biggr)^{2}\langle\hat{\chi}'^{2}(t-r)\rangle-\biggl(\frac{e}{4\pi r}\biggr)^{2}\frac{1}{2r}\langle\bigl\{\hat{\chi}'(t-r),\,\hat{\chi}(t-r)\bigr\}\rangle\,,\label{E:kedkgffd}
\end{align}
for $t>r$, where a prime denotes taking the derivative with respect to the function's argument. If we compute the energy flow associated \eqref{E:kedkgffd}, across any spherical surface of   radius $r$ centered at the atom, we get
\begin{align}\label{E:deeksgbkdr}
	\int_{\partial V}\!d\mathcal{A}\;\frac{1}{2}\langle\bigl\{\partial_{t}\hat{\phi}_{\textsc{r}}(\bm{x},t),\,\partial_{r}\hat{\phi}_{\textsc{r}}(\bm{x},t)\bigr\}\rangle&=-2m\gamma\,\langle\hat{\chi}'^{2}(t-r)\rangle-\frac{m\gamma}{r}\,\langle\bigl\{\hat{\chi}'(t-r),\,\hat{\chi}(t-r)\bigr\}\rangle\notag\\
	&=P_{\gamma}(t-r)-\frac{m\gamma}{r}\,\partial_{t}\langle\hat{\chi}^{2}(t-r)\rangle\,,
\end{align}
where we have used the substitution $e^2=8\pi\gamma m$. The first term has a special significance. From~\cite{HCSH18,QRadVac,FDRSq}, we know that $P_{\gamma}(t)$ is the energy the atom's internal degree of freedom loses to the surrounding field due to damping. Thus the first term tells that the energy lost by the atom at time $t-r$ takes time $r$ to propagate to a location at a distance $r$ away from the atom. It is the only contribution in \eqref{E:deeksgbkdr} that may survive at spatial infinity.

The contribution from the cross terms need a little more algebraic manipulations, but it gives
\begin{align}
	&\quad\frac{1}{2}\langle\bigl\{\partial_{t}\hat{\phi}_{\mathrm{h}}(\bm{x},t),\,\partial_{r}\hat{\phi}_{\textsc{r}}(\bm{x},t)\bigr\}\rangle+\frac{1}{2}\langle\bigl\{\partial_{t}\hat{\phi}_{\textsc{r}}(\bm{x},t),\,\partial_{r}\hat{\phi}_{\mathrm{h}}(\bm{x},t)\bigr\}\rangle\notag\\
	&=\frac{1}{4\pi r^{2}}\,P_{\xi}(t-r)-\frac{e^{2}}{4\pi r^{2}}\frac{\partial}{\partial t}\int_{0}^{t-r}\!ds\;G_{\mathrm{R}}^{(\chi)}(t-r-s)\,G_{\mathrm{H},0}^{(\phi)}(\bm{x},t;\bm{0},s)\,,\label{E:diufidfckk}
\end{align}
in which the integral in the last expression is nothing but
\begin{equation}
	e^{2}\int_{0}^{t-r}\!ds\;G_{\mathrm{R}}^{(\chi)}(t-r-s)\,G_{\mathrm{H},0}^{(\phi)}(\bm{x},t;\bm{0},s)=\frac{e}{2}\langle\bigl\{\hat{\chi}(t-r),\,\hat{\phi}_{\mathrm{h}}(\bm{x},t)\bigr\}\rangle\,.
\end{equation}
Here $P_{\xi}(t)$ is the power the free quantum field fluctuations deliver to the atom's internal dynamics at time $t$ at the atom's location. Then we find the incoming energy flow given by 
\begin{align}
	&\quad\int_{\partial V}\!d\mathcal{A}\;\biggl[\frac{1}{2}\langle\bigl\{\partial_{t}\hat{\phi}_{\mathrm{h}}(\bm{x},t),\,\partial_{r}\hat{\phi}_{\textsc{r}}(\bm{x},t)\bigr\}\rangle+\frac{1}{2}\langle\bigl\{\partial_{t}\hat{\phi}_{\textsc{r}}(\bm{x},t),\,\partial_{r}\hat{\phi}_{\mathrm{h}}(\bm{x},t)\bigr\}\rangle\biggr]\notag\\
	&=P_{\xi}(t-r)-\frac{\partial}{\partial t}\Bigl[\frac{e}{2}\langle\bigl\{\hat{\chi}(t-r),\,\hat{\phi}_{\mathrm{h}}(\bm{x},t)\bigr\}\rangle\Bigr]\,.\label{E:ujdfuder}
\end{align}
Combining \eqref{E:deeksgbkdr} and \eqref{E:ujdfuder}, we thus have 
\begin{equation}\label{E:plrtjfgd}
	\langle\Delta\hat{T}_{tr}\rangle=\frac{1}{4\pi r^{2}}\biggl\{P_{\xi}(t-r)+P_{\gamma}(t-r)-\frac{\partial}{\partial t}\Bigl[\frac{m\gamma}{r}\,\langle\hat{\chi}^{2}(t-r)\rangle+\frac{e}{2}\langle\bigl\{\hat{\chi}(t-r),\,\hat{\phi}_{\mathrm{h}}(\bm{x},t)\bigr\}\rangle\Bigr]\biggr\}\,,
\end{equation}
From the relaxation of the reduced dynamics of the atom's internal degree of freedom, outlined in the beginning of Sec.~\ref{S:eoired}, we  learn that the internal dynamics will reach equilibrium, where $P_{\xi}(t)+P_{\gamma}(t)=0$ for $t\gg\gamma^{-1}$. Hence, from \eqref{E:plrtjfgd}, we easily conclude that, for a fixed $r$, when $t-r$ is much greater than the relaxation time scale, the sum of the dominant terms $P_{\xi}(t-r)+P_{\gamma}(t-r)$ vanishes. At large distance away from the atom, the remaining terms are very small, decaying at least like $1/r^3$. It is consistent with our results in Sec.~\ref{S:vbkdf}, where we explicitly show that they actually give zero at late times. Alternatively we may deduce the same conclusion here, since 1) $\langle\hat{\chi}^{2}(t)\rangle$ will approach a constant at late times {as the  atom's internal dynamics  asymptotically reaches an equilibrium,} and 2) with the help of the explicit expression
\begin{equation*}
    \frac{e}{2}\langle\bigl\{\hat{\chi}(t-r),\,\hat{\phi}_{\mathrm{h}}(\bm{x},t)\bigr\}\rangle=-\frac{e^2}{4\pi r}\int_0^{\infty}\!\frac{d\omega}{2\pi}\;\coth\frac{\beta\omega}{2}\,\cosh2\eta\int_0^{t-r}\!ds\;G_R^{(\chi)}(t-r-s)\,\frac{\sin\omega r}{4\pi r^2}\,\cos\omega(t-s)\,,
\end{equation*}
it is not difficult to see that the time derivative of the last term inside the square brackets will vanish at late times.

{Thus, the detector away from the atom will not measure any radiation energy flux from the stationary atom at late times even though the atom is coupled to a nonstationary squeezed state.}

\subsection{General behavior of field energy density \texorpdfstring{$\langle\Delta\hat{T}_{tt} \rangle$}{}}

It is also of interest to examine the change of the field energy density at distances far away from the atom resulting from their mutual interactions.  Conceptually, the atom's internal dynamics will send out to spatial infinity  spherical waves centered at the location of the atom. Due to the quantum nature of the internal dynamics, this radiation wave in general has a random phase and its magnitude is inversely proportional to the distance to the atom. Following our previous discussion, a detector at large but fixed distance away from the atom will receive a net outward energy flux in the beginning, and then it will find the magnitude of the flux rapidly falls off to zero with time. After all these activities settle down, will the earlier energy flux leave any footprint in the space surrounding the atom, say, by shifting the local field energy density, albeit almost imperceptibly? This is what we try to find out in this section.

In the same manner, we will re-write $\langle\Delta T_{tt}\rangle$, following \eqref{E:ieusgloet}. Other than the overall factor $1/2$, we first show
\begin{align*}
	\frac{1}{2}\langle\bigl\{\partial_{t}\hat{\phi}_{\textsc{r}}(\bm{x},t),\,\partial_{t}\hat{\phi}_{\textsc{r}}(\bm{x},t)\bigr\}\rangle+\frac{1}{2}\langle\bigl\{\partial_{r}\hat{\phi}_{\textsc{r}}(\bm{x},t),\,\partial_{r}\hat{\phi}_{\textsc{r}}(\bm{x},t)\bigr\}\rangle=-2\times\frac{1}{4\pi r^{2}}\biggl\{P_{\gamma}(t-r)+\frac{\partial}{\partial r}\Bigl[\frac{m\gamma}{r}\,\langle\hat{\chi}^{2}(t-r)\rangle\Bigr]\biggr\}\,,
\end{align*}
where $\partial_{t}\chi(t-r)=-\partial_{r}\chi(t-r)$. Here due to the minus sign up front, this contribution tends to grow to a positive value at late times. We further note that the same expression $P_{\gamma}(t-r)$ also appears in the component \eqref{E:deeksgbkdr} of $\langle\Delta \hat{T}_{tr}(t)\rangle$ that is purely caused by the radiation field. Thus, we see the outward energy flux due to \eqref{E:deeksgbkdr} imparts field energy into the space around the atom.

For the cross terms, we obtain
\begin{align}
	&\quad2\times\frac{1}{2}\,\langle\bigl\{\partial_{t}\hat{\phi}_{\mathrm{h}}(\bm{x},t),\,\partial_{t}\hat{\phi}_{\textsc{r}}(\bm{x},t)\bigr\}\rangle+2\times\frac{1}{2}\,\langle\bigl\{\partial_{r}\hat{\phi}_{\mathrm{h}}(\bm{x},t),\,\partial_{r}\hat{\phi}_{\textsc{r}}(\bm{x},t)\bigr\}\rangle\notag\\
	&=-2\times\frac{1}{4\pi r^{2}}\biggl\{P_{\xi}(t-r)+\frac{\partial}{\partial r}\Bigl[\frac{e}{2}\langle\bigl\{\hat{\chi}(t-r),\,\hat{\phi}_{\mathrm{h}}(\bm{x},t)\bigr\}\rangle\Bigr]\biggr\}\,.
\end{align}
Following the same argument, the inward energy flux from \eqref{E:diufidfckk} on the other hand is prone to take away the field energy in the surrounding space.

Together we thus find that the net field energy density outside the atom is given by
\begin{align}
	\langle\Delta \hat{T}_{tt}\rangle=-\frac{1}{4\pi r^{2}}\biggl\{P_{\xi}(t-r)+P_{\gamma}(t-r)+\frac{\partial}{\partial r}\Bigl[\frac{m\gamma}{r}\,\langle\hat{\chi}^{2}(t-r)\rangle+\frac{e}{2}\langle\bigl\{\hat{\chi}(t-r),\,\hat{\phi}_{\mathrm{h}}(\bm{x},t)\bigr\}\rangle\Bigr]\biggr\}\,,\label{E:eoiddd}
\end{align}
for $t>r>0$. This looks very similar to \eqref{E:plrtjfgd}, and can be the consequence of the continuity equation. The dominant term in Eq.~\eqref{E:eoiddd} will vanish at late times, as a consequences of the relaxation of the internal dynamics of the atom, and the behavior of \eqref{E:plrtjfgd} due to the appearance of $t-r$. Following our earlier arguments, the remaining term, $\langle\bigl\{\hat{\chi}(t-r),\,\hat{\phi}_{\mathrm{h}}(\bm{x},t)\bigr\}\rangle$, on the other hand, becomes a constant at late times, which falls off at least like $1/r^3$, as is explicitly shown in Sec.~\ref{S:kvckdfd}.

Indeed, Eqs.~\eqref{E:plrtjfgd} and \eqref{E:eoiddd} enable us to verify the continuity equation
\begin{align}
	\frac{1}{r^{2}}\frac{\partial}{\partial r}\Bigl(r^{2}\langle\Delta \hat{T}_{tr}\rangle\Bigr)&=\frac{1}{4\pi r^{2}}\biggl\{\partial_{r}P_{\xi}(t-r)+\partial_{r}P_{\gamma}(t-r)-\frac{\partial^{2}}{\partial t\partial r}\Bigl[\frac{m\gamma}{r}\,\langle\hat{\chi}^{2}(t-r)\rangle+\frac{e}{2}\langle\bigl\{\hat{\chi}(t-r),\,\hat{\phi}_{\mathrm{h}}(\bm{x},t)\bigr\}\rangle\Bigr]\biggr\}\notag\\
	&=\frac{1}{4\pi r^{2}}\biggl\{-\partial_{t}P_{\xi}(t-r)-\partial_{t}P_{\gamma}(t-r)-\frac{\partial^{2}}{\partial t\partial r}\Bigl[\frac{m\gamma}{r}\,\langle\hat{\chi}^{2}(t-r)\rangle+\frac{e}{2}\langle\bigl\{\hat{\chi}(t-r),\,\hat{\phi}_{\mathrm{h}}(\bm{x},t)\bigr\}\rangle\Bigr]\biggr\}\notag\\
	&=\frac{\partial}{\partial t}\langle\Delta\hat{T}_{tt}\rangle\,,
\end{align}
for $r>0$. To include $r=0$, the location of the atom, we need the form \eqref{E:kdkkde} which also takes into account the energy flow into and out of the atom, from the scalar field perspective.

From the considerations presented so far we may now see better how the radiation flux, generated by the internal dynamics of the atom, propagates outward and at the same time intakes the field energy in space outside the atom. Meanwhile,  due to the remarkable correlation between the quantum radiation field and the free quantum field, there exists an inward flux. On its way toward the atom, it pulls out field energy stored in the proximity of the atom. From this hindsight, it can be understood that the net power $e\,\phi(\bm{0},t)\dot{\chi}(t)$ delivered by the quantum field to the atom is not equal to the rate of work done to the field $e\,\dot{\phi}(\bm{0},t)\chi(t)$, as can be inferred from \eqref{E:egdfhys}. Otherwise $\langle\Delta \hat{T}_{tr}\rangle$ in \eqref{E:plrtjfgd} will simply be proportional to $P_{\xi}(t-r)+P_{\gamma}(t-r)$. Their difference accounts for the field energy density stored in the space  for the configuration we have studied.
	
In summary,  the results in this section tell us that at late times the observer will not measure any net energy flow associated with the radiation emitted from the atom driven by the squeezed quantum field, but the observer can still detect a constant radiation energy density, which is related to the squeeze parameter. However, a restrictive condition is that the residual radiation energy density is of the near-field nature, and it  falls off with the distance to the atom like $1/r^3$. Thus its detection can be difficult at large distances, unless the squeeze parameter is large.

Even with these difficulties, we can still locally identify the squeezing via the response of the atom interacting with the squeezed field.

In the next section we shall discuss how the squeeze parameter depends on the parametric process the quantum field has experienced, such that we may acquire certain information about the process once we have measured the squeeze parameter.

\section{functional dependence of the squeeze parameter on the parametric process}\label{S:rthgsdfg}
Now we turn to how we may possibly extract from the behavior of the squeeze parameter the information of the parametric process that occurs earlier. To be specific, consider the simple case that a massless quantum Klein-Gordon field in flat spacetime undergoes a parametric process such that the frequency of mode $\bm{k}$ transits smoothly from one constant value $\omega_{i}$ for $0\leq t\leq t_{a}$ to another constant $\omega_{f}$ for $t\geq t_{b}>t_{a}$.

We would like first to show that the quantum field in the out-region behaves as if it is in its squeezed thermal state with a time-independent squeeze parameter $\zeta_{\bm{k}}$ if the initial state of the field at $t=0$ is a thermal state. Then we express the squeeze parameter in terms of the Bogoliubov coefficients,  a common tool to treat the dynamics of the quantum field in a parametric (time-varying) process. We will also link the squeeze parameter to the fundamental solutions of the equation of motion of the field, in which  information of the parametric process is embedded.

We work with the Heisenberg picture, so the field remains in its initial state at $t=0$. Formally we can expand the field operator $\hat{\phi}(x)$ in terms of different sets of mode functions. In the out-region, two convenient choices are
\begin{align}
	u_{\bm{k}}^{\textsc{in}}(x)&=\frac{1}{\sqrt{2\omega_{i}}}\,e^{i\bm{k}\cdot\bm{x}}\Bigl[d_{\bm{k}}^{(1)}(t)-i\,\omega_{i}\,d_{\bm{k}}^{(2)}(t)\Bigr]\,,&u_{\bm{k}}^{\textsc{out}}(x)&=\frac{1}{\sqrt{2\omega_{f}}}\,e^{i\bm{k}\cdot\bm{x}}e^{-i\omega_{f}t}\,,
\end{align}
where $d_{\bm{k}}^{(i)}(t)$ satisfies the equation $\ddot{d}_{\bm{k}}^{(i)}+\omega^{2}(t)\,d_{\bm{k}}^{(i)}(t)=0$, and $u_{\bm{k}}^{\textsc{out}}(x)$ is the standard  plane-wave mode function in the out-region, while $u_{\bm{k}}^{\textsc{in}}(x)$ represents the mode function which evolves from the plane-wave mode function in the in-region.  Thus the field operator may have the expansion
\begin{align}\label{E:dksusdfk}
	\hat{\phi}(x)&=\begin{cases}
		\displaystyle\sum_{\bm{k}}\hat{a}_{\bm{k}}^{\vphantom{\dagger}}u_{\bm{k}}^{\textsc{in}}(x)+\hat{a}_{\bm{k}}^{\dagger}u_{\bm{k}}^{\textsc{in}*}(x)\,,&\vphantom{\biggl|}\\
		\displaystyle\sum_{\bm{k}}\hat{b}_{\bm{k}}^{\vphantom{\dagger}}u_{\bm{k}}^{\textsc{out}}(x)+\hat{b}_{\bm{k}}^{\dagger}u_{\bm{k}}^{\textsc{out}*}(x)\,,&
	\end{cases}&\sum_{\bm{k}}&=\int\!\frac{d^{3}\bm{k}}{(2\pi)^{\frac{3}{2}}}\,,
\end{align}
in the out-region.

We further suppose $(\hat{b}_{\bm{k}}^{\vphantom{\dagger}},\hat{b}_{\bm{k}}^{\dagger})$ are related to $(\hat{a}_{\bm{k}}^{\vphantom{\dagger}},\hat{a}_{\bm{k}}^{\dagger})$ by
\begin{equation}\label{E:qosnzkqewr}
	\hat{b}_{+\bm{k}}^{\vphantom{\dagger}}=\alpha_{+\bm{k}}^{\vphantom{\dagger}}\,\hat{a}_{+\bm{k}}^{\vphantom{\dagger}}+\beta_{-\bm{k}}^{\vphantom{\dagger}*}\,\hat{a}_{-\bm{k}}^{\dagger}\,,
\end{equation}
whence the completeness condition implies that the Bogoliubov coefficients $\alpha_{\bm{k}}$, $\beta_{\bm{k}}$ can be parametrized by
\begin{align}\label{E:lkgdkf}
    \alpha_{\bm{k}}&=\cosh\eta_{\bm{k}}\,,&\beta_{-\bm{k}}^{\vphantom{\dagger}*}&=-e^{i\theta_{\bm{k}}}\,\sinh\eta_{\bm{k}}\,,
\end{align}
such that 
\begin{equation}\label{E:fuedhfser}
    \hat{b}_{+\bm{k}}^{\vphantom{\dagger}}=\cosh\eta_{\bm{k}}\,\hat{a}_{+\bm{k}}^{\vphantom{\dagger}}-e^{i\theta_{\bm{k}}}\,\sinh\eta_{\bm{k}}\,\hat{a}_{-\bm{k}}^{\dagger}=\hat{S}_{2}^{\dagger}(\zeta_{\bm{k}})\,\hat{a}_{+\bm{k}}^{\vphantom{\dagger}}\,\hat{S}_{2}^{\vphantom{\dagger}}(\zeta_{\bm{k}})\,.
\end{equation}
The parametrization \eqref{E:lkgdkf} can be alternatively implemented by the two-mode squeeze operator $\hat{S}_{2}^{\dagger}(\zeta_{\bm{k}})$. We summarize its properties in Appendix~\ref{S:eoueisf}.

Then a quantity of the field like the Hadamard function in the out-region can be cast into
\begin{align}
	&\quad\frac{1}{2}\langle\textsc{in}\vert\bigl\{\hat{\phi}(x),\,\hat{\phi}(x')\bigr\}\vert\textsc{in}\rangle\notag\\
	&=\sum_{\bm{k}}\frac{1}{2\omega_{f}}\biggl[\,\frac{1}{2}\langle\textsc{in}\vert\bigl\{\hat{b}_{\bm{k}}^{\vphantom{\dagger}},\,\hat{b}_{\bm{k}}^{\vphantom{\dagger}}\bigr\}\vert\textsc{in}\rangle\,e^{i\bm{k}\cdot(\bm{x}+\bm{x}')}e^{-i\omega_{f}(t+t')}+\frac{1}{2}\langle\textsc{in}\vert\bigl\{\hat{b}_{\bm{k}}^{\vphantom{\dagger}},\,\hat{b}_{-\bm{k}}^{\vphantom{\dagger}}\bigr\}\vert\textsc{in}\rangle\,e^{i\bm{k}\cdot(\bm{x}-\bm{x}')}e^{-i\omega_{f}(t+t')}\biggr.\notag\\
	&\qquad\qquad\quad\quad+\biggl.\frac{1}{2}\langle\textsc{in}\vert\bigl\{\hat{b}_{\bm{k}}^{\vphantom{\dagger}},\,\hat{b}_{\bm{k}}^{\dagger}\bigr\}\vert\textsc{in}\rangle\,e^{i\bm{k}\cdot(\bm{x}-\bm{x}')}e^{-i\omega_{f}(t-t')}+\frac{1}{2}\langle\textsc{in}\vert\bigl\{\hat{b}_{\bm{k}}^{\vphantom{\dagger}},\,\hat{b}_{-\bm{k}}^{\dagger}\bigr\}\vert\textsc{in}\rangle\,e^{i\bm{k}\cdot(\bm{x}+\bm{x}')}e^{-i\omega_{f}(t-t')}+\text{C.C.}\biggr]\notag\\
	&=\sum_{\bm{k}}\frac{1}{2\omega_{f}}\,e^{i\bm{k}\cdot(\bm{x}-\bm{x}')}\biggl[\,\frac{1}{2}\langle\zeta_{\bm{k},\textsc{in}}^{\textsc{tmsq}}\vert\bigl\{\hat{a}_{\bm{k}}^{\vphantom{\dagger}},\,\hat{a}_{-\bm{k}}^{\vphantom{\dagger}}\bigr\}\vert\zeta_{\bm{k},\textsc{in}}^{\textsc{tmsq}}\rangle\,e^{-i\omega_{f}(t+t')}+\frac{1}{2}\langle\zeta_{\bm{k},\textsc{in}}^{\textsc{tmsq}}\vert\bigl\{\hat{a}_{\bm{k}}^{\vphantom{\dagger}},\,\hat{a}_{\bm{k}}^{\dagger}\bigr\}\vert\zeta_{\bm{k},\textsc{in}}^{\textsc{tmsq}}\rangle\,e^{-i\omega_{f}(t-t')}\biggr.\notag\\
	&\qquad\qquad\qquad\qquad\qquad+\biggl.\text{C.C.}\biggr]\,,\label{E:yrsvear}
\end{align}
where $\lvert\zeta_{\bm{k},\textsc{in}}^{\textsc{tmsq}}\rangle=\hat{S}_{2}^{\vphantom{\dagger}}(\zeta_{\bm{k}})\,\lvert\textsc{in}\rangle$ is the two-mode squeezed state of the initial in-state $\lvert\textsc{in}\rangle$. Thus Eq.~\eqref{E:yrsvear} gives a Hadamard function in the two-mode squeezed in-state. Although here we use the pure state form, the result can be easily adapted for a mixed state.

On the other hand, the same Hadamard function can be expressed in terms of the in-mode functions,  by the expansion 
\begin{align}
	&\quad\frac{1}{2}\langle\textsc{in}\vert\bigl\{\hat{\phi}(x),\,\hat{\phi}(x')\bigr\}\vert\textsc{in}\rangle\notag\\
	&=\sum_{\bm{k}}\frac{1}{2\omega_{i}}e^{i\bm{k}\cdot(\bm{x}-\bm{x}')}\Bigl(N_{\bm{k}}^{(\beta)}+\frac{1}{2}\Bigr)\Bigl[d_{\bm{k}}^{(1)}(t)-i\,\omega_{i}\,d_{\bm{k}}^{(2)}(t)\Bigr]\Bigl[d_{\bm{k}}^{(1)}(t')+i\,\omega_{i}\,d_{\bm{k}}^{(2)}(t')\Bigr]+\text{C.C.}\,.\label{E:yroerudfj}
\end{align}
If the in-state is a thermal state, then $\langle\textsc{in}\vert\bigl\{\hat{a}_{\bm{k}}^{\vphantom{\dagger}},\,\hat{a}_{\bm{k}}^{\dagger}\bigr\}\vert\textsc{in}\rangle$ {is understood in terms of the trace average, and gives} $2N_{\bm{k}}^{(\beta)}+1$, with $N_{\bm{k}}^{(\beta)}$ being the average number density of the thermal state at temperature $\beta^{-1}$,
\begin{equation}
    N_{\bm{k}}^{(\beta)}=\frac{1}{e^{\beta\omega_i}-1}\,.
\end{equation}
Observe the structural similarities between \eqref{E:yrsvear} and \eqref{E:yroerudfj}.

To make them more revealing, we evaluate the expectation values on the righthand side of \eqref{E:yrsvear},
\begin{align*}
	\langle\textsc{in}\vert\bigl\{\hat{b}_{+\bm{k}}^{\vphantom{\dagger}},\,\hat{b}_{+\bm{k}}^{\vphantom{\dagger}}\bigr\}\vert\textsc{in}\rangle&=0\,,&\langle\textsc{in}\vert\bigl\{\hat{b}_{+\bm{k}}^{\vphantom{\dagger}},\,\hat{b}_{-\bm{k}}^{\vphantom{\dagger}}\bigr\}\vert\textsc{in}\rangle&=\bigl(\alpha_{+\bm{k}}^{\vphantom{\dagger}}\beta_{+\bm{k}}^{\vphantom{\dagger}*}+\alpha_{+\bm{k}}^{\vphantom{\dagger}}\beta_{-\bm{k}}^{\vphantom{\dagger}*}\bigr)\bigl(2N_{\bm{k}}^{(\beta)}+1\bigr)\,,\\
	\langle\textsc{in}\vert\bigl\{\hat{b}_{+\bm{k}}^{\vphantom{\dagger}},\,\hat{b}_{-\bm{k}}^{\dagger}\bigr\}\vert\textsc{in}\rangle&=0\,,&\langle\textsc{in}\vert\bigl\{\hat{b}_{+\bm{k}}^{\vphantom{\dagger}},\,\hat{b}_{+\bm{k}}^{\dagger}\bigr\}\vert\textsc{in}\rangle&=\bigl(\lvert\alpha_{+\bm{k}}^{\vphantom{\dagger}}\rvert^{2}+\lvert\beta_{-\bm{k}}^{\vphantom{\dagger}}\rvert^{2}\bigr)\bigl(2N_{\bm{k}}^{(\beta)}+1\bigr)\,.
\end{align*}
Thus, Eq.~\eqref{E:yrsvear} becomes
\begin{align}
	&\quad\frac{1}{2}\langle\textsc{in}\vert\bigl\{\hat{\phi}(x),\,\hat{\phi}(x')\bigr\}\vert\textsc{in}\rangle\notag\\
    &=\sum_{\bm{k}}\frac{1}{2\omega_{f}}\,e^{i\bm{k}\cdot(\bm{x}-\bm{x}')}\Bigl(N_{\bm{k}}^{(\beta)}+\frac{1}{2}\Bigr)\Bigl[\,2\alpha_{\bm{k}}^{\vphantom{\dagger}}\beta_{\bm{k}}^{\vphantom{\dagger}*}\,e^{-i\omega_{f}(t+t')}+\bigl(\lvert\alpha_{\bm{k}}^{\vphantom{\dagger}}\rvert^{2}+\lvert\beta_{\bm{k}}^{\vphantom{\dagger}}\rvert\bigr)\,e^{-i\omega_{f}(t-t')}+\text{C.C.}\Bigr]\,.\label{E:dkaudkf}
\end{align}
Comparing this equation with \eqref{E:yroerudfj} we find
\begin{align}
	\lvert\alpha_{\bm{k}}^{\vphantom{\dagger}}\rvert^{2}+\lvert\beta_{\bm{k}}^{\vphantom{\dagger}}\rvert^{2}&=\frac{1}{2}\,\biggl[\frac{\omega_{f}}{\omega_{i}}\,d_{\bm{k}}^{(1)2}(t_{f})+\omega_{f}\omega_{i}\,d_{\bm{k}}^{(2)2}(t_{f})+\frac{1}{\omega_{f}\omega_{i}}\,\dot{d}_{\bm{k}}^{(1)2}(t_{f})+\frac{\omega_{i}}{\omega_{f}}\,\dot{d}_{\bm{k}}^{(2)2}(t_{f})\Bigr]\,,\label{E:dckfgue}\\
	2\alpha_{\bm{k}}^{\vphantom{\dagger}}\beta_{\bm{k}}^{\vphantom{\dagger}*}&=\frac{1}{2\omega_{i}\omega_{f}}\,\Bigl[+i\,\omega_{f}\,d_{\bm{k}}^{(1)}(t_{f})+\omega_{i}\omega_{f}\,d_{\bm{k}}^{(2)}(t_{f})-\dot{d}_{\bm{k}}^{(1)}(t_{f})+i\,\omega_{i}\,\dot{d}_{\bm{k}}^{(2)}(t_{f})\Bigr]\notag\\
	&\qquad\qquad\qquad\qquad\qquad\times\Bigl[-i\,\omega_{f}\,d_{\bm{k}}^{(1)}(t_{f})+\omega_{i}\omega_{f}\,d_{\bm{k}}^{(2)}(t_{f})+\dot{d}_{\bm{k}}^{(1)}(t_{f})+i\,\omega_{i}\,\dot{d}_{\bm{k}}^{(2)}(t_{f})\Bigr]\,.\label{E:mnbjwie}
\end{align}
These results easily enable us to find the Bogoliubov coefficients $\alpha_{\bm{k}}$, $\beta_{\bm{k}}$ once we have the fundamental solutions $d_{\bm{k}}^{(i)}(t)$ with $i=1$, 2.

Following the same arguments leading to \eqref{E:yrsvear} and \eqref{E:yroerudfj}, if we compute $\langle\hat{\phi}^2(x)\rangle$, $\langle\hat{\pi}^2(x)\rangle$ and $\langle\bigl\{\hat{\phi}(x),\,\hat{\pi}(x)\bigr\}\rangle$ in the out-region, we obtain that, for each mode
\begin{align}
	\cosh2\eta_{\bm{k}}&=\frac{1}{2}\biggl[\frac{1}{\omega_{f}\omega_{i}}\,\dot{d}_{\bm{k}}^{(1)2}(t)+\frac{\omega_{i}}{\omega_{f}}\,\dot{d}_{\bm{k}}^{(2)2}(t)+\frac{\omega_{f}}{\omega_{i}}\,d_{\bm{k}}^{(1)2}(t)+\omega_{f}\omega_{i}\,d_{\bm{k}}^{(2)2}(t)\biggr]\,,\label{E:kbsdkskdu1}\\
	\cos(\theta_{\bm{k}}-2\omega_{f}t)\,\sinh2\eta_{\bm{k}}&=\frac{1}{2}\biggl[\frac{1}{\omega_{f}\omega_{i}}\,\dot{d}_{\bm{k}}^{(1)2}(t)+\frac{\omega_{i}}{\omega_{f}}\,\dot{d}_{\bm{k}}^{(2)2}(t)-\frac{\omega_{f}}{\omega_{i}}\,d_{\bm{k}}^{(1)2}(t)-\omega_{f}\omega_{i}\,d_{\bm{k}}^{(2)2}(t)\biggr]\,,\label{E:kbsdkskdu2}\\
	\sin(\theta_{\bm{k}}-2\omega_{f}t)\,\sinh2\eta_{\bm{k}}&=-\biggl[\frac{1}{\omega_{i}}\,d_{\bm{k}}^{(1)}(t)\dot{d}_{\bm{k}}^{(1)}(t)+\omega_{i}\,d_{\bm{k}}^{(2)}(t)\dot{d}_{\bm{k}}^{(2)}(t)\biggr]\,,\label{E:kbsdkskdu3}
\end{align}
explicit relations between the squeeze parameters and the fundamental solutions. Here $\hat{\pi}(x)$ is the canonical momentum conjugated to $\hat{\phi}(x)$. Eqs.~\eqref{E:kbsdkskdu1}--\eqref{E:kbsdkskdu3} clearly tell how squeezing may dynamically arise from the parametric process of the field.

As a consistency check, we substitute the coefficients $\alpha_{\bm{k}}$, $\beta_{\bm{k}}$ on the righthand side by \eqref{E:qosnzkqewr}, and obtain
\begin{align}\label{E:oweoqalw}
	\lvert\alpha_{\bm{k}}^{\vphantom{\dagger}}\rvert^{2}+\lvert\beta_{\bm{k}}^{\vphantom{\dagger}}\rvert&=\cosh2\eta_{\bm{k}}\,,&2\alpha_{\bm{k}}^{\vphantom{\dagger}}\beta_{\bm{k}}^{\vphantom{\dagger}*}&=-e^{i\theta_{\bm{k}}}\sinh2\eta_{\bm{k}}\,.
\end{align}
In fact, \eqref{E:dckfgue} and \eqref{E:mnbjwie} only determine $\alpha_{\bm{k}}$, $\beta_{\bm{k}}$ up to a phase factor or a rotation, which we have been ignoring. For example, from \eqref{E:mnbjwie}, we may let
\begin{align}
	\alpha_{\bm{k}}&=\frac{1}{2\sqrt{\omega_{i}\omega_{f}}}\Bigl[+i\,\omega_{f}\,d_{\bm{k}}^{(1)}(t_{f})+\omega_{i}\omega_{f}\,d_{\bm{k}}^{(2)}(t_{f})-\dot{d}_{\bm{k}}^{(1)}(t_{f})+i\,\omega_{i}\,\dot{d}_{\bm{k}}^{(2)}(t_{f})\Bigr]\,,\\
	\beta_{\bm{k}}&=\frac{1}{2\sqrt{\omega_{i}\omega_{f}}}\Bigl[-i\,\omega_{f}\,d_{\bm{k}}^{(1)}(t_{f})+\omega_{i}\omega_{f}\,d_{\bm{k}}^{(2)}(t_{f})+\dot{d}_{\bm{k}}^{(1)}(t_{f})+i\,\omega_{i}\,\dot{d}_{\bm{k}}^{(2)}(t_{f})\Bigr]\,,
\end{align}
and then we can directly show that \eqref{E:dckfgue} is recovered and 
\begin{align}
	\lvert\alpha_{\bm{k}}^{\vphantom{\dagger}}\rvert^{2}-\lvert\beta_{\bm{k}}^{\vphantom{\dagger}}\rvert^{2}=d_{\bm{k}}^{(1)}(t_{f})\,\dot{d}_{\bm{k}}^{(2)}(t_{f})-\dot{d}_{\bm{k}}^{(1)}(t_{f})d_{\bm{k}}^{(2)}(t_{f})=1\,.
\end{align}
However, obviously in this case $\alpha_{\bm{k}}$ is not real, as \eqref{E:lkgdkf} implies.

\begin{figure}
	\centering
	\includegraphics[width=0.95\textwidth]{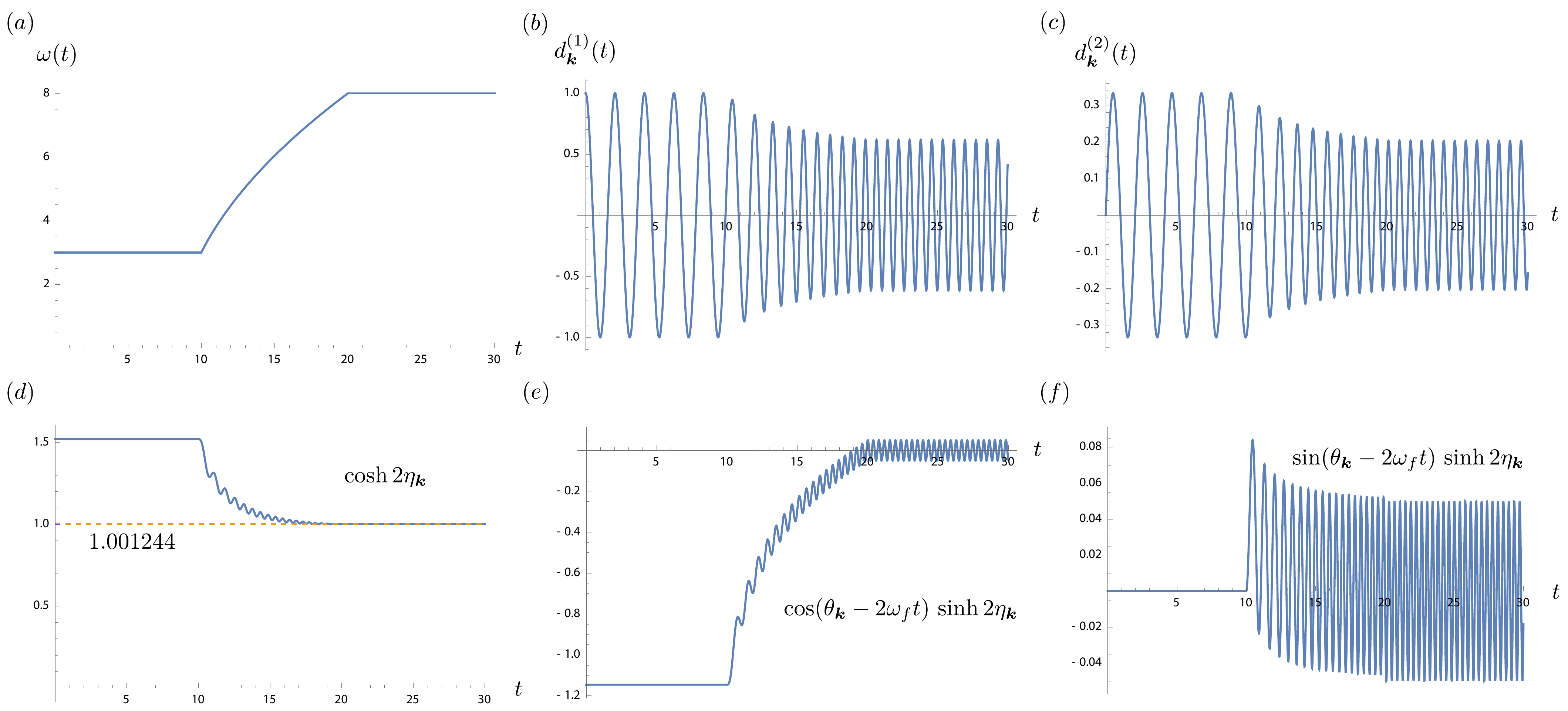}
	\caption{An example of a parametric process described by \eqref{E:fgbkdfss}. We choose $\omega_{i}=3$, $\omega_{f}=8$, $t_{a}=10$ and $t_{b}=20$.}\label{Fi:example1}
\end{figure}

We may recover the missing phase by putting back the rotation operator we have ignored all along. To be more specific, for example, let the rotation operator $\hat{R}$ be given by
\begin{equation}
	\hat{R}(\Phi_{i})=\exp\Bigl[i\Phi_{i}\bigl(\hat{a}_{i}^{\dagger}\hat{a}_{i}^{\vphantom{\dagger}}+\frac{1}{2}\bigr)\Bigr]\,.
\end{equation}
We find $\hat{R}^{\dagger}(\Phi_{i})\,\hat{a}_{1}\hat{R}(\Phi_{1})=\hat{a}_{1}\,e^{i\Phi_{1}}$. Then,  depending on the order of the rotation operator and the squeeze operator, we may have either
\begin{align*}
	\hat{S}_{2}^{\dagger}(\zeta)\hat{R}^{\dagger}(\Phi_{i})\,\hat{a}_{1}\hat{R}(\Phi_{i})\hat{S}_{2}^{\vphantom{\dagger}}(\zeta)&=e^{i\Phi_{1}}\cosh\eta\,\hat{a}_{1}-e^{i\Phi_{1}}e^{i\theta}\sinh\eta\,\hat{a}_{2}\,,
	\intertext{or}
	\hat{R}^{\dagger}(\Phi_{i})\hat{S}_{2}^{\dagger}(\zeta)\,\hat{a}_{1}\hat{S}_{2}^{\vphantom{\dagger}}(\zeta)\hat{R}(\Phi_{i})&=e^{i\Phi_{1}}\cosh\eta\,\hat{a}_{1}-e^{i\Phi_{2}}e^{i\theta}\sinh\eta\,\hat{a}_{2}\,.
\end{align*}
We see that there is always a phase ambiguity. In both cases, from \eqref{E:lkgdkf} and \eqref{E:fuedhfser}, we find $\alpha_{\bm{k}}$ will be complex in general, but in principle, we may factor out the overall phase factor for each mode to render $\alpha_{\bm{k}}$ real.

Before we proceed further with our analysis, let us examine a few illustrative examples. 
\paragraph{\textbf{Case 1}}
Consider the parametric process in which the squared frequency $\omega^{2}(t)$ varies with time according to
\begin{equation}\label{E:fgbkdfss}
	\omega^{2}(t)=\begin{cases}
					\omega_{i}^{2}\,,&0\leq t\leq t_{a}\,,\\
					\omega_{i}^{2}+\bigl(\omega_{f}^{2}-\omega_{i}^{2}\bigr)\,\dfrac{t-t_{a}}{t_{b}-t_{a}}\,,&t_{a}\leq t\leq t_{b}\,,\\
					\omega_{f}^{2}\,,&t\geq t_{b}\,,
				\end{cases}
\end{equation}
That is, $\omega^{2}(t)$ is a piecewise-continuous function of time. The time evolution of $d_{\bm{k}}^{(1)}(t)$, $d_{\bm{k}}^{(2)}(t)$ are shown in Fig.~\ref{Fi:example1}-($b$) and -($c$), where we observe that the oscillation amplitudes of the two fundamental solutions change by different amounts, implying the occurrence of quantum squeezing. Notice in Fig.~\ref{Fi:example1}-($d$), as $t\geq t_{b}$ the squeeze parameter $\eta_{\bm{k}}$ becomes a constant,  squeezing is very small because $\cosh2\eta_{\bm{k}}\sim1$. This small squeezing results from the slow transition rate in the parametric process. The plots ($e$) and ($f$) show oscillations of frequency $2\omega_{f}$, consistent with expectation. However, from these plots it is hard to tell whether $\theta_{\bm{k}}$ is time independent. We will take another approach to show it later.

\begin{figure}
	\centering
	\includegraphics[width=0.95\textwidth]{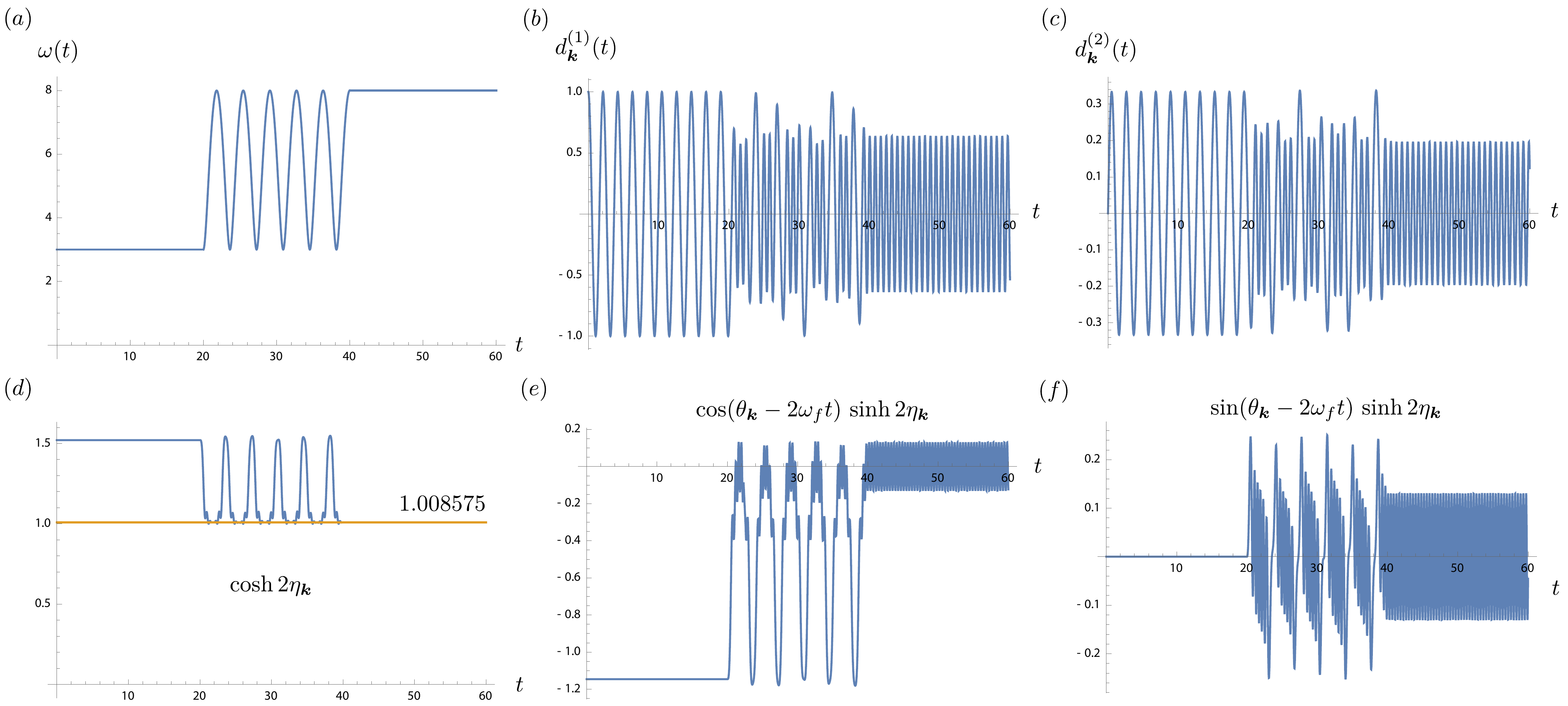}
	\caption{An example of a parametric process described by \eqref{E:opwesndkwe}. We choose $\omega_{i}=3$, $\omega_{f}=8$, $t_{a}=10$, $t_{b}=20$ and $n=11$.}\label{Fi:example2}
\end{figure}

\paragraph{\textbf{Case 2}} In the second example, we choose a non-monotonic parametric process,
\begin{equation}\label{E:opwesndkwe}
	\omega^{2}(t)=\begin{cases}
					\omega_{i}^{2}\,,&0\leq t\leq t_{a}\,,\\
					\omega_{i}^{2}+\bigl(\omega_{f}^{2}-\omega_{i}^{2}\bigr)\,\sin^{2}\Bigl[\dfrac{t-t_{a}}{t_{b}-t_{a}}\dfrac{n\pi}{2}\Bigr]\,,&t_{a}\leq t\leq t_{b}\,,\\
					\omega_{f}^{2}\,,&t\geq t_{b}\,.
				\end{cases}
\end{equation}
Here $n$,  an odd integer, gives the number of oscillations of the transition process. Thus the frequency variation is not monotonic. The corresponding plots are shown in Fig.~\ref{Fi:example2}. Since both processes, though continuous, are not smooth, it allows us to clearly see when the transitions start and end. Compared with Case 1, the non-monotonic transition introduces tumultuous behavior in the fundamental solutions $d_{\bm{k}}^{(i)}$ during the transition, in Fig.~\ref{Fi:example2}-($b$) and -($c$), but right after $t=t_{b}$, the out mode immediately oscillates at frequency $\omega_{f}$. Again,  we can see the time-independence of $\eta_{\bm{k}}$ in the out-region.

For the simple case we consider here, let us carry out some analysis about the time dependence of the squeeze parameter in the out-region. We may write $d_{\bm{k}}^{(i)}(t)$ for $t\geq t_{b}$ as
\begin{align}
	d_{\bm{k}}^{(1)}(t)=d_{\bm{k}}^{(1)}(t_{b})\,\cos\omega_{f}(t-t_{b})+d_{\bm{k}}^{(2)}(t_{b})\,\frac{1}{\omega_{f}}\,\sin\omega_{f}(t-t_{b})\,,\label{E:ejersnj1}\\
	d_{\bm{k}}^{(2)}(t)=d_{\bm{k}}^{(1)}(t_{b})\,\frac{1}{\omega_{f}}\,\sin\omega_{f}(t-t_{b})+d_{\bm{k}}^{(2)}(t_{b})\,\cos\omega_{f}(t-t_{b})\label{E:ejersnj2}\,.
\end{align}
The amplitudes of $d_{\bm{k}}^{(i)}(t)$ in the out-region are determined by
\begin{align}
	&d_{\bm{k}}^{(1)}(t): &&\Biggl[d_{\bm{k}}^{(1)2}(t_{b})+\frac{d_{\bm{k}}^{(2)2}(t_{b})}{\omega_{f}^{2}}\Biggr]^{\frac{1}{2}}\,,&&\text{and}&&d_{\bm{k}}^{(2)}(t): &&\Biggl[\frac{d_{\bm{k}}^{(1)2}(t_{b})}{\omega_{f}^{2}}+d_{\bm{k}}^{(2)2}(t_{b})\Biggr]^{\frac{1}{2}}\,,
\end{align}
which in turns are determined by the values of the fundamental solution at the end of the parametric process. We readily find that $d_{\bm{k}}^{(1)}(t)\dot{d}_{\bm{k}}^{(1)}(t)-\dot{d}_{\bm{k}}^{(1)}(t)d_{\bm{k}}^{(1)}(t)=1$ for all $t$ even for the parametric process via the Wronskian of the differential equation the mode function satisfies.

According to Eqs.~\eqref{E:mnbjwie}, we can show
\begin{align}\label{E:jbkiwuew}
	\lvert\alpha_{\bm{k}}^{\vphantom{\dagger}}(t)\rvert^{2}+\lvert\beta_{\bm{k}}^{\vphantom{\dagger}}(t)\rvert^{2}=\lvert\alpha_{\bm{k}}^{\vphantom{\dagger}}(t_b)\rvert^{2}+\lvert\beta_{\bm{k}}^{\vphantom{\dagger}}(t_b)\rvert^{2}
\end{align}
for all $t\geq t_{b}$. Since $\lvert\alpha_{\bm{k}}^{\vphantom{\dagger}}\rvert^{2}+\lvert\beta_{\bm{k}}^{\vphantom{\dagger}}\rvert^{2}$ is proportional to $\cosh2\eta_{\bm{k}}$, Eq.~\eqref{E:jbkiwuew} then shows that $\eta_{\bm{k}}$ is a time-independent constant for all $t\geq t_{b}$. On the other hand, with the help of \eqref{E:ejersnj1} and \eqref{E:ejersnj2}, we find
\begin{figure}
	\centering
	\includegraphics[width=0.95\textwidth]{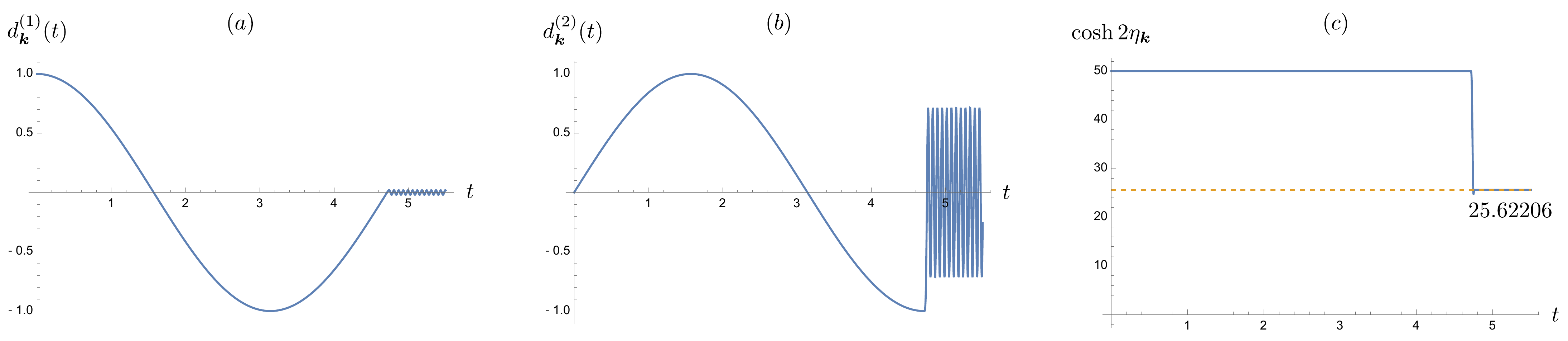}
	\caption{An example of a parametric process described by \eqref{E:opwesndkwe}, but with $\omega_{i}=1$, $\omega_{f}=100$, $t_{a}=3\pi/2$, $t_{b}=3\pi/2+0.05$ and $n=1$.}\label{Fi:example4}
\end{figure}\begin{align}\label{E:uewiugfdew}
	\frac{1}{\omega_{f}\omega_{i}}\,\dot{d}_{\bm{k}}^{(1)2}(t)+\frac{\omega_{i}}{\omega_{f}}\,\dot{d}_{\bm{k}}^{(2)2}(t)-\frac{\omega_{f}}{\omega_{i}}\,d_{\bm{k}}^{(1)2}(t)-\omega_{f}\omega_{i}\,d_{\bm{k}}^{(2)2}(t)&=A_{\bm{k}}\,\sin(2\omega_{f}t_{b}-2\omega_{f}t)+B_{\bm{k}}\,\cos(2\omega_{f}t_{b}-2\omega_{f}t)\notag\\
	&=\sqrt{A_{\bm{k}}^{2}+B_{\bm{k}}^{2}}\,\cos(\vartheta_{\bm{k}}+2\omega_{f}t_{b}-2\omega_{f}t)\,,
\end{align}
with
\begin{align}
	A_{\bm{k}}&=2\biggl[\frac{1}{\omega_{i}}\,d_{\bm{k}}^{(1)}(t_{b})\,\dot{d}_{\bm{k}}^{(1)}(t_{b})+\omega_{i}\,d_{\bm{k}}^{(2)}(t)\,\dot{d}_{\bm{k}}^{(2)}(t)\biggr]\,,\label{E:eigsvdj1}\\
	B_{\bm{k}}&=\biggl[\frac{1}{\omega_{f}\omega_{i}}\,\dot{d}_{\bm{k}}^{(1)2}(t_{b})+\frac{\omega_{i}}{\omega_{f}}\,\dot{d}_{\bm{k}}^{(2)2}(t_{b})-\frac{\omega_{f}}{\omega_{i}}\,d_{\bm{k}}^{(1)2}(t_{b})-\omega_{f}\omega_{i}\,d_{\bm{k}}^{(2)2}(t_{b})\biggr]\,.\label{E:eigsvdj2}
\end{align}
and
\begin{align}
	\cos\vartheta_{\bm{k}}&=\frac{B_{\bm{k}}}{\sqrt{A_{\bm{k}}^{2}+B_{\bm{k}}^{2}}}\,,&\sin\vartheta_{\bm{k}}&=-\frac{A_{\bm{k}}}{\sqrt{A_{\bm{k}}^{2}+B_{\bm{k}}^{2}}}\,.
\end{align}
From \eqref{E:kbsdkskdu2} and \eqref{E:kbsdkskdu3}, we see $\sqrt{A_{\bm{k}}^{2}+B_{\bm{k}}^{2}}$ is proportional to $\sinh2\eta_{\bm{k}}$, and thus is a time-independent constant. We immediately can identify $\vartheta_{\bm{k}}$ in Eqs.~\eqref{E:uewiugfdew}--\eqref{E:eigsvdj2} to be the same $\theta_{\bm{k}}$ in \eqref{E:kbsdkskdu1}--\eqref{E:kbsdkskdu3}. Finally, we may note that \eqref{E:uewiugfdew} still looks slightly different from the lefthand sides of \eqref{E:kbsdkskdu1}--\eqref{E:kbsdkskdu3} in the arguments of the trigonometric functions. It results from the choice of the out mode function in \eqref{E:fuedhfser}. The reason for such a choice is that for an observer in the out-region, he has no reference to identify the origin of time coordinate. In addition, the choice of a fixed time origin at most amounts to an absolute phase, which in most cases is of no significance. However, in the current case we are comparing two formalisms. So for   consistency's sake, we may define the origin of the time coordinate in the out-region at $t_{b}$ rather than $0$. Simply shifting $t$ in \eqref{E:fuedhfser} to $t\to t-t_{b}$,   then \eqref{E:uewiugfdew} will look the same as those in \eqref{E:kbsdkskdu1}--\eqref{E:kbsdkskdu3}. At this point, we have shown that in the out-region an observer may report on having experienced a quantum field in a squeezed thermal state, with a time-independent squeeze parameter.

\begin{figure}
	\centering
	\includegraphics[width=0.95\textwidth]{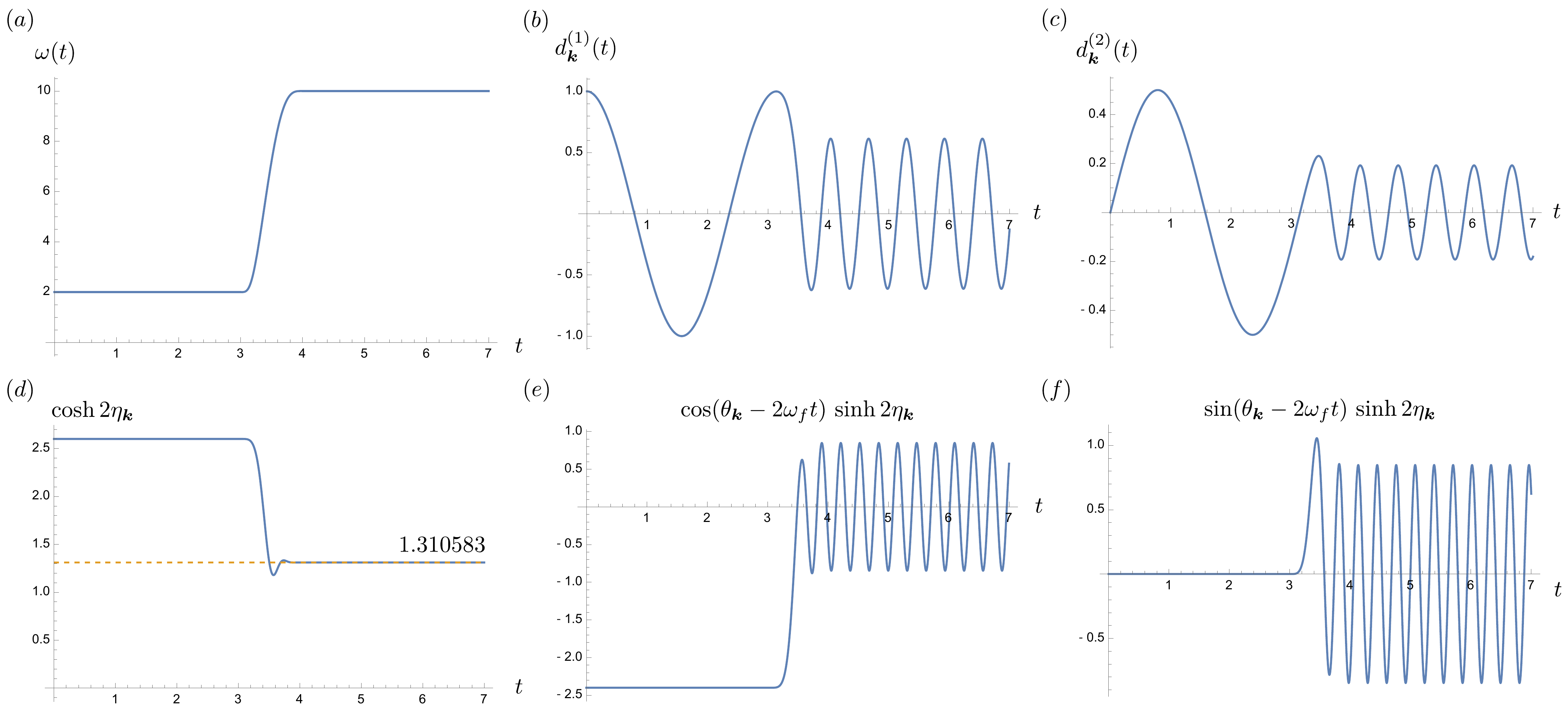}
	\caption{An example of a parametric process described by \eqref{E:dgsjehrs}, but with $\omega_{i}=2$, $\omega_{f}=10$, $t_{a}=3$, and $t_{b}=4$.}\label{Fi:example5}
\end{figure}

The squeeze parameter $\eta_{\bm{k}}$ we showed earlier is quite small because $\cosh2\eta_{\bm{k}}\sim1$. This results from the fact that the parametric processes in the previous cases vary mildly. Now we show a case similar to that in Fig.~\ref{Fi:example2} but with a much sharper transition with $\omega_{i}=1$, $\omega_{f}=100$, $t_{a}=3\pi/2$, $t_{b}=3\pi/2+0.05$ and $n=1$. The plot for $\cosh2\eta_{\bm{k}}$ is shown in Fig.~\ref{Fi:example4}. We see in this case $\cosh2\eta_{\bm{k}}=25.62206$ in the out-region, much larger than the previous two cases. Therefore, it is consistent with the understanding that to generate large squeezing, the parametric process had better not be adiabatic.  This is corroborated by our prior knowledge of cosmological particle production.

Finally,  as a contrast to the previous piecewise-continuous parametric processes, we consider a sufficiently smooth $\omega^{2}(t)$, such as the thrice differentiable function $\omega^{2}(t)$ 
\begin{equation}\label{E:dgsjehrs}
	\omega^{2}(t)=-\frac{(t-t_{a})^{4}}{(t_{b}-t_{a})^{7}}\bigl[20t^{3}-10(7t_{b}^{\vphantom{2}}-t_{a}^{\vphantom{2}})\,t^{2}+4(21t_{b}^{2}-7t_{b}^{\vphantom{2}}t_{a}^{\vphantom{2}}+t_{a}^{2})\,t-(35t_{b}^{3}-21t_{b}^{2}t_{a}^{\vphantom{2}}+7t_{b}^{\vphantom{2}}t_{a}^{2}-t_{a}^{3})\bigr]\,,
\end{equation}
between $t=t_{a}$ and $t=t_{b}$. The corresponding results are shown in Fig.~\ref{Fi:example5}. We do not discern any difference in the generic behavior, so for the quantities of interest in our present  study, the piecewise-continuous $\omega^{2}(t)$ suffices.

The fact that the squeeze parameter becomes time-independent in the out-region implies that this result is time-translation invariant. To be more precise, if we shift the parametric process along the time axis, then as long as the conditions that 1) the observation time $t$ is still in the out-region of the shifted process and 2) the initial time $t_i$ remains in the in-region of the shifted process, are satisfied, the observer at time $t$ will still see the same squeeze parameter. In other words, the observer in the out-region cannot extract from the squeeze parameter any information about when the parametric process starts or ends. At first, it sounds odd that there exists such a time-translation  invariance for a nonequilibrium, nonstationary process. We show this in Appendix~\ref{S:eerudj}; however, after seeing these examples, this result may not appear that dubious. The full evolution from the initial time $t_i$ to the observation time $t$ is not invariant under time translation, but after the parametric process has ended at $t_b$, the mode functions in the out-region reverts to  a  sinusoidal time dependence. The time translation in a sinusoidal function of time amounts to a phase shift.  Thus, if the  quantity of interest is independent of this phase shift, it appears to possess time-translation invariance.

\begin{figure}
	\centering
	\includegraphics[width=0.95\textwidth]{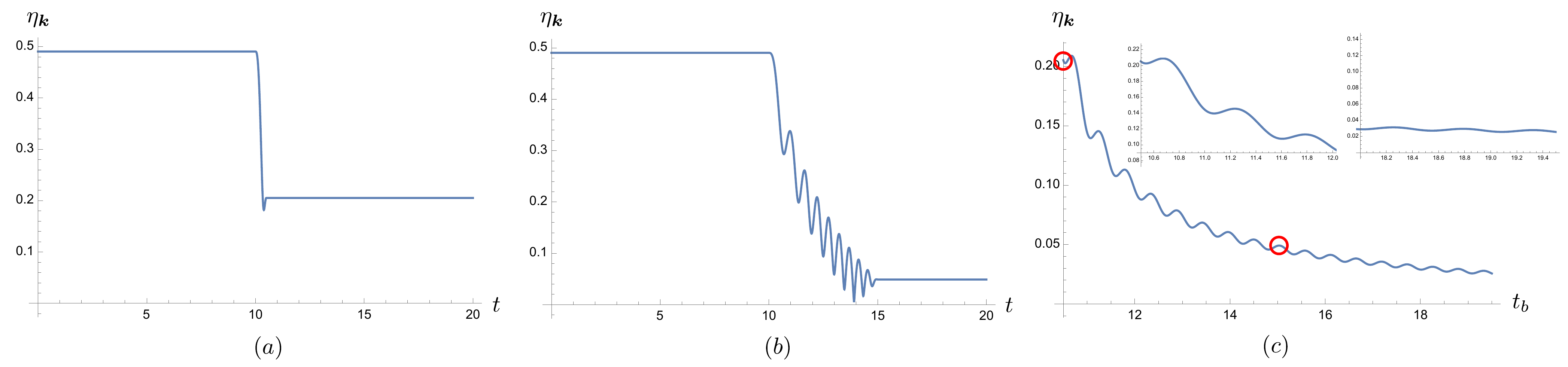}
	\caption{In (c) we show the $t_{b}$ dependence of the squeeze parameter $\eta_{\bm{k}}$. We choose the parametric process described by \eqref{E:fgbkdfss} to have $\omega_{i}=3$, $\omega_{f}=8$, $t_{i}=0$, $t_{a}=10$. A fictitious detector is placed at $t=20$, and we vary $t_{b}$ from 10.5 to 19.5. For comparison, we include the time dependence of $\eta_{\bm{k}}$ when (a) $t_{b}=10.5$ and (b) $t_{b}=15$. These two cases are highlighted in (c) by red circles.}\label{Fi:squeezeTb}
\end{figure}

Although in the current configuration the detector in the out-region will sense the same squeeze parameter when the parametric process is shifted along the time axis, the measured results in the detector still depend on the duration $t_{b}-t_{a}$ and the functional form of the process. In fact this can be expected from \eqref{E:kbsdkskdu1}--\eqref{E:kbsdkskdu3}, where the squeeze parameters are expressed in terms of the fundamental solutions, which in turn depend on the functional form of the parametric process in their equation of motion.

For illustrations, Fig.~\ref{Fi:squeezeTb}-($c$) shows the dependence of the squeeze parameter $\eta_{\bm{k}}$ on the duration of the parametric process, given by Eq.~\eqref{E:fgbkdfss}. We fix the starting time $t_a$ of the parametric process and the moment $t$ the measurement in the out-region is performed. We find that the squeeze parameter $\eta_{\bm{k}}$, defined in \eqref{E:kbsdkskdu1}, is oscillatory but decreases with increasing ending time $t_{b}$. Except for the small oscillations, the curve in ($c$) gives the general trend that the squeezing (as manifested, e.g., in particle pair production), is subdued with a slower transition rate or longer transition duration $t_b-t_a$. This example shows that this kind of parametric process gives a lower production of particles as it moves toward the adiabatic regime. The mild oscillations may be related to the kinks at $t_{a}$ and $t_{b}$ due to the nonsmoothness of $\omega(t)$. This may be seen from the observation that the oscillations, as shown in the blow-ups in Fig.~\ref{Fi:squeezeTb}-($c$), appear shallower with larger $t_{b}-t_{a}$ because the kinks are less abrupt. This argument also find its support from Fig.~\ref{Fi:squeezeff}-($b$) and -($d$).

In Fig.~\ref{Fi:squeezeff}, the dependence of the squeeze parameter $\eta_{\bm{k}}$ on the functional form of the parametric process is shown. We choose three different parametric processes of $\omega^2(t)$ for examples. In Fig.~\ref{Fi:squeezeff}-($a$), the process has the piecewise-continuous form given by Eq.~\eqref{E:fgbkdfss}, so it gives a same result as Fig.~\ref{Fi:squeezeTb}-(c). Fig.~\ref{Fi:squeezeff}-($b$) is described by the smooth transition, Eq.~\eqref{E:dgsjehrs}. We have the same process, Eq.~\eqref{E:opwesndkwe}, for Fig.~\ref{Fi:squeezeff}-($c$) and -($d$) but with different choices of $n$. We choose $n=11$ for ($c$) and $n=1$ for ($d$). Thus in plot ($c$) the transition does not monotonically vary with time. 
\begin{figure}
	\centering
	\includegraphics[width=0.99\textwidth]{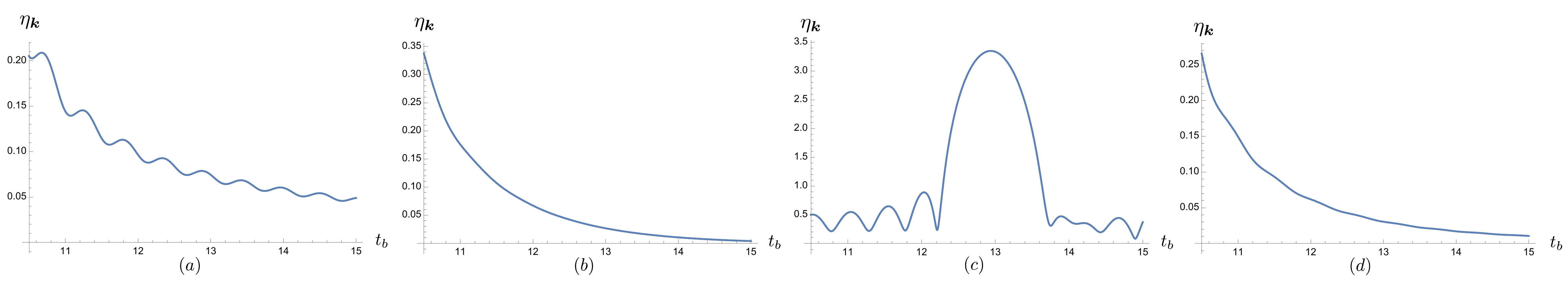}
	\caption{The dependence of the squeeze parameter $\eta$ on the functional form of $\omega^2(t)$. The functional form of the parametric process is described by Eq.~\eqref{E:fgbkdfss} in (a), and Eq.~\eqref{E:dgsjehrs} in (b). The same Eq.~\eqref{E:opwesndkwe} is used in (c) with $n=11$, and (d) with $n=1$. The relevant parameters are $\omega_{i}=3$, $\omega_{f}=8$, $t_{i}=0$, and $t_{a}=10$. The measurement is carried out at $t=20$, and we vary $t_{b}$ from 10.5 to 15.}\label{Fi:squeezeff}
\end{figure}

Since both Fig.~\ref{Fi:squeezeff}-($b$) and -($d$) are associated with smooth parametric processes, when we prolong the duration of the parametric processes by fixing $t_{a}$ and increasing $t_b$, we see that the value of $\eta_{\bm{k}}$ monotonically decreases without any ripples. Thus, together with the behavior of the curve in Fig.~\ref{Fi:squeezeTb}-($c$), they imply that the ripples in Fig.~\ref{Fi:squeezeff}-($a$) may originate from the non-smoothness of the parametric process at the transition times $t_a$ and $t_b$.

Among the plots in Fig.~\ref{Fi:squeezeff}, Fig.~\ref{Fi:squeezeff}-($c$) shows a more interesting behavior. The transition described by \eqref{E:opwesndkwe} with $n=11$ is sinusoidal, as shown in Fig.~\ref{Fi:example2}-(a). The squeeze parameter $\eta_{\bm{k}}$ in Fig.~\ref{Fi:squeezeff}-(c) then reveals more structures, and may signify the presence of resonance, shown by a large peak. It seems to imply that the particle production can be greatly enhanced if the transition duration is tuned to the right value for fixed $\omega_{i}$ and $\omega_{f}$, as in parametric resonance. This  may constitute  a mechanism to generate stronger squeezing, in addition to the common one  in a runaway setting~\cite{GP85,HH22a}. Another unusual feature  is that $\eta_{\bm{k}}$ in Fig.~\ref{Fi:squeezeff}-(c) is not significantly reduced when we have a larger $t_{b}$.

In fact, this resonance feature can be traced back to parametric instability. If we write the equation of motion for a parametric oscillator with the frequency given by \eqref{E:opwesndkwe} into the canonical form
\begin{equation}\label{E:shdskrrs}
	x''(\tau)+\bigl[a-2q\,\cos(2\tau)\bigr]\,x(\tau)=0\,,
\end{equation}
where a prime represents taking the derivative with respect to $\tau$,  $\Omega\, t=2\tau$, and
\begin{align}
	a&=\frac{4A^{2}}{\Omega^{2}}\,,&q&=\frac{2B^{2}}{\Omega^{2}}\,,&\Omega&=\frac{n\pi}{t_{b}-t_{a}}\,,&A^{2}&=\frac{\omega_{f}^{2}+\omega_{i}^{2}}{2}\,,&B^{2}&=\frac{\omega_{f}^{2}-\omega_{i}^{2}}{2}\,.
\end{align}
Now compare the stability diagram Fig.~\ref{Fi:squeezeIns}-($a$) of \eqref{E:shdskrrs} with Fig.~\ref{Fi:squeezeff}-(c). We immediately see that the portion of the curve in Fig.~\ref{Fi:squeezeIns}-(b) that corresponds to exceptionally high squeezing is essentially located within the unstable region (white area) of the stability diagram in Fig.~\ref{Fi:squeezeIns}-(a). Thus, the large squeezing in the mode driven by a non-monotonic $\omega(t)$ is caused by instability in parametric resonance.

\begin{figure}
	\centering
	\includegraphics[width=0.99\textwidth]{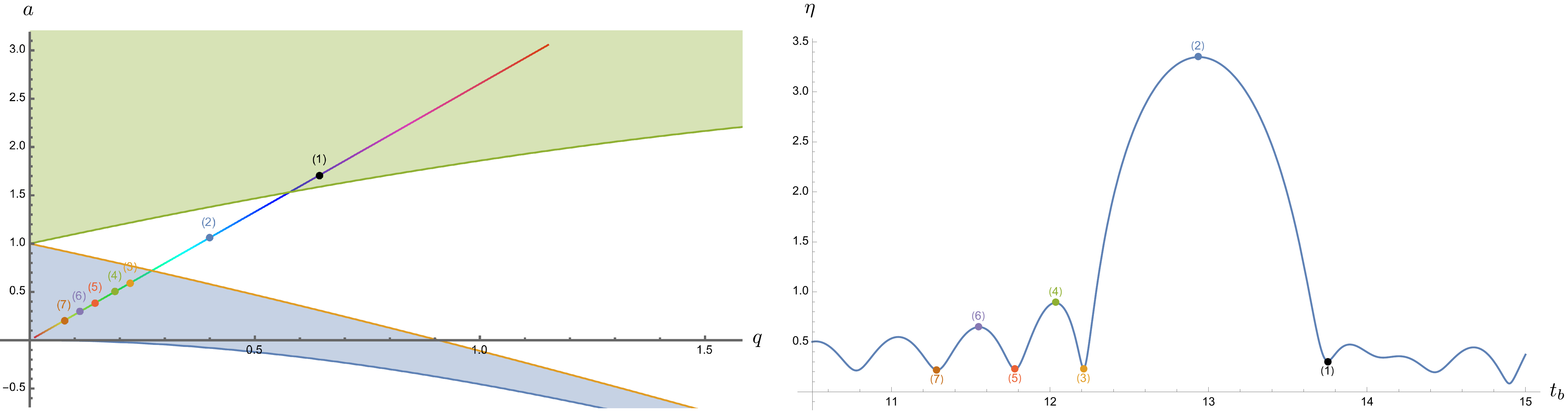}
	\caption{Comparison between the stability diagram of \eqref{E:shdskrrs} on the left with Fig.~\ref{Fi:squeezeff}-(c), shown on the right here. The motion of the parametric oscillator \eqref{E:shdskrrs} is stable in color-shaded regions. We highlight a few peaks an valley of the curve in (b), and the corresponding points are shown in the stability diagram on the left. The straight line on the left panel with varying hue from red, yellow, green, cyan, blue, and magenta corresponds to the time axis $t_{b}$ on the right panel from $t_{b}=10.5$ to 15. The parameters in this plot are the same as in Fig.~\ref{Fi:squeezeff}.}\label{Fi:squeezeIns}
\end{figure}

The above simple examples illustrate a point that the dependence of the squeeze parameter $\eta_{\bm{k}}$ on the duration of the parametric process does reflect qualitative features of the parametric process. In fact, we can apply the same idea to the spectral dependence of the squeeze parameter for various functional forms of the parametric processes. These will serve as templates from which we may extract qualitative information about the parametric process the quantum field has experienced.

%\textcolor{red}{(so far we have not explained how we can choose mode-independent squeeze parameters?)}

\section{Summary}\label{S:etudguei}

A major task of theoretical cosmology is to find ways to deduce the state of the early universe from relics observed today such as features of primordial radiation and matter contents. In addition to particle creation at the Planck time and structure formation after the GUT (grand unified theory) time we add here another fundamental quantum process: quantum radiation from atoms induced by of quantum fields.

%{With in mind the motivation how nonMarkovian in quantum field~\cite{nMCos} may provide a means to search for information about the early universe, in this paper we set up a simplified configuration toward this goal.} 

A massless quantum field in Minkowski space is subjected to a parametric process which varies the frequencies of the normal modes of the field from one constant value to another over a finite time interval. The initial (in-) state of the field will become squeezed after the process. In the out-region we couple an atom to this field in the out-state; the response of the atom certainly depends on this squeezing. Meanwhile, the radiation generated by the interacting atom will send out signals outwards, such that an observer or a probe at a much later time may still be able to identify the squeezing via the signatures in the radiation.

{Once we have identified the squeezing, locally or remotely, we then ask how the squeezing may depend on the details of the parametric process the field had been subjected to. If we can work out templates that relate these quantities, then measurements of certain features of squeezing can tell us something about the parametric process.}

The salient points in our results are summarized below. In the first part of this paper we show the following:
\begin{itemize}
    \item The radiation field at a location far away from the atom looks stationary; its nonstationary component decays with time exponentially fast.
    \item The  net energy flow cancels at late times like the case discussed in~\cite{QRadVac}. These features are of particular interest considering that the atom that emits this radiation is coupled to a squeezed field, which is nonstationary by nature. However they are consistent with the fact that the atom's internal dynamics relaxes in time.
    \item This implies that we are unable to measure the extent of squeezing by measuring the net radiation energy flow at a location  far away from the atom.
    \item On the other hand, one can receive residual radiation energy density at late times, which is a time-independent constant and is related to the squeeze parameter. But it is of a near-field nature, so the observer cannot be located too far away from the atom.
\end{itemize}
Note that there is always an ambient free component of the same squeezed quantum field everywhere, in addition to the above radiation component generated by the atom's internal motion. 

 In the second part of this paper we focus on the dependence of the squeeze parameter on the parametric process  and obtain these results:
\begin{itemize}
    \item Formally it can be shown that the squeeze parameter depends on the evolution of the field in the parametric process.
    \item In the current configuration, for a given parametric process, {\it the squeeze parameter depends only on the duration of the process}, it does not depend on the starting or the ending time of the process.
    \item In general, for a {\it monotonically varying process}, the value of the squeeze parameter decreases with increasing duration of the process.
    \item This implies that, for an adiabatic parametric process, the squeezing tends to be very small, but it can be quite {\it significant for non-adiabatic parametric processes}. These results are consistent with studies of cosmological particle creation in the 70s as parametric amplification of vacuum fluctuations.
    \item If the parametric process changes with time sinusoidally, then the dependence of the squeeze parameter on the duration of the process shows interesting additional structures. For certain lengths of the process, the squeeze parameter can have unusually large values.
    \item This non-monotonic behavior turns out to be related to {\it parametric instability}. The resulting large squeeze parameter is caused by the choice of the parameter that falls within the unstable regime of the parametric process.
\end{itemize}
Our results in this second part indicate that not much information about a monotonic parametric process can be gleaned off from the squeeze parameter. On the other hand, {\it for a non-monotonic parametric process, the squeeze parameter shows nontrivial dependence on the duration of the process}. The latter is expected to be the typical case in nature in which various scales are involved in a transition.  Moreover, to conform to realistic measurements, we can {\it apply similar considerations to examine the mode dependence of the squeeze parameter}. Thus the current scenario seems to offer a viable means to extract partial information about the parametric process a quantum field has undergone.

As a final remark, we comment on applying the discussions in Sec.~\ref{S:eiutie} to the case of the atom coupled to quantized electromagnetic fields. To begin with, in conventional sense, such a configuration has a supra-Ohmic, Markovian dynamics. It is inherently unstable, as discussed in~\cite{HH22}. It has runaway solutions, so it makes no sense to discuss the late time  behavior, relaxation, energy balance or the fluctuation-dissipation relation of such a system. Although we may render such a system to have stable dynamics by resorting to order reduction~\cite{Sp00} via the critical manifold argument, this approach is not very satisfactory from the viewpoint of  open systems because order reduction only asymmetrically changes the behavior of the dissipation on the atom side; it does not accordingly modify the fluctuation noise force on the atom. Thus the resulting nonequilibrium fluctuation-dissipation relation associated with the atom's internal dynamics will not take the elegant form we usually see for the Ohmic, Markovian dynamics. Furthermore this relation will depend on the parameters of the reduced system, and thus loses its universality for  interacting linear systems and for some nonlinear systems. 

To restore the stable dynamics of the atom's internal dynamics~\cite{HH22} and the beauty of the associated nonequilibrium fluctuation-dissipation relation, it is probably easiest if we generalize the Markovian spectrum of the quantized electromagnetic field to the non-Markovian one.  However this introduces additional complexity to the analysis presented in Sec.~\ref{S:eiutie} because the radiation field will not have a simple local (apart from retardation) form like \eqref{E:ieusgloet}. Thus it is not clear yet whether the rather general properties we expounded in Sec.~\ref{S:eiutie} for a quantum scalar field also convey to a quantized electromagnetic field, a topic saved for future investigations.\\

\textbf{Acknowledgment} J.-T. Hsiang is supported by the Ministry of Science and Technology of Taiwan, R.O.C. under Grant No.~MOST 111-2811-M-008-022. B.-L. Hu appreciates the kind hospitality of Prof. Chong-Sun Chu of the National Center for Theoretical Sciences and Prof. Hsiang-nan Li of the Academia Sinica, R.O.C. in the finishing stage of this paper.

\appendix
\section{two-mode squeezed state}\label{S:eoueisf}
Here we outline the properties of the two-mode squeezed operator $\hat{S}_2(\zeta)$ associated mode 1 and 2,
\begin{equation}\label{E:gvsjhfgd}
	\hat{S}_{2}(\zeta)=\exp\Bigl[\zeta^{*}\hat{a}_{1}^{\vphantom{\dagger}}\hat{a}_{2}^{\vphantom{\dagger}}-\zeta\,\hat{a}_{1}^{\dagger}\hat{a}_{2}^{\dagger}\Bigr]\,,
\end{equation}
such that the two-mode squeezed thermal state is defined by
\begin{equation}
	\hat{\rho}_{\textsc{tmst}}=\hat{S}_{2}^{\vphantom{\dagger}}(\zeta)\,\hat{\rho}_{\beta}\,\hat{S}_{2}^{\dagger}(\zeta)\,.
\end{equation}
where $\hat{\rho}_{\beta}$ is a thermal state. The creation and annihilation operators satisfy the standard commutation relation $[\hat{a},\hat{a}^{\dagger}]=1$. It is convenient to put the squeeze parameter $\zeta$ into a polar form $\zeta=\eta\,e^{i\theta}$, with $\eta\in\mathbb{R}^{+}$ and $0\leq\theta<2\pi$. We thus have
\begin{align}
	\hat{S}_{2}^{\dagger}\,\hat{a}_{1}\,\hat{S}_{2}^{\vphantom{\dagger}}&=\cosh\eta\,\hat{a}_{1}^{\vphantom{\dagger}}-e^{+i\theta}\sinh\eta\,\hat{a}_{2}^{\dagger}\,,&\hat{S}_{2}^{\dagger}\,\hat{a}_{2}\,\hat{S}_{2}^{\vphantom{\dagger}}&=\cosh\eta\,\hat{a}_{2}^{\vphantom{\dagger}}-e^{+i\theta}\sinh\eta\,\hat{a}_{1}^{\dagger}\,.
\end{align}
so that
\begin{align}
	\langle \hat{a}_{i}^{2}\rangle_{\textsc{tmst}}&=\operatorname{Tr}\Bigl\{\rho_{\beta}\,\hat{S}_{2}^{\dagger}\,\hat{a}_{i}^{2}\,\hat{S}_{2}^{\vphantom{\dagger}}\Bigr\}=0\,,&\langle \hat{a}_{i}^{\dagger2}\rangle_{\textsc{tmst}}&=0\,,\label{E:ghvsjg1}\\
	\langle \hat{a}_{1}^{\vphantom{\dagger}}\hat{a}_{1}^{\dagger}\rangle_{\textsc{tmst}}&=\bigl(\bar{n}_{1}+1\bigr)\,\cosh^{2}\eta+\bar{n}_{2}\,\sinh^{2}\eta\,,&\langle \hat{a}_{1}^{\dagger}\hat{a}_{1}^{\vphantom{\dagger}}\rangle_{\textsc{tmst}}&=\bar{n}_{1}\,\cosh^{2}\eta+\bigl(\bar{n}_{2}+1\bigr)\,\sinh^{2}\eta\,,\\
	\langle \hat{a}_{2}^{\vphantom{\dagger}}\hat{a}_{2}^{\dagger}\rangle_{\textsc{tmst}}&=\bigl(\bar{n}_{2}+1\bigr)\,\cosh^{2}\eta+\bar{n}_{1}\,\sinh^{2}\eta\,,&\langle \hat{a}_{2}^{\dagger}a_{2}^{\vphantom{\dagger}}\rangle_{\textsc{tmst}}&=\bar{n}_{2}\,\cosh^{2}\eta+\bigl(\bar{n}_{1}+1\bigr)\,\sinh^{2}\eta\,,\\
	\langle \hat{a_{1}}\hat{a}_{2}\rangle_{\textsc{tmst}}&=-\frac{e^{+i\phi}}{2}\bigl(\bar{n}_{1}+\bar{n}_{2}+1\bigr)\,\sinh2\eta\,,&\langle \hat{a}_{1}^{\dagger}a_{2}^{\dagger}\rangle_{\textsc{tmst}}&=-\frac{e^{-i\phi}}{2}\bigl(\bar{n}_{1}+\bar{n}_{2}+1\bigr)\,\sinh2\eta\,,\label{E:ghvsjg4}
\end{align}
where $\bar{n}_i$ is the mean particle number of the thermal state associated with mode $i$,
\begin{equation}
    \bar{n}_i=\operatorname{Tr}\bigl\{\hat{\rho}_\beta\,\hat{a}_{i}^{\dagger}\hat{a}_{i}^{\vphantom{\dagger}}\bigr\}
\end{equation}

It is interesting to compare with the single-mode squeezed thermal state $\hat{\rho}_{\textsc{st}}$, defined by  
\begin{equation}
	\hat{\rho}_{\textsc{st}}=\hat{S}(\zeta)\,\hat{\rho}_{\beta}\,\hat{S}^{\dagger}(\zeta)\,.
\end{equation}
Here $\hat{S}(\zeta)$ is the squeezed operator
\begin{equation}
	\hat{S}(\zeta)=\exp\Bigl[\frac{1}{2}\,\zeta^{*}\hat{a}^{2}-\frac{1}{2}\,\zeta\,\hat{a}^{\dagger2}\Bigr]\,.
\end{equation}
We then find
\begin{align}
	\langle \hat{a}\rangle_{\textsc{st}}&=0=\langle a^{\dagger}\rangle_{\textsc{st}}\,,\\
	\langle \hat{a}^{2}\rangle_{\textsc{st}}&=-e^{+i\theta}\,\sinh2\eta\,\Bigl(\bar{n}+\frac{1}{2}\Bigr)\,,\qquad\qquad\qquad\langle a^{\dagger2}\rangle_{\textsc{st}}=-e^{-i\theta}\,\sinh2\eta\,\Bigl(\bar{n}+\frac{1}{2}\Bigr)\,,\\
	\langle \hat{a}^{\dagger}\hat{a}\rangle_{\textsc{st}}&=\cosh2\eta\,\Bigl(\bar{n}+\frac{1}{2}\Bigr)-\frac{1}{2}=\cosh2\eta\,\bar{n}+\sinh^{2}\eta\,,\label{E:gbkf}
\end{align}
where $\bar{n}$ is the mean particle number of the thermal state. Both the two single-mode squeezed state and the two-squeezed state are connected. We first write two modes in terms of their normal modes $i=\pm$,
\begin{align}
	\hat{a}_{1}&=\frac{\hat{a}_{+}+\hat{a}_{-}}{\sqrt{2}}\,,&\hat{a}_{2}&=\frac{\hat{a}_{+}-\hat{a}_{-}}{\sqrt{2}}\,,
\end{align}
with $[\hat{a}_{+},\hat{a}_{-}]=0$. Then we can write the two-mode squeeze operator $\hat{S}_{2}$, \eqref{E:gvsjhfgd}, as
\begin{align}
	\hat{S}_{2}&=\exp\Bigl[\frac{\zeta^{*}}{2}\bigl(\hat{a}_{+}^{\vphantom{\dagger}}+\hat{a}_{-}^{\vphantom{\dagger}}\bigr)\bigl(\hat{a}_{+}^{\vphantom{\dagger}}-\hat{a}_{-}^{\vphantom{\dagger}}\bigr)-\frac{\zeta}{2}\,\bigl(\hat{a}_{+}^{\dagger}+\hat{a}_{-}^{\dagger}\bigr)\bigl(\hat{a}_{+}^{\dagger}-\hat{a}_{-}^{\dagger}\bigr)\Bigr]\notag\\
	&=\exp\Bigl[\frac{\zeta^{*}}{2}\,\hat{a}_{+}^{\vphantom{\dagger}2}-\frac{\zeta}{2}\,\hat{a}_{+}^{\dagger2}\Bigr]\times\exp\Bigl[-\frac{\zeta^{*}}{2}\,\hat{a}_{-}^{\vphantom{\dagger}2}+\frac{\zeta}{2}\,\hat{a}_{-}^{\dagger2}\Bigr]\notag\\
	&=\hat{S}_{\hat{a}_{+}}(\zeta)\times\hat{S}_{\hat{a}_{-}}(-\zeta)\,.\label{E:dkgueusd}
\end{align}
That is, in terms of the normal modes, it can be decomposed into a product of two single-mode squeezed operators.

In the current setting the connection is even closer because the pair of modes in $\hat{S}_2$ have the opposite momenta $\pm\bm{k}$. Since the two-mode squeeze operator $\hat{S}_{2}$, \eqref{E:gvsjhfgd}, symmetrically contains the annihilation and the creation operators of both modes, during the mode counting, contributions from each pair $(\bm{k},-\bm{k})$ will appear twice, and the resulting expressions may look like that given by two single-mode squeezed state. For example, let us compute the Hadamard function of the scalar field $\phi$ in the two-mode squeezed thermal state in spatially isotropic spacetime,
\begin{align}
	G_{\mathrm{H},0}^{(\phi)}(x,x')&=\int\!\frac{d^{3}\bm{k}}{(2\pi)^{\frac{3}{2}}}\frac{1}{\sqrt{\smash[b]2\omega}}\!\int\!\frac{d^{3}\bm{k}'}{(2\pi)^{\frac{3}{2}}}\frac{1}{\sqrt{2\omega'}}\biggl[\,\frac{1}{2}\,\langle\bigl\{\hat{a}_{\bm{k}}^{\vphantom{\dagger}},\,\hat{a}_{\bm{k}'}^{\vphantom{\dagger}}\bigr\}\rangle_{\textsc{tmst}}\,e^{i\bm{k}\cdot\bm{x}+i\bm{k}'\cdot\bm{x}'}e^{-i\omega t-i\omega't'}\biggr.\notag\\
	&\qquad\qquad\qquad\qquad\qquad\qquad\qquad\quad+\biggl.\frac{1}{2}\,\langle\bigl\{\hat{a}_{\bm{k}}^{\vphantom{\dagger}},\,\hat{a}_{\bm{k}'}^{\dagger}\bigr\}\rangle_{\textsc{tmst}}\,e^{i\bm{k}\cdot\bm{x}-i\bm{k}'\cdot\bm{x}'}e^{-i\omega t+i\omega't'}+\text{H.C.}\biggr]\notag\\
	&=\int\!\frac{d^{3}\bm{k}}{(2\pi)^{3}}\frac{1}{2\omega}\,\biggl[\,\frac{1}{2}\,\langle\bigl\{\hat{a}_{\bm{k}}^{\vphantom{\dagger}},\,\hat{a}_{\bm{k}}^{\vphantom{\dagger}}\bigr\}\rangle_{\textsc{tmst}}\,e^{i\bm{k}\cdot(\bm{x}+\bm{x}')}e^{-i\omega(t+t')}+\frac{1}{2}\,\langle\bigl\{\hat{a}_{\bm{k}}^{\vphantom{\dagger}},\,\hat{a}_{-\bm{k}}^{\vphantom{\dagger}}\bigr\}\rangle_{\textsc{tmst}}\,e^{i\bm{k}\cdot(\bm{x}-\bm{x}')}e^{-i\omega (t+t')}\biggr.\notag\\
	&\qquad\qquad\qquad+\frac{1}{2}\,\langle\bigl\{\hat{a}_{\bm{k}}^{\vphantom{\dagger}},\,\hat{a}_{\bm{k}}^{\dagger}\bigr\}\rangle_{\textsc{tmst}}\,e^{i\bm{k}\cdot(\bm{x}-\bm{x}')}e^{-i\omega (t-t')}+\frac{1}{2}\,\langle\bigl\{\hat{a}_{\bm{k}}^{\vphantom{\dagger}},\,\hat{a}_{-\bm{k}}^{\dagger}\bigr\}\rangle_{\textsc{tmst}}\,e^{i\bm{k}\cdot(\bm{x}+\bm{x}')}e^{-i\omega (t-t')}\notag\\
    &\qquad\qquad\qquad\qquad\qquad\qquad+\biggl.\text{H.C.}\biggr]\notag\\
	&=\int\!\frac{d^{3}\bm{k}}{(2\pi)^{3}}\frac{1}{2\omega}\;e^{i\bm{k}\cdot(\bm{x}-\bm{x}')}\biggl[\bigl(n_{\omega}+\frac{1}{2}\bigr)\,\cosh2\eta_{\omega}\,e^{-i\omega(t-t')}-e^{i\theta_{\omega}}\bigl(n_{\omega}+\frac{1}{2}\bigr)\,\sinh2\eta_{\omega}\,e^{-i\omega(t+t')}\biggr.\notag\\
    &\qquad\qquad\qquad\qquad\qquad\qquad+\biggl.\text{H.C.}\biggr]\,,
\end{align}
with $\omega=\lvert\bm{k}\rvert$ and the help of \eqref{E:ghvsjg1}--\eqref{E:ghvsjg4}. This is the same as the Hadamard function of the field in the squeezed thermal state, consistent with \eqref{E:dkgueusd}.

\section{late-time behavior of \texorpdfstring{$\langle\Delta\hat{T}_{\mu\nu} \rangle$}{}}\label{S:vbkdf}
\subsection{Late-Time Energy Flux Density \texorpdfstring{$\langle\Delta\hat{T}_{tr} \rangle$}{}}
Here we will examine the late-time net radiation flux of the field at distance $r$ sufficiently far away from the atom, so that only the dominant contribution, independent $r$, will be of our interest. The energy flux density is given by
\begin{equation}
    \langle\Delta\hat{T}_{rt}(x)\rangle = \lim_{x' \to x} \frac{\partial^2}{\partial r \partial t'}\Bigl[  G^{(\phi)}_{\mathrm{H}} (x,x') - G^{(\phi)}_{\mathrm{H}, 0}(x,x') \Bigr]\,.
\end{equation}
 We first look at the stationary component.

\subsubsection{stationary component \texorpdfstring{$\langle\Delta\hat{T}_{tr} \rangle_{\textsc{st}}$}{}}
We will assume that $t\gg r$, but place no other restrictions on $r$ for the moment. Following the decomposition in \eqref{E:kfdgbdf}, the stationary part of the contribution purely from the radiation field is given by
\begin{equation}
    \lim_{x' \to x} \frac{\partial^2}{\partial r \partial t'} \frac{1}{2} \langle \bigl\{ \hat{\phi}_{\textsc{br}}(x), \hat{\phi}_{\textsc{br}}(x') \bigr\}\rangle_{\textsc{st}}= -e^2\int_{-\infty}^{\infty}\!\frac{d\omega}{2\pi}\;\omega^2 \,\widetilde{G}^{(\chi)}_{\mathrm{H}}(\omega)\,\lvert\widetilde{G}^{(\phi)}_{\mathrm{R}, 0}(\bm{x}; \omega)\rvert^2\,,\label{E:rutdb}
\end{equation}
with the help of
\begin{align}
    \frac{\partial}{\partial r}\widetilde{G}^{(\phi)}_{\mathrm{R}, 0}(\bm{x}; \omega)&=\Bigl(i\omega-\frac{1}{r}\Bigr)\,\widetilde{G}^{(\phi)}_{\mathrm{R}, 0}(\bm{x}; \omega)\,,\label{E:dksdjs1}\\
    \frac{\partial}{\partial r}\widetilde{G}^{(\phi),\textsc{st}}_{\mathrm{H}, 0}(\bm{x}; \omega)&=\cosh2\eta\,\coth\frac{\beta\omega}{2}\Bigl[\omega\,\operatorname{Re}\widetilde{G}^{(\phi)}_{\mathrm{R}, 0}(\bm{x}; \omega)-\frac{1}{r}\,\operatorname{Im}\widetilde{G}^{(\phi)}_{\mathrm{R}, 0}(\bm{x}; \omega)\Bigr]\,,\label{E:dksdjs2}
\end{align}
where the stationary part of the free-field Hadamard function is
\begin{equation}
    \widetilde{G}^{(\phi),\textsc{st}}_{\mathrm{H}, 0}(\bm{x},\bm{0}; \omega)\equiv\widetilde{G}^{(\phi),\textsc{st}}_{\mathrm{H}, 0}(\bm{x}; \omega)=\cosh2\eta\,\coth\frac{\beta\omega}{2}\,\frac{\sin\omega r}{4\pi r}\,.
\end{equation}
To arrive at Eq.~\eqref{E:rutdb}, we observe the term proportional to $r^{-1}$ is odd in $\omega$, thus, vanishing, and that $\operatorname{Re} \widetilde{G}^{(\chi)}_{\mathrm{R}}(\omega)$ is an even function of $\omega$. At time greater than the relaxation time scale we may make the identification 
\begin{equation}\label{E:thytsds}
    \widetilde{G}^{(\chi)}_{\mathrm{H}}(\omega)=\cosh 2\eta \, \coth \frac{\beta \omega}{2}\,\operatorname{Im}\widetilde{G}^{(\chi)}_{\mathrm{R}}(\omega)\,,
\end{equation}
by the nonequilibrium fluctuation-dissipation relation~\cite{HCSH18,FDRSq,AHH22} of the internal dynamics of the atom if initially it is coupled to a squeezed thermal field.

On the other hand, for the cross-term, containing the correlation between the radiation field and the free field at large distance from the atom, its stationary part is
\begin{align}
    &\quad \lim_{x' \to x} \frac{\partial^2}{\partial r \partial t'} \frac{1}{2} \Bigl[\langle \bigl\{ \hat{\phi}_{\mathrm{h}}(x), \hat{\phi}_{\textsc{br}}(x')\bigr\}\rangle_{\textsc{st}} + \langle \bigl\{ \hat{\phi}_{\textsc{br}}(x), \hat{\phi}_{\mathrm{h}}(x')\big\}\rangle_{\textsc{st}} \Bigr]\notag\\
    &=-i\,e^2 \int_{-\infty}^{\infty}\!\frac{d\omega}{2\pi}\; \omega^2\,\rho^{(\textsc{st})}(\omega)\,\widetilde{G}^{(\chi)}_{\mathrm{R}}(\omega)\,\lvert\widetilde{G}^{(\phi)}_{\mathrm{R}, 0}(\bm{x}; \omega)\rvert^2\,.\label{E:mxcs}
\end{align}
Hereafter it is convenient to introduce the shorthand notations
\begin{align}
    &\rho^{(\textsc{st})}(\omega)=\cosh2\eta \,\coth\frac{\beta\omega}{2}\,, &&\text{and} &\rho^{(\textsc{ns})}(\omega)=\sinh2\eta \,\coth\frac{\beta\omega}{2}\,.
\end{align}
Since the contribution from $\operatorname{Re}\widetilde{G}^{(\chi)}_{\mathrm{R}}(\omega)$ is an even function of $\omega$, we immediately see it is the opposite of \eqref{E:rutdb} by \eqref{E:thytsds}, and thus cancels with \eqref{E:rutdb}. We already know from the case of the unsqueezed thermal state that the corresponding contributions of \eqref{E:rutdb} and \eqref{E:mxcs} cancel with one another. Therefore here we have shown that at late times the stationary part of the net energy flux density far away from the atom vanishes even though the field is initially in the squeezed thermal state, a nonstationary configuration. 

Here it is worth emphasizing the significance of vanishing stationary component of the net radiation flux. This is a unique feature of self-consistent quantum dynamics. The cancellation between \eqref{E:rutdb} and \eqref{E:mxcs} is not possible if the radiated field $\phi_{\textsc{r}}(x)$ is not correlated with the free field $\phi_{\mathrm{h}}(x)$. This correlation is established because the atom that sends out the radiation field is driven by a free field, as seen by the Langevin equation \eqref{E:bvkesfds}. It still holds when the field state is a vacuum state~\cite{QRadVac}, so a pure classical system will not have such correlation. Secondly our system is distinct from the classical driven dipole because the latter has a net outward radiation flux supplied by the external driving agent and is then independent of the field state. This is best seen from the component \eqref{E:rutdb} of the flux that is solely composed of the radiation field. This flux is field-state dependent because the internal dynamics of the atom is driven by the quantum fluctuations of the field, rather than an external agent.

\subsubsection{nonstationary component \texorpdfstring{$\langle\Delta\hat{T}_{tr} \rangle_{\textsc{ns}}$}{}}
Now we turn to the nonstationary contribution in $\langle\Delta\hat{T}_{rt}(x)\rangle$. The component purely from the radiation field is given by
\begin{align}
    &\quad\lim_{x' \to x} \frac{\partial^2}{\partial r \partial t'} \frac{1}{2}\langle \bigl\{ \hat{\phi}_{\textsc{br}}(x), \hat{\phi}_{\textsc{br}}(x') \bigr\}\rangle_{\textsc{ns}}\notag\\
    &=i\,e^4\int_0^{\infty}\!\frac{d\omega}{2\pi} \; \frac{\omega^2}{4\pi}\,\rho^{(\textsc{ns})}(\omega)\,\Bigl(i\omega-\frac{1}{r}\Bigr)\Bigl[\widetilde{G}^{(\chi)}_{\mathrm{R}}(\omega)\,\widetilde{G}^{(\phi)}_{\mathrm{R}, 0}(\bm{x}; \omega)\Bigr]^2e^{-i2\omega t+i\theta}+\text{C.C.}\,,\label{E:dkgb1}
\end{align}
while the counterpart from the cross term is
\begin{align}
     &\quad\lim_{x' \to x} \frac{\partial^2}{\partial r \partial t'} \frac{1}{2} \Bigl[\langle \big\{ \hat{\phi}_{\mathrm{h}}(x), \hat{\phi}_{\textsc{br}}(x')\big\}\rangle_{\textsc{ns}}+\langle \big\{ \hat{\phi}_{\textsc{br}}(x), \hat{\phi}_{\mathrm{h}}(x')\big\}\rangle_{\textsc{ns}} \Bigr]\label{E:dkgb2}\\
     &=i\,e^2\int_0^{\infty}\!\frac{d\omega}{2\pi}\;\omega\,\rho^{(\textsc{ns})}(\omega)\,\widetilde{G}_{\mathrm{R}}^{(\chi)}(\omega)\,\widetilde{G}_{\mathrm{R},0}^{(\phi)}(\bm{x};\omega)\Bigl[\omega\,\widetilde{G}_{\mathrm{R},0}^{(\phi)}(\bm{x};\omega)-\frac{2}{r}\,\operatorname{Im}\widetilde{G}_{\mathrm{R},0}^{(\phi)}(\bm{x};\omega)\Bigr]\,e^{-i2\omega t+i\theta}+\text{C.C.}\,.\notag
\end{align}
Contrary to the stationary contribution, in general, \eqref{E:dkgb1} does not cancel \eqref{E:dkgb2}. This can be explicitly shown by examining the sum of the expressions which are proportional to $\omega$ inside the curly brackets in both equations. We find these terms together give
\begin{align}
    &\quad i\,e^2\int_0^{\infty}\!\frac{d\omega}{2\pi}\;\omega^2\rho^{(\textsc{ns})}(\omega)\,\widetilde{G}_{\mathrm{R}}^{(\chi)}(\omega)\,\Bigl[\widetilde{G}_{\mathrm{R},0}^{(\phi)}(\bm{x};\omega)\Bigr]^2\Bigl[i\,e^2\,\frac{\omega}{4\pi}\widetilde{G}_{\mathrm{R}}^{(\chi)}(\omega)+1\Bigr]\,e^{-i2\omega t+i\theta}+\text{C.C.}\notag\\
    &=i\,e^2\int_0^{\infty}\!\frac{d\omega}{2\pi}\;\omega^2\rho^{(\textsc{ns})}(\omega)\,\operatorname{Re}\widetilde{G}_{\mathrm{R}}^{(\chi)}(\omega)\,\frac{\widetilde{G}_{\mathrm{R}}^{(\chi)\hphantom{*}}(\omega)}{\widetilde{G}_{\mathrm{R}}^{(\chi)*}(\omega)}\Bigl[\widetilde{G}_{\mathrm{R},0}^{(\phi)}(\bm{x};\omega)\Bigr]^2e^{-i2\omega t+i\theta}+\text{C.C.}\,,\label{E:tueherw}
\end{align}   
where we have used the fact that $e^2=8\pi\gamma m$ and
\begin{align}
    \operatorname{Im}\widetilde{G}_{\mathrm{R}}^{(\chi)}(\omega)&=2m\gamma\omega\,\lvert\widetilde{G}_{\mathrm{R}}^{(\chi)}(\omega)\rvert^2\,.\label{E:fkgues}
\end{align}
Obviously \eqref{E:tueherw} does not vanish in general. If the squeeze angle $\theta$ is really a mode- and time-independent constant, we may set $\theta=0$ without loss of generality and further simplify the nonstationary contribution, \eqref{E:tueherw}, in $\langle\Delta\hat{T}_{rt}(x)\rangle$ to
\begin{align}
    i\,e^2\int_{0}^{\infty}\!\frac{d\omega}{2\pi}\;\omega^2\rho^{(\textsc{ns})}(\omega)\,\Bigl[\widetilde{G}_{\mathrm{R},0}^{(\phi)}(\bm{x};\omega)\Bigr]^2\,\frac{\widetilde{G}_{\mathrm{R}}^{(\chi)\hphantom{*}}(\omega)}{\widetilde{G}_{\mathrm{R}}^{(\chi)*}(\omega)}\,\operatorname{Re}\widetilde{G}_{\mathrm{R}}^{(\chi)}(\omega)\,e^{-i2\omega t}\,.\label{E:kgbdje}
\end{align}
Eq.~\eqref{E:tueherw}, or the special case Eq.~\eqref{E:kgbdje}, in general is time dependent and is not identically zero.

For completeness, let us write down the sum of the terms proportional to $1/r$ in the square brackets in \eqref{E:dkgb1} and \eqref{E:dkgb2}, which is
\begin{align}
    &\quad-i\,\frac{e^2}{r}\!\int_{0}^{\infty}\!\frac{d\omega}{2\pi}\;\omega\,\rho^{(\textsc{ns})}(\omega)\,\widetilde{G}_{\mathrm{R}}^{(\chi)}(\omega)\,\widetilde{G}_{\mathrm{R},0}^{(\phi)}(\bm{x};\omega)\Bigl[e^2\frac{\omega}{4\pi}\,\widetilde{G}_{\mathrm{R}}^{(\chi)}(\omega)\,\widetilde{G}_{\mathrm{R},0}^{(\phi)}(\bm{x};\omega)+2\operatorname{Im}\widetilde{G}_{\mathrm{R},0}^{(\phi)}(\bm{x};\omega)\Bigr]\,e^{-i2\omega t+i\theta}\notag\\
    &=-\frac{e^2}{r}\!\int_{0}^{\infty}\!\frac{d\omega}{2\pi}\;\omega\,\rho^{(\textsc{ns})}(\omega)\,\biggl\{\Bigl[\widetilde{G}_{\mathrm{R},0}^{(\phi)}(\bm{x};\omega)\Bigr]^2\frac{\widetilde{G}_{\mathrm{R}}^{(\chi)\hphantom{*}}(\omega)}{\widetilde{G}_{\mathrm{R}}^{(\chi)*}(\omega)}\,\operatorname{Re}\widetilde{G}_{\mathrm{R}}^{(\chi)}(\omega)+\widetilde{G}_{\mathrm{R}}^{(\chi)}(\omega)\,\lvert\widetilde{G}_{\mathrm{R},0}^{(\phi)}(\bm{x};\omega)\rvert^2\biggr\}\,e^{-i2\omega t+i\theta}\,,
\end{align}
plus its complex conjugate. Here again we have used \eqref{E:fkgues} to recast the expression into a more compact form. This will be useful later to check the consistency via the continuity equation.

The sum of \eqref{E:dkgb1} and \eqref{E:dkgb2} gives the only contribution of the net radiation flux that is likely to survive after the relaxation time scale. However, in Appendix~\ref{S:evsfdjg}, we can argue that although Eqs.~\eqref{E:dkgb1} and~\eqref{E:dkgb2} do not cancel in general, their sum still decays with time and furthermore it actually falls off to zero exponentially fast, proportional to $e^{-2\gamma t}$.

Thus combining the discussions about the stationary component, we show that at late times no net radiation flux exudes to spatial infinity from a static atom even though the internal degree of freedom of the atom is driven by the quantum fluctuations of the squeezed field caused by its parametric process. At late times it leaves no trace of detectable signal to a detector at distance much greater than the typical scales in the atom's internal dynamics, except for the case of extremely large squeezing where we might have the chance to see a trace of the net flux since it will take a longer time to decay. However, the information of the squeeze parameter $\eta$ can still be read out from the asymptotic equilibrium state of the atom's internal dynamics.

\subsection{late-time field energy density \texorpdfstring{$\langle\Delta\hat{T}_{tt} \rangle$}{}}\label{S:kvckdfd}

The energy density is given by
\begin{equation}
    \langle\Delta\hat{T}_{tt}(x)\rangle = \lim_{x' \to x} \frac{1}{2}\biggl( \frac{\partial^2}{\partial t \partial t'} + \frac{\partial^2}{\partial r \partial r'}\biggr) \Big[  G^{(\phi)}_H (x,x') - G^{(\phi)}_{H, 0}(x,x') \Big]\,.\label{E:kfbeie}
\end{equation}
That is, we subtract off the energy density of the free field in the squeezed state, and Eq.~\eqref{E:kfbeie} tells us the change in energy density due to the atom-field interaction.

\subsubsection{stationary component \texorpdfstring{$\langle\Delta\hat{T}_{tt} \rangle_{\textsc{st}}$}{}}
We first consider the stationary component of \eqref{E:kfbeie}. Following the decomposition \eqref{E:kfdgbdf}, the part purely caused by the radiation field is given by
\begin{align}
    &\quad\lim_{x' \to x} \frac{1}{2}\biggl( \frac{\partial^2}{\partial t \partial t'} + \frac{\partial^2}{\partial r \partial r'}\biggr) \frac{1}{2}\,\langle \big\{ \hat{\phi}_{\textsc{br}}(x), \hat{\phi}_{\textsc{br}}(x')\big\}\rangle_{\textsc{st}}\notag\\
    &=e^2\int_{-\infty}^{\infty}\!\frac{d\omega}{2\pi}\;\rho^{(\textsc{st})}(\omega)\,\Bigl(\omega^2+\frac{1}{2r^2}\Bigr)\,\operatorname{Im}\widetilde{G}_{\mathrm{R}}^{(\chi)}(\omega)\,\lvert\widetilde{G}_{\mathrm{R},0}^{(\phi)}(\bm{x};\omega)\rvert^2\,.\label{E:kddkkws}
\end{align}
The contribution from the cross term is then     
\begin{align}
     &\quad\lim_{x' \to x} \frac{1}{2}\biggl( \frac{\partial^2}{\partial t \partial t'} + \frac{\partial^2}{\partial r \partial r'}\biggr) \frac{1}{2} \Bigl[\langle \big\{ \hat{\phi}_{\mathrm{h}}(x), \hat{\phi}_{\textsc{br}}(x')\big\}\rangle_{\textsc{st}}+ \langle \big\{ \hat{\phi}_{\textsc{br}}(x), \hat{\phi}_{\mathrm{h}}(x')\big\}\rangle_{\textsc{st}} \Bigr]\label{E:oeits}\\
     &=-e^2\int_{-\infty}^{\infty}\!\frac{d\omega}{2\pi}\;\rho^{(\textsc{st})}(\omega)\,\biggl\{\Bigl(\omega^2+\frac{1}{2r^2}\Bigr)\,\lvert\widetilde{G}_{\mathrm{R},0}^{(\phi)}(\bm{x};\omega)\rvert^2\,\operatorname{Im}\widetilde{G}_{\mathrm{R}}^{(\chi)}(\omega)+\frac{\omega}{r}\,\operatorname{Re}\Bigl[\widetilde{G}_{\mathrm{R}}^{(\chi)}(\omega)\,\widetilde{G}_{\mathrm{R},0}^{(\phi)\,2}(\bm{x};\omega)\Bigr]\biggr.\notag\\
     &\qquad\qquad\qquad\qquad\qquad\qquad-\biggl.\frac{1}{2r^2}\,\operatorname{Im}\Bigl[\widetilde{G}_{\mathrm{R}}^{(\chi)}(\omega)\,\widetilde{G}_{\mathrm{R},0}^{(\phi)\,2}(\bm{x};\omega)\Bigr]\biggr\}\,,\notag
\end{align}
with the help of \eqref{E:dksdjs1} and \eqref{E:dksdjs2} to re-express the term
\begin{align*}
    2\operatorname{Re}\Bigl[\widetilde{G}_{\mathrm{R}}^{(\chi)}(\omega)\,\widetilde{G}_{\mathrm{R},0}^{(\phi)}(\bm{x};\omega)\Bigr]\,\operatorname{Im}\widetilde{G}_{\mathrm{R},0}^{(\phi)}(\bm{x};\omega)=-\lvert\widetilde{G}_{\mathrm{R},0}^{(\phi)}(\bm{x};\omega)\rvert^2\,\operatorname{Im}\widetilde{G}_{\mathrm{R}}^{(\chi)}(\omega)+\operatorname{Im}\Bigl[\widetilde{G}_{\mathrm{R}}^{(\chi)}(\omega)\,\widetilde{G}_{\mathrm{R},0}^{(\phi)\,2}(\bm{x};\omega)\Bigr]\,.
\end{align*}
Adding both contributions together gives
\begin{align}
    \langle\Delta\hat{T}_{tt}\rangle_{\textsc{st}}=-e^2\int_{-\infty}^{\infty}\!\frac{d\omega}{2\pi}\;\rho^{(\textsc{st})}(\omega)\,\biggl\{\frac{\omega}{r}\,\operatorname{Re}\Bigl[\widetilde{G}_{\mathrm{R}}^{(\chi)}(\omega)\,\widetilde{G}_{\mathrm{R},0}^{(\phi)\,2}(\bm{x};\omega)\Bigr]-\frac{1}{2r^2}\operatorname{Im}\Bigl[\widetilde{G}_{\mathrm{R}}^{(\chi)}(\omega)\,\widetilde{G}_{\mathrm{R},0}^{(\phi)\,2}(\bm{x};\omega)\Bigr]\biggr\}\,,\label{E:gbkweur}
\end{align}
at late times and at large distance away from the atom. The dominant contribution of the stationary component of the local field energy density vanishes, so Eq.~\eqref{E:gbkweur} is subdominant since it is proportional to $1/r^3$. It comes from the correlation between the radiation field and the free field at the location far away from the atom, and it decays faster with $r$. Note that there are two type of $1/r^3$ contributions in the cross terms \eqref{E:oeits} but one of them cancels with its counterpart in \eqref{E:kddkkws}. The residual energy density \eqref{E:gbkweur} has relatively short range by nature. We note it is proportional to $e^2$, which is proportional to the damping constant $\gamma$. Since this is a time-independent constant, it means that this may be an unambiguous aftereffect of the transient peregrinating radiation field.

Next we turn to the nonstationary component.

\subsubsection{nonstationary component \texorpdfstring{$\langle\Delta\hat{T}_{tt} \rangle_{\textsc{ns}}$}{}}
We first examine the late-time contribution purely from the radiation field
\begin{align}\label{E:dkgb3}
    &\quad\lim_{x' \to x} \frac{1}{2}\biggl( \frac{\partial^2}{\partial t \partial t'} + \frac{\partial^2}{\partial r \partial r'}\biggr) \frac{1}{2}\,\langle \big\{ \hat{\phi}_{\textsc{br}}(x), \hat{\phi}_{\textsc{br}}(x')\big\}\rangle_{\textsc{ns}}\notag\\
    &=e^4\int_{0}^{\infty}\!\frac{d\omega}{2\pi}\;\frac{\omega}{4\pi}\,\rho^{(\textsc{ns})}(\omega)\,\Bigl(\omega^2+i\,\frac{\omega}{r}-\frac{1}{2r^2}\Bigr)\,\Bigl[\widetilde{G}_{\mathrm{R}}^{(\chi)}(\omega)\,\widetilde{G}_{\mathrm{R},0}^{(\phi)}(\bm{x};\omega)\Bigr]^2e^{-i2\omega t+i\theta}+\text{C.C.}\,.
\end{align}
On the other hand, the corresponding contribution from the cross term gives
\begin{align}
     &\quad\lim_{x' \to x} \frac{1}{2}\biggl( \frac{\partial^2}{\partial t \partial t'} + \frac{\partial^2}{\partial r \partial r'}\biggr) \frac{1}{2} \Bigl[\langle \big\{ \hat{\phi}_{\mathrm{h}}(x), \hat{\phi}_{\textsc{br}}(x')\big\}\rangle_{\textsc{ns}}+ \langle \big\{ \hat{\phi}_{\textsc{br}}(x), \hat{\phi}_{\mathrm{h}}(x')\big\}\rangle_{\textsc{ns}} \Bigr]\notag\\
     &=-e^2\int_{0}^{\infty}\!\frac{d\omega}{2\pi}\;\rho^{(\textsc{ns})}(\omega)\,\Bigl\{i\,\Bigl(\omega^2+i\,\frac{\omega}{r}\Bigr)\widetilde{G}_{\mathrm{R}}^{(\chi)}(\omega)\,\Bigl[\widetilde{G}_{\mathrm{R},0}^{(\phi)}(\bm{x};\omega)\Bigr]^2\biggr.\notag\\
     &\qquad\qquad\qquad\qquad\qquad\qquad\qquad\quad+\biggl.\frac{1}{r^2}\,\widetilde{G}_{\mathrm{R}}^{(\chi)}(\omega)\,\widetilde{G}_{\mathrm{R},0}^{(\phi)}(\bm{x};\omega)\operatorname{Im}\widetilde{G}_{\mathrm{R},0}^{(\phi)}(\bm{x};\omega)\Bigr\}\,e^{-i2\omega t+i\theta}+\text{C.C.}\,,\label{E:dkgb4}
\end{align}
Now we will use the same trick \eqref{E:fkgues} to combine both contributions, and we obtain
\begin{align}
    \langle\Delta\hat{T}_{tt}\rangle_{\textsc{ns}}&=-i\,e^2\int_{0}^{\infty}\!\frac{d\omega}{2\pi}\;\rho^{(\textsc{ns})}(\omega)\,\Bigl\{-i\,\Bigl(\omega^2+i\,\frac{\omega}{r}-\frac{1}{2r^2}\Bigr)\Bigl[\widetilde{G}_{\mathrm{R},0}^{(\phi)}(\bm{x};\omega)\Bigr]^2\frac{\widetilde{G}_{\mathrm{R}}^{(\chi)\hphantom{*}}(\omega)}{\widetilde{G}_{\mathrm{R}}^{(\chi)*}(\omega)}\,\operatorname{Re}\widetilde{G}_{\mathrm{R}}^{(\chi)}(\omega)\Bigr.\notag\\
    &\qquad\qquad\qquad\qquad\qquad\qquad\qquad\quad+\Bigr.\frac{1}{2r^2}\,\widetilde{G}_{\mathrm{R}}^{(\chi)}(\omega)\,\lvert\widetilde{G}_{\mathrm{R},0}^{(\phi)}(\bm{x};\omega)\rvert^2\Bigr\}\,e^{-i2\omega t+i\theta}+\text{C.C.}\,.\label{E:rkteds}
\end{align}   
The dominant term in \eqref{E:rkteds} has the same form as \eqref{E:tueherw}, and thus will vanish at late times at distance far away from the atom.

Therefore at late times, $\langle\Delta\hat{T}_{tt}\rangle$ settles down to a constant value, whose value decays like $1/r^3$ away from the atom.

\subsection{continuity equation}
Here in passing we would like to examine the consistency of our results of the nonstationary components of  $\langle\Delta \hat{T}_{\mu t}\rangle$ in terms of the continuity equation $\partial^{\mu}T_{\mu t}=0$. It turns out convenient to list the previous results for the relevant components
\begin{align*}
    \langle \hat{T}_{rt}\rangle^{(\textsc{rr})}_{\textsc{ns}}&=i\,e^4\int_0^{\infty}\!\frac{d\omega}{2\pi} \; \frac{\omega^2}{4\pi}\,\rho^{(\textsc{ns})}(\omega)\,\Bigl(i\omega-\frac{1}{r}\Bigr)\Bigl[\widetilde{G}^{(\chi)}_{\mathrm{R}}(\omega)\,\widetilde{G}^{(\phi)}_{\mathrm{R}, 0}(\bm{x}; \omega)\Bigr]^2e^{-i2\omega t+i\theta}+\text{C.C.}\,,\\
    \langle\hat{T}_{rt}\rangle^{(\textsc{hr})}_{\textsc{ns}}&=i\,e^2\int_0^{\infty}\!\frac{d\omega}{2\pi}\;\omega\,\rho^{(\textsc{ns})}(\omega)\,\widetilde{G}_{\mathrm{R}}^{(\chi)}(\omega)\,\widetilde{G}_{\mathrm{R},0}^{(\phi)}(\bm{x};\omega)\Bigl[\omega\,\widetilde{G}_{\mathrm{R},0}^{(\phi)}(\bm{x};\omega)-\frac{2}{r}\,\operatorname{Im}\widetilde{G}_{\mathrm{R},0}^{(\phi)}(\bm{x};\omega)\Bigr]\,e^{-i2\omega t+i\theta}+\text{C.C.}\,,\\
    \langle\hat{T}_{tt}\rangle^{(\textsc{rr})}_{\textsc{ns}}&=e^4\int_{0}^{\infty}\!\frac{d\omega}{2\pi}\;\frac{\omega}{4\pi}\,\rho^{(\textsc{ns})}(\omega)\,\Bigl(\omega^2+i\,\frac{\omega}{r}-\frac{1}{2r^2}\Bigr)\,\Bigl[\widetilde{G}_{\mathrm{R}}^{(\chi)}(\omega)\,\widetilde{G}_{\mathrm{R},0}^{(\phi)}(\bm{x};\omega)\Bigr]^2e^{-i2\omega t+i\theta}+\text{C.C.}\,,\notag\\
    \langle \hat{T}_{tt}\rangle^{(\textsc{hr})}_{\textsc{ns}}&=-e^2\int_{0}^{\infty}\!\frac{d\omega}{2\pi}\;\rho^{(\textsc{ns})}(\omega)\,\widetilde{G}_{\mathrm{R}}^{(\chi)}(\omega)\,\Bigl\{i\,\Bigl(\omega^2+i\,\frac{\omega}{r}\Bigr)\Bigl[\widetilde{G}_{\mathrm{R},0}^{(\phi)}(\bm{x};\omega)\Bigr]^2\biggr.\notag\\
     &\qquad\qquad\qquad\qquad\qquad\qquad\qquad\quad+\biggl.\frac{1}{r^2}\,
\widetilde{G}_{\mathrm{R},0}^{(\phi)}(\bm{x};\omega)\operatorname{Im}\widetilde{G}_{\mathrm{R},0}^{(\phi)}(\bm{x};\omega)\Bigr\}\,e^{-i2\omega t+i\theta}+\text{C.C.}\,.
\end{align*}
The superscript $(\mathrm{RR})$ represents the contributions purely from the radiation field like \eqref{E:dkgb1} and \eqref{E:dkgb3} while the superscript $(\mathrm{HR})$ denotes those from the cross terms like \eqref{E:dkgb2} and \eqref{E:dkgb4}.  For our configuration, we do not have $T_{\vartheta t}$ and $T_{\varphi t}$ where $\vartheta$ and $\varphi$ are respectively the polar angle, and azimuthal angle, so explicitly the continuity equation can be put in the form
\begin{equation}
    \partial^{\mu}\hat{T}_{\mu t}=0=\partial_tT_{tt}-\frac{1}{r^2}\,\partial_r\bigl(r^2T_{rt}\bigr)\,.
\end{equation}
Thus we find
\begin{align*}
    \frac{1}{r^2}\,\partial_r\bigl(r^2\langle \hat{T}_{rt}\rangle^{(\textsc{rr})}_{\textsc{ns}}\bigr)&=-2i\,e^4\int_0^{\infty}\!\frac{d\omega}{2\pi} \; \frac{\omega^2}{4\pi}\,\rho^{(\textsc{ns})}(\omega)\Bigl(\omega^2+i\,\frac{\omega}{r}-\frac{1}{2r^2}\Bigr)\Bigl[\widetilde{G}^{(\chi)}_{\mathrm{R}}(\omega)\,\widetilde{G}^{(\phi)}_{\mathrm{R}, 0}(\bm{x}; \omega)\Bigr]^2e^{-i2\omega t+i\theta}+\text{C.C.}\,,
    \intertext{which is equal to $\partial_t \langle\hat{T}_{tt}\rangle^{(\textsc{rr})}_{\textsc{ns}}$, and}
    \frac{1}{r^2}\,\partial_r\bigl(r^2\langle \hat{T}_{rt}\rangle^{(\textsc{hr})}_{\textsc{ns}}\bigr)&=2e^2\int_{0}^{\infty}\!\frac{d\omega}{2\pi}\;\rho^{(\textsc{ns})}(\omega)\,\widetilde{G}_{\mathrm{R}}^{(\chi)}(\omega)\,\Bigl\{i\,\Bigl(\omega^2+i\,\frac{\omega}{r}\Bigr)\Bigl[\widetilde{G}_{\mathrm{R},0}^{(\phi)}(\bm{x};\omega)\Bigr]^2\biggr.\notag\\
     &\qquad\qquad\qquad\qquad\qquad\qquad\qquad\quad+\biggl.\frac{1}{r^2}\,
\widetilde{G}_{\mathrm{R},0}^{(\phi)}(\bm{x};\omega)\operatorname{Im}\widetilde{G}_{\mathrm{R},0}^{(\phi)}(\bm{x};\omega)\Bigr\}\,e^{-i2\omega t+i\theta}+\text{C.C.}\,.
\end{align*}
It is exactly $\partial_t \langle\hat{T}_{tt}\rangle^{(\textsc{hr})}_{\textsc{ns}}$. Thus the continuity equation is satisfied.

\section{late-time behavior of the nonstationary contribution in \texorpdfstring{$\langle\Delta\hat{T}_{rt} \rangle$}{}}\label{S:evsfdjg}
It turns out that the nonstationary contribution in $\langle\Delta\hat{T}_{rt}(x)\rangle$ still decays with time. To begin with, it proves useful to quote the late-time expressions of the energy exchange between the atom and the surrounding field~\cite{HCSH18,QRadVac}. For times greater than the relation time, the nonstationary contribution of the energy flow into the atom from field has the form
\begin{align}
    P_{\xi}^{(\textsc{ns})}(t)&=i\,e^2\int_0^{\infty}\!\frac{d\omega}{2\pi}\;\frac{\omega^2}{4\pi}\,\sinh2\eta\,\coth\frac{\beta\omega}{2}\biggl\{\widetilde{G}_{\mathrm{R}}^{(\chi)}(\omega)\,e^{-i2\omega t+i\theta}-\text{C.C.}\biggr\}\,,\label{E:eusii1}
    \intertext{and the corresponding contribution of the energy flow out of the atom due to friction is}
    P_{\gamma}^{(\textsc{ns})}(t)&=-e^4\int_0^{\infty}\!\frac{d\omega}{2\pi}\;\frac{\omega^3}{(4\pi)^2}\,\sinh2\eta\,\coth\frac{\beta\omega}{2}\,\biggl\{\Bigl[\widetilde{G}_{\mathrm{R}}^{(\chi)}(\omega)\Bigr]^2\,e^{-i2\omega t+i\theta}+\text{C.C.}\biggr\}\,.\label{E:eusii2}
\end{align}
For comparison, let us put the terms proportional to $\omega$ inside of the curly brackets of \eqref{E:dkgb2} and \eqref{E:dkgb1} here,
\begin{align}
    \eqref{E:dkgb2}&:&&i\,e^2\int_0^{\infty}\!\frac{d\omega}{2\pi}\;\frac{\omega^2}{(4\pi)^2r^2}\,\sinh2\eta\,\coth\frac{\beta\omega}{2}\biggl\{\widetilde{G}_{\mathrm{R}}^{(\chi)}(\omega)\,e^{-i2\omega (t-r)+i\theta}-\text{C.C.}\biggr\}\,,\label{E:eusii3}\\
    \eqref{E:dkgb1}&:&&-e^4\int_0^{\infty}\!\frac{d\omega}{2\pi} \; \frac{\omega^3}{(4\pi)^3r^2}\,\sinh2\eta \,\coth\frac{\beta\omega}{2}\biggl\{\Bigl[\widetilde{G}^{(\chi)}_{\mathrm{R}}(\omega)\Bigr]^2\,e^{-i2\omega (t-r)+i\theta}+\text{C.C.}\biggr\}\,.\label{E:eusii4}
\end{align}
We see correspondence between \eqref{E:eusii1} and \eqref{E:eusii3}, as well as between \eqref{E:eusii2} and \eqref{E:eusii4}. For example, the integrand of \eqref{E:eusii3} is $\dfrac{e^{i2\omega r}}{4\pi r^2}$ times the integrand of \eqref{E:eusii1}. The same applies to the \eqref{E:eusii2}--\eqref{E:eusii4} pair. For a fix $r$ and at times $t\gg r$, we may shift or re-define $t$ in \eqref{E:eusii3} and \eqref{E:eusii4} by $t-r\to t$ such that the integrals in both equations are essentially proportional to those in \eqref{E:eusii1} and \eqref{E:eusii2}. We have numerically shown in~\cite{AHH22} that \eqref{E:eusii1} and \eqref{E:eusii2} vanish as $t\to \infty$, so we conclude that the terms proportional to $\omega$ inside of the curly brackets of \eqref{E:dkgb1} and \eqref{E:dkgb2} will also vanish at late times. For example, as shown in Fig.~\ref{Fi:nonst} generated by numerical calculations, the contribution in \eqref{E:dkgb1} actually decays exponentially fast to zero. 
\begin{figure}
	\centering
	\includegraphics[width=0.55\textwidth]{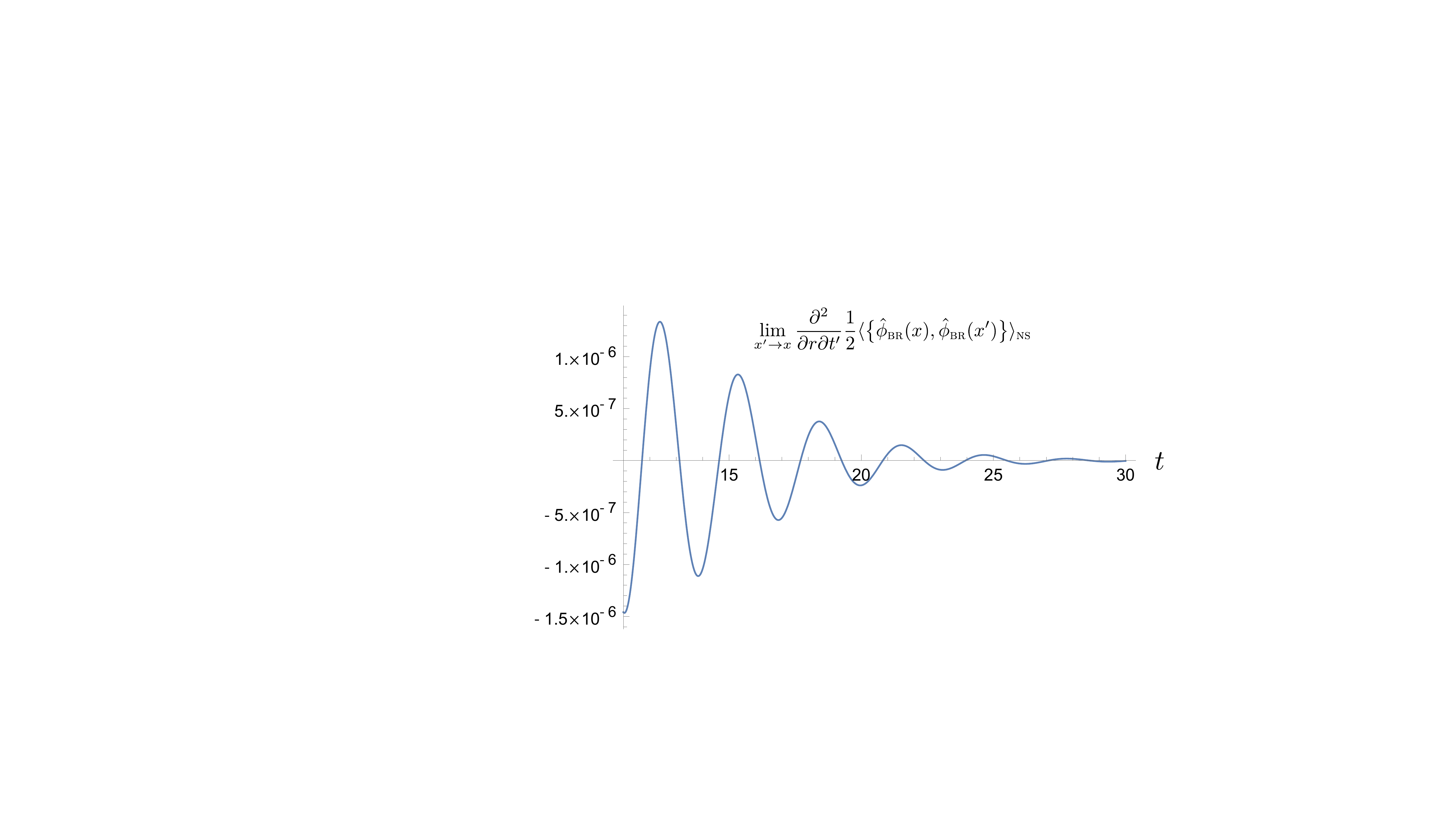}
	\caption{The time dependence of Eq.~\eqref{E:dkgb1}, one of the nonstationary component of the radiation flux. The curve falls off to zero exponentially fast. Here we choose $\beta^{-1}=0$, $\gamma=0.2$, $r=10$ and $\omega_{\textsc{r}}=1$.}\label{Fi:nonst}
\end{figure}

For terms which are proportional to $1/r$ inside of the curly brackets of \eqref{E:dkgb2} and \eqref{E:dkgb1}, we immediately see that if we take the time derivative of \eqref{E:dkgb1}, we obtain an expression which is $2/r$ times \eqref{E:eusii4}. Likewise, the time derivative of \eqref{E:dkgb2} gives,
\begin{align}
    \frac{2}{r}\times i\,e^2\int_0^{\infty}\!\frac{d\omega}{2\pi}\;\frac{\omega^2}{(4\pi)^2r^2}\,\sinh2\eta\,\coth\frac{\beta\omega}{2}\biggl\{\widetilde{G}_{\mathrm{R}}^{(\chi)}(\omega)\,\Bigl[e^{-i2\omega (t-r)+i\theta}-e^{-i2\omega t+i\theta}\Bigr]-\text{C.C.}\biggr\}\,.
\end{align}
Following the previous arguments, for a fixed $r$, it vanishes in the limit $t\to\infty$ too. Thus we have demonstrated that the time derivatives of the terms proportional to $1/r$ inside of the curly brackets of \eqref{E:dkgb2} and \eqref{E:dkgb1} vanishes, so these terms must be constants for sufficiently large time. On the other hand, these terms are time dependent and are not sign definite, so the asymptotic constants must be zero.

Putting these results together, we reach the conclusion that the nonstationary terms of the net radiation flux vanishes eventually at large distance away from the atom. However it does tell how slowly the nonstationary term decay. In fact, at least for the zero temperature limit $\beta\to\infty$, we can carry out the above integrals analytically, and the brute-force calculations show that for fixed $r$, these integrals give results that decay exponentially fast to zero in a form proportional to $\sinh 2\eta\,e^{-2\gamma t}$. Thus at late times, no net radiation energy flux leaks to the spatial infinity from the atom even though the internal degree of freedom of the atom is coupled to a nonstationary squeezed quantum field.

\section{time-translational invariance of the squeeze parameter in the out-region}\label{S:eerudj}
Next we argue that if we shift the parametric process by $\Delta$ along the time axis, as shown in Fig.~\ref{Fi:paraProc3d}, then we will have
\begin{equation}\label{E:kdfgsbksxc}
	\mathsf{d}_{i}^{(\textsc{ii})}(t+\Delta,t_{a}+\Delta)=d_{i}^{(\textsc{ii})}(t,t_{a})
\end{equation}
where $\mathsf{d}_{i}^{(\textsc{ii})}(t+\Delta,t_{a}+\Delta)$ on sans serif style is the fundamental solution in-region I\!I for the shifted case described by the orange curve in Fig.~\ref{Fi:paraProc3d}.

Since the fundamental solutions $\mathsf{d}_{i}^{(\textsc{ii})}(y,y_{a})$ in-region I\!I for the orange curve satisfies the equation of motion
\begin{align}
	\ddot{\mathsf{d}}_{i}^{(\textsc{ii})}(y,y_{a})+\omega^{2}(y)\,\mathsf{d}_{i}^{(\textsc{ii})}(y,y_{a})=0\,,
\end{align}
with the standard initial conditions given at $y_{a}=t_{a}+\Delta$, while the fundamental solutions $d_{i}^{(\textsc{ii})}(y,y_{a})$ in-region I\!I for the blue curve satisfies the equation of motion
\begin{align}
	\ddot{d}_{i}^{(\textsc{ii})}(t,t_{a})+\omega^{2}(t)\,d_{i}^{(\textsc{ii})}(t,t_{a})=0\,,
\end{align}
with the standard initial conditions given at $t_{a}$, we immediately see that both cases are related by $y=t+\Delta$, and thus we arrive at \eqref{E:kdfgsbksxc} and in particular,
\begin{equation}\label{E:niww}
	\mathsf{d}_{i}^{(\textsc{ii})}(t_{b}+\Delta,t_{a}+\Delta)=d_{i}^{(\textsc{ii})}(t_{b},t_{a})\,.
\end{equation}
Later we will need this result.

Now we would like to explicitly show that an detector, in the out-region with $t>t_{b}+\Delta$, will find the same squeeze parameter for both cases described in Fig.~\ref{Fi:paraProc3d}, independent of the shift $\Delta$, even if it carries out the measurement at the same time $t$. Note that in both cases the motion starts at the same $t_{i}=0$. Following the previous discussions on the construction of the fundamental solutions, we carry out a similar construction of the fundamental solution for the shifted process. At $t>t_{b}+\Delta$. In particular, we focus on the moment $t=t_{b}+\Delta$ since afterward the squeeze parameter is a constant. Before we start, we first note that $\mathsf{d}_{i}^{(\textsc{i})}(t)$ is essentially the same as $d_{i}^{(\textsc{i})}(t)$, except that the former applies to the time interval $0=t_{i}\leq t\leq t_{a}+\Delta$ while the latter only to $0=t_{i}\leq t\leq t_{a}$. With this recognition, we can write $\mathsf{d}_{i}^{(\textsc{i})}(t_{a}+\Delta)$ as
\begin{equation}\label{E:dkuewskd}
	\mathsf{d}_{i}^{(\textsc{i})}(t_{a}+\Delta)=d_{1}^{(\textsc{i})}(\Delta)\,d_{i}^{(\textsc{i})}(t_{a})+d_{2}^{(\textsc{i})}(\Delta)\,\dot{d}_{i}^{(\textsc{i})}(t_{a})\,.
\end{equation}
This will be convenient later. Now we will write $\mathsf{d}_{i}(t_{b}+\Delta,0)$ in terms $d_{i}^{(\textsc{x})}$ with $\mathrm{X}=\mathrm{I}$, I\!I, and I\!I\!I of the unshifted process. We show the calculations explicitly for $\mathsf{d}_{1}(t_{b}+\Delta)$, and the result for $\mathsf{d}_{2}(t_{b}+\Delta)$ will follow similarly. Pretty straightforwardly, we have

\begin{figure}
	\centering
	\includegraphics[width=0.7\textwidth]{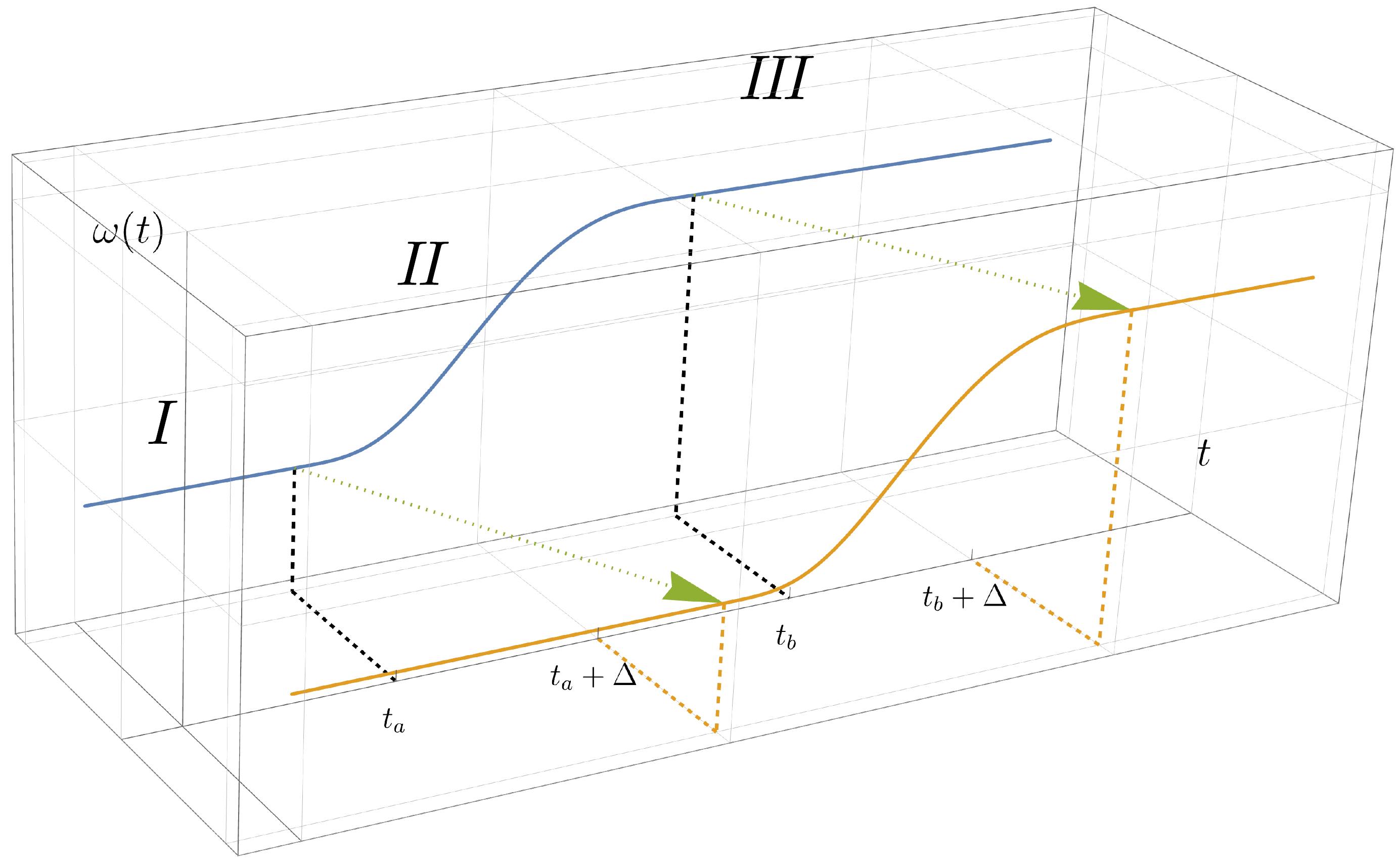}
	\caption{The parametric process described by a piecewise continuous function. The transition begins at $t_{a}$, and ends at $t_{b}$. Between $t_{i}$ and $t_{a}$, region I, the frequency $\omega(t)$ takes on a constant value $\omega_{i}$ and is given by another constant $\omega_{f}$ after $t\geq t_{b}$. We will consider a shift $\Delta$ of the parametric process along the timeline while maintaining the functional form, the duration $t_{b}-t_{a}$ of the process and the initial time $t_{i}$ of the motion unchanged. We displace these two cases by a horizontal displacement for discernibility.}\label{Fi:paraProc3d}
\end{figure}

\begin{align}
	\mathsf{d}_{1}(t_{b}+\Delta,0)&=\mathsf{d}_{1}^{(\textsc{ii})}(t_{b}+\Delta,t_{a}+\Delta)\,\mathsf{d}_{1}^{(\textsc{i})}(t_{a}+\Delta)+\mathsf{d}_{2}^{(\textsc{ii})}(t_{b}+\Delta,t_{a}+\Delta)\,\dot{\mathsf{d}}_{1}^{(\textsc{i})}(t_{a}+\Delta)\label{E:bcnmmwek}\\
	&=d_{1}^{(\textsc{ii})}(t_{b},t_{a})\Bigl[d_{1}^{(\textsc{i})}(\Delta)\,d_{1}^{(\textsc{i})}(t_{a})+d_{2}^{(\textsc{i})}(\Delta)\,\dot{d}_{1}^{(\textsc{i})}(t_{a})\Bigr]+d_{2}^{(\textsc{ii})}(t_{b},t_{a})\Bigl[\dot{d}_{1}^{(\textsc{i})}(\Delta)\,d_{1}^{(\textsc{i})}(t_{a})+\dot{d}_{2}^{(\textsc{i})}(\Delta)\,\dot{d}_{1}^{(\textsc{i})}(t_{a})\Bigr]\notag
\end{align}
with the help of \eqref{E:kdfgsbksxc} and \eqref{E:dkuewskd}. We use a trick that only applies to the undamped harmonic oscillator. in-region I, we have the identities
\begin{align}
	\dot{d}_{2}^{(\textsc{i})}(t)&=d_{1}^{(\textsc{i})}(t)\,,&&\Rightarrow&d_{2}^{(\textsc{i})}(\Delta)\,\dot{d}_{1}^{(\textsc{i})}(t_{a})&=d_{2}^{(\textsc{i})}(t_{a})\,\dot{d}_{1}^{(\textsc{i})}(\Delta)\,,
\end{align}
and it allows us to re-write \eqref{E:bcnmmwek},
\begin{align}
	\mathsf{d}_{1}(t_{b}+\Delta,0)&=d_{1}^{(\textsc{ii})}(t_{b},t_{a})\Bigl[d_{1}^{(\textsc{i})}(\Delta)\,d_{1}^{(\textsc{i})}(t_{a})+\dot{d}_{1}^{(\textsc{i})}(\Delta)\,d_{2}^{(\textsc{i})}(t_{a})\Bigr]+d_{2}^{(\textsc{ii})}(t_{b},t_{a})\Bigl[\dot{d}_{1}^{(\textsc{i})}(\Delta)\,d_{1}^{(\textsc{i})}(t_{a})+d_{1}^{(\textsc{i})}(\Delta)\,\dot{d}_{1}^{(\textsc{i})}(t_{a})\Bigr]\notag\\
	&=\Bigl[d_{1}^{(\textsc{ii})}(t_{b},t_{a})\,d_{1}^{(\textsc{i})}(t_{a})+d_{2}^{(\textsc{ii})}(t_{b},t_{a})\,\dot{d}_{1}^{(\textsc{i})}(t_{a})\Bigr]\,d_{1}^{(\textsc{i})}(\Delta)+\Bigl[d_{1}^{(\textsc{ii})}(t_{b},t_{a})\,d_{2}^{(\textsc{i})}(t_{a})+d_{2}^{(\textsc{ii})}(t_{b},t_{a})\,d_{1}^{(\textsc{i})}(t_{a})\Bigr]\,\dot{d}_{1}^{(\textsc{i})}(\Delta)\notag\\
	&=d_{1}(t_{b},0)\,d_{1}^{(\textsc{i})}(\Delta)+d_{2}(t_{b},0)\,\dot{d}_{1}^{(\textsc{i})}(\Delta)\,.
\end{align}
Similarly for $\mathsf{d}_{2}(t_{b}+\Delta,0)$, we have
\begin{equation}
	\mathsf{d}_{2}(t_{b}+\Delta,0)=d_{1}(t_{b},0)\,d_{2}^{(\textsc{i})}(\Delta)+d_{2}(t_{b},0)\,\dot{d}_{2}^{(\textsc{i})}(\Delta)\,.
\end{equation}
Now we plug these results into the expressions in \eqref{E:kbsdkskdu1}--\eqref{E:kbsdkskdu3} to find the corresponding squeeze parameter in-region I\!I\!I,
\begin{align}
	&\frac{1}{\omega_{f}\omega_{i}}\,\dot{\mathsf{d}}_{\bm{k}}^{(1)2}(t_{b}+\Delta)+\frac{\omega_{i}}{\omega_{f}}\,\dot{\mathsf{d}}_{\bm{k}}^{(2)2}(t_{b}+\Delta)+\frac{\omega_{f}}{\omega_{i}}\,\mathsf{d}_{\bm{k}}^{(1)2}(t_{b}+\Delta)+\omega_{f}\omega_{i}\,\mathsf{d}_{\bm{k}}^{(2)2}(t_{b}+\Delta)\notag\\
	&\qquad\qquad\qquad\qquad\qquad=\frac{1}{\omega_{f}\omega_{i}}\,\dot{d}_{\bm{k}}^{(1)2}(t_{b})+\frac{\omega_{i}}{\omega_{f}}\,\dot{d}_{\bm{k}}^{(2)2}(t_{b})+\frac{\omega_{f}}{\omega_{i}}\,d_{\bm{k}}^{(1)2}(t_{b})+\omega_{f}\omega_{i}\,d_{\bm{k}}^{(2)2}(t_{b})\,,\\
	&\frac{1}{\omega_{f}\omega_{i}}\,\dot{\mathsf{d}}_{\bm{k}}^{(1)2}(t_{b}+\Delta)+\frac{\omega_{i}}{\omega_{f}}\,\dot{\mathsf{d}}_{\bm{k}}^{(2)2}(t_{b}+\Delta)-\frac{\omega_{f}}{\omega_{i}}\,\mathsf{d}_{\bm{k}}^{(1)2}(t_{b}+\Delta)-\omega_{f}\omega_{i}\,\mathsf{d}_{\bm{k}}^{(2)2}(t_{b}+\Delta)\notag\\
	&\qquad\qquad\qquad\qquad\qquad=\frac{1}{\omega_{f}\omega_{i}}\,\dot{d}_{\bm{k}}^{(1)2}(t_{b})+\frac{\omega_{i}}{\omega_{f}}\,\dot{d}_{\bm{k}}^{(2)2}(t_{b})-\frac{\omega_{f}}{\omega_{i}}\,d_{\bm{k}}^{(1)2}(t_{b})-\omega_{f}\omega_{i}\,d_{\bm{k}}^{(2)2}(t_{b})\,,\\
	&\frac{1}{\omega_{i}}\,\mathsf{d}_{\bm{k}}^{(1)}(t_{b}+\Delta)\dot{\mathsf{d}}_{\bm{k}}^{(1)}(t_{b}+\Delta)+\omega_{i}\,\mathsf{d}_{\bm{k}}^{(2)}(t_{b}+\Delta)\dot{\mathsf{d}}_{\bm{k}}^{(2)}(t_{b}+\Delta)\notag\\
	&\qquad\qquad\qquad\qquad\qquad=\frac{1}{\omega_{i}}\,d_{\bm{k}}^{(1)}(t_{b})\dot{d}_{\bm{k}}^{(1)}(t_{b})+\omega_{i}\,d_{\bm{k}}^{(2)}(t_{b})\dot{d}_{\bm{k}}^{(2)}(t_{b})\,,
\end{align}
due to the facts that
\begin{align*}
	d_{1}^{(\textsc{i})}(\Delta)\,\dot{d}_{1}^{(\textsc{i})}(\Delta)&=-\omega_{i}^{2}d_{2}^{(\textsc{i})}(\Delta)\,\dot{d}_{2}^{(\textsc{i})}(\Delta)\,,&d_{1}^{(\textsc{i})2}(\Delta)+\omega_{i}^{2}d_{2}^{(\textsc{i})2}(\Delta)&=1\,,&\frac{1}{\omega_{i}^{2}}\,\dot{d}_{1}^{(\textsc{i})2}(\Delta)+\dot{d}_{2}^{(\textsc{i})2}(\Delta)&=1\,.
\end{align*}
The results for the shifted process are the same as those for un-shifted process. Thus both cases generate the same squeezing at $t\geq t_{b}+\Delta$.


\newpage

\begin{thebibliography}{99}


\bibitem{Unr76} 
	W. G. Unruh, \textit{Notes on black-hole evaporation}, Phys. Rev. D \textbf{14}, 870 (1976).

\bibitem{QRadVac} 
	J.-T. Hsiang, and B. L. Hu, \textit{Atom-field interaction: From vacuum fluctuations to quantum radiation and quantum dissipation or radiation reaction}, Physics \textbf{1}, 430 (2019).

\bibitem{DeW79}  
	B. S. DeWitt, \textsl{General Relativity: An Einstein Centenary Survey}, Edited by S. W. Hawking and W. Israel (Cambridge Press, Cambridge, UK, 1979). 

\bibitem{RadReact}
	J. R. Ackerhalt, P. L. Knight, and J. H. Eberly, \textit{Radiation reaction and radiative frequency shifts}, Phys. Rev. Lett. \textbf{30}, 456 (1973).

\bibitem{RadReact1}
	P. W. Milonni, and W. A. Smith, \textit{Radiation reaction and vacuum fluctuations in spontaneous emission}, Phys. Rev. A 1975, \textbf{11}, 814. [CrossRef]

\bibitem{RadReact2}
	J. Dalibard, J. Dupont-Roc, and C. Cohen-Tannodji, \textit{Vacuum fluctuations and radiation reaction: Identification of their respective contributions}, J. Phys. (Paris) \textbf{43}, 1617 (1982).

\bibitem{RadReact3}
	J. Dalibard, J. Dupont-Roc, and C. Cohen-Tannodji, \textit{Dynamics of a small system coupled to a reservoir: Reservoir fluctuations and self-reaction}, J. Phys. (Paris) \textbf{45}, 637 (1984). 

\bibitem{Jackson}
    J. D. Jackson, \textsl{Classical Electrodynamics, 2nd Ed.} (Wiley, New York, 1975).

\bibitem{Rohrich}
    F. Rohrlich, \textsl{Classical Charged Particles—Foundation of their Theories} (Westview, Colorado, 1990).

\bibitem{JH1} 
	P. R. Johnson, and B. L. Hu, \textit{Stochastic theory of relativistic particles moving in a quantum field: Scalar Abraham-Lorentz-Dirac-Langevin equation, radiation reaction, and vacuum fluctuations}, Phys. Rev. D \textbf{65}, 065015 (2002).

\bibitem{JHUnr}
 	P. R. Johnson, and B. L. Hu, \textit{Unruh effect in a uniformly accelerated charge: From quantum fluctuations to classical radiation}, Found. Phys. \textbf{35}, 1117 (2005).

\bibitem{QRadCoh} 
    J.-T. Hsiang, and B. L. Hu, \textit{Quantum radiation and dissipation in relation to classical radiation and radiation reaction}, Phys. Rev. D \textbf{106}, 045002 (2022).

%\bibitem{QRadCos}    

\bibitem{SqBooks} E.g., 
	P. D. Drummond and Z. Ficek, \textsl{Quantum Squeezing} (Springer-Verlag, Berlin, Heidelberg, 2004).

\bibitem{GriSid} 
	L. P. Grishchuk, and Y. V. Sidorov, \textit{Squeezed quantum states of relic gravitons and primordial density fluctuations}. Phys. Rev. D \textbf{42}, 3413 (1990).

\bibitem{HKM94} 
	B. L. Hu, G. Kang and A. Matacz,  \textit{Squeezed vacuua and the quantum statistics of cosmological particle creation}, Int. J. Mod. Phys. A\textbf{9}, 991 (1994).

\bibitem{StrFor} 
	S. Weinberg, \textsl{Cosmology} (Oxford University Press, 2008)

\bibitem{nMCos} 
	J. T. Hsiang, and B. L. Hu, \textit{NonMarkovianity in Cosmology: Memories kept in a quantum field}, Ann. Phys. \textbf{434}, 168656 (2021).

\bibitem{FDRSq} 
	J. T. Hsiang, and B. L. Hu, \textit{Fluctuation-dissipation relation for a quantum Brownian oscillator in a parametrically squeezed thermal field}, Ann. Phys. \textbf{433}, (2021).

\bibitem{Parker} 
	L. Parker, \textit{Quantized fields and particle creation in expanding universes. I}, Phys. Rev. \textbf{183}, 1057 (1969).

\bibitem{Zeldovich} 
	Ya B. Zel'dovich, \textit{Particle production in cosmology}, JETP Lett. \textbf{12}, 443 (1970).

\bibitem{AHH22}
	O. Ar{\i}soy, J. T. Hsiang, and B. L. Hu, \textit{Quantum-parametric-oscillator heat engines in squeezed thermal baths: foundational theoretical issues}, Phys. Rev. E \textbf{105}, 014108 (2022).
	
\bibitem{HCSH18}
	J. T. Hsiang, C. H. Chou, Y. Suba{\c s}{\i}, and B. L. Hu, \textit{Quantum thermodynamics from the nonequilibrium dynamics of open systems: Energy, heat capacity, and the third law}, Phys. Rev. E \textbf{97}, 012135 (2018).
	
\bibitem{HH22}
	J. T. Hsiang, and B. L. Hu, \textit{Non-Markovian Abraham-Lorenz-Dirac equation: radiation reaction without pathology}, Phys. Rev. D \textbf{106}, 125108 (2022).

\bibitem{GP85}
	A. H. Guth, and S.-Y. Pi, \textit{Quantum mechanics of the scalar field in the new inflationary universe}, Phys. Rev. D \textbf{32}, 1899 (1985).

\bibitem{HH22a}
	J. T. Hsiang, and B. L. Hu, \textit{No intrinsic decoherence of inflationary cosmological perturbations}, Universe \textbf{8}, 27 (2022).
	 
\bibitem{Sp00}
	H. Spohn, \textit{The critical manifold of the Lorentz–Dirac equation}, Europhys. Lett. \textbf{49}, 287 (2000).
\end{thebibliography}
\end{document}